\RequirePackage{pdf14}

\documentclass[a4paper,11pt]{article}
\pdfoutput=1

\usepackage{array,mathrsfs,amsfonts,amsmath,yfonts,dsfont,bbm,colonequals,amscd,overpic}
\usepackage{relsize,suffix,mathtools,cancel,bbm,tikz-cd}
\usepackage{subcaption}
\usepackage{multirow}

\usetikzlibrary{positioning}
\usetikzlibrary{chains}
\usetikzlibrary{arrows,fit,decorations.pathreplacing}
\tikzstyle{every picture}+=[remember picture]
\tikzstyle{na} = [baseline=-.5ex]

\graphicspath{ {Figures/} }

\usepackage{jheppub} 
\usepackage{enumerate}

\usepackage[T1]{fontenc} 
\usepackage{subcaption}

\newcommand{\be}{\begin{equation}}
\newcommand{\ee}{\end{equation}}

\newcommand{\bea}{\begin{eqnarray}}
\newcommand{\eea}{\end{eqnarray}}

\newcommand{\ba}{\begin{equation}\begin{aligned}}
\newcommand{\ea}{\end{aligned}\end{equation}}

\newcommand{\zb}{\bar{z}}
\newcommand{\wb}{\bar{w}}

\newcommand{\Disc}{\textup{Disc}}

\newcommand{\Df}{\Delta_{\phi}}
\newcommand{\cG}{\mathcal{G}}

\newcommand{\cO}{\mathcal{O}}
\newcommand{\dDisc}{\mathrm{dDisc}}

\colorlet{darkblue}{blue!70!black}
\colorlet{darkgreen}{green!70!black}
\colorlet{darkred}{red!70!black}

\abstract{
We develop an analytic approach to the four-point crossing equation in CFT, for general spacetime dimension. In a unitary CFT, the crossing equation  (for, say, the $s$- and $t$-channel expansions)
can be thought of as a vector equation 
in an infinite-dimensional space of complex analytic functions in two variables, which satisfy a boundedness condition at infinity. 
We identify a useful basis for this space of functions, consisting of  the set of s- {\it and} t-channel conformal blocks  of double-twist operators in mean field theory.
We describe two independent algorithms to construct the {\it dual} basis of linear functionals, and work out explicitly many examples. Our basis of functionals  appears to be closely related to the CFT dispersion relation recently derived by Carmi and Caron-Huot.
}

\title{\boldmath \LARGE
A Basis of Analytic Functionals for CFTs\\in General Dimension
 }

\author[a,b]{Dalimil Maz\'a\v{c},}
\author[b]{Leonardo Rastelli,}
\author[c]{Xinan Zhou}

\affiliation[a]{Simons Center for Geometry and Physics, Stony Brook University, \\Stony Brook, NY 11794, U.S.A.}
\affiliation[b]{C. N. Yang Institute for Theoretical Physics, Stony Brook University, \\Stony Brook, NY 11794, U.S.A.}
\affiliation[c]{Princeton Center for Theoretical Science, Princeton University, \\Princeton, NJ 08544, U.S.A.}

\emailAdd{dalimil.mazac@stonybrook.edu}
\emailAdd{leonardo.rastelli@stonybrook.edu}
\emailAdd{xinanz@princeton.edu}

\begin{document}
\maketitle
\flushbottom


\section{Introduction}\label{sec:Introduction}

A decade after its modern renaissance \cite{Rattazzi:2008pe}, the
conformal bootstrap program continues to undergo rapid development. 
The numerical bootstrap (see \cite{Poland:2018epd} for a recent review) has achieved extraordinary  sophistication, 
surpassing in precision all other theoretical methods to determine critical exponents.\footnote{See, e.g., \cite{O2-prep} for upcoming work on the world's most precise calculation of critical exponents in the three-dimensional $O(2)$ model.}
Our analytic understanding of the bootstrap equations is also rapidly improving. Many recent developments concern properties of CFTs in the Lorentzian regime \cite{Komargodski:2012ek,Fitzpatrick:2012yx,Alday:2015eya,Alday:2015ewa,Alday:2016njk,Caron-Huot:2017vep,Li:2017lmh,Costa:2017twz,Kravchuk:2018htv,Kologlu:2019bco}. The central new tool here is Caron-Huot's \cite{Caron-Huot:2017vep}
Lorentzian inversion formula (LIF). The LIF expresses the ``coefficient function'' $c (\Delta, J)$ of the conformal partial wave expansion of the four-point function ${\cal G}$ in a certain OPE channel in terms of the ``double-discontinuities'' of ${\cal G}$ around the singularities of the  other two channels.  One of the most important consequences of the formula is the fact that the CFT data are analytic in the spin variable $J$.\footnote{Another recent development on the analytic bootstrap front involves application of Tauberian theorems to constrain spectral and OPE asymptotics \cite{Qiao:2017xif,Mukhametzhanov:2018zja,Mukhametzhanov:2019pzy,Ganguly:2019ksp,Pal:2019yhz}.}
 
 A parallel analytic development has been the construction of exact bootstrap functionals \cite{Mazac:2016qev,Mazac:2018mdx,Mazac:2018ycv,Mazac:2018qmi,Kaviraj:2018tfd,Mazac:2018biw}. The highlight of this approach has been the derivation of exact bootstrap bounds for CFTs in one dimension \cite{Mazac:2016qev,Mazac:2018mdx,Mazac:2018ycv}. Remarkably, one can also reinterpret  the analytic results of \cite{Mazac:2016qev,Mazac:2018mdx,Mazac:2018ycv} in the context  of the modular bootstrap and of the sphere packing problem in Euclidean geometry \cite{Hartman:2019pcd}, leading for example to a CFT rederivation of the famous result of Viazovska \cite{Viazovska} that the $E_8$ lattice gives the optimal sphere packing in eight dimensions. 
 Work on analytic functionals has so far been limited to 
 situations where a single cross-ratio is involved, i.e. four-point functions one-dimensional CFT and two-point functions in {\it boundary} CFT in any dimension \cite{Kaviraj:2018tfd,Mazac:2018biw}.\footnote{See however the very recent paper \cite{Paulos:2019gtx}. We briefly comment on its relation with our approach in the Discussion section.} 
 
In this paper, we develop analytic functionals for four-point functions in a CFT in general dimension $d >1$. 
While this is {\it a priori} a more complicated setup, 
the presence of two independent cross-ratios will allow for more flexible complex-analytic manipulations,
and lead to somewhat simpler functionals than in the one-dimensional case. Despite these simplifications, our analysis will be quite  technical. In the rest of this introduction we summarize 
 the main logic of the paper.

We focus  for simplicity on four-point functions of non-necessarily identical scalar operators of equal conformal dimension $\Delta_\phi$. We write it as
\be
\langle\phi_1(x_1)\phi_2(x_2)\phi_3(x_3)\phi_4(x_4)\rangle = (|x_{13}||x_{24}|)^{-2\Df}\, ,
\mathcal{G}(z,\bar{z})\,,
\label{eq:4ptFunctionIntro}
\ee
with $z$ and $\bar z$  the usual cross ratios. The CFT data are constrained by crossing equations, which is another name for equality of independent OPEs in their common region of overlap. In this paper, we will focus on understanding the crossing equation relating the s- and t-channel OPE of \eqref{eq:4ptFunctionIntro}
\begin{equation} \label{OPE}
{\cal G}(z,\zb) =  \sum\limits_{\cO}f_{12\cO}f_{34\cO}G^{s}_{\Delta_{\cO},J_\mathcal{O}}(z,\bar{z})=
\sum\limits_{\mathcal{P}}f_{23\mathcal{P}}f_{41\mathcal{P}}G^{t}_{\Delta_{\mathcal{P},J_\mathcal{P}}}(z,\bar{z})\,.
\end{equation}

In a unitary CFT, this equation holds as a function of \emph{independent complex} variables $z$ and $\zb$ in a certain domain in $\mathbb{C}^2$. $\mathcal{G}(z,\bar{z})$ can be analytically continued to a symmetric function of $z$ and $\bar z$ which is complex analytic in both variables in this domain. Unitarity further implies that $\cG(z,\zb)$ is bounded by a constant away from $z,\zb=0,1$. Our result is most easily stated for a more restricted class of four-point functions, which satisfy a stronger boundedness condition as $z, \bar z \to \infty$, namely 
$|\mathcal{G}(z,\bar{z})| \lesssim |z|^{-\frac{1}{2} -\epsilon}| \bar z|^{-\frac{1}{2} -\epsilon}$. In this paper, we will collect evidence that such a  ``superbounded'' four-point function\footnote{In the terminology that we introduce below, symmetric functions $f (z, \bar z)$ with suitable analyticity properties belong to the function space  ${\cal V}$ if they are just bounded,  and to the space ${\cal U} \subset {\cal V}$ if they are superbounded.} 
can be expanded as
\ba \label{master}
\cG(z,\zb) &= \sum\limits_{n,\ell}
\left\{\alpha^{s}_{n,\ell}[\cG]G^{s}_{\Delta_{n,\ell},\ell}(z,\zb) +
\beta^{s}_{n,\ell}[\cG]\partial_{\Delta}G^{s}_{\Delta_{n,\ell},\ell}(z,\zb)\right\}+\\
&\phantom{,}+\sum\limits_{n,\ell}
\left\{\alpha^{t}_{n,\ell}[\cG]G^{t}_{\Delta_{n,\ell},\ell}(z,\zb) +
\beta^{t}_{n,\ell}[\cG]\partial_{\Delta}G^{t}_{\Delta_{n,\ell},\ell}(z,\zb)\right\}\,.
\ea
The sums run  over
all non-negative integers $n$ and $\ell$. $G^{s}_{\Delta,\ell}$ and $G^{t}_{\Delta,\ell}$  are the s- and t-channel conformal blocks for exchanged operator of dimension $\Delta$ and spin $\ell$, and $\Delta_{n,\ell}$ denotes  the  ``double-trace'' dimension
$
\Delta_{n,\ell} = 2\Df + 2n+\ell\,.
$
In other terms, we claim that  s- and t-channel double-trace blocks and their derivatives with respect to $\Delta$ form a basis for superbounded four-point functions.\footnote{While we do not provide a rigorous proof in this work, this claim has since been demonstrated in the more recent work \cite{Caron-Huot:2020adz} using the conformal dispersion relation of reference \cite{DispersionRelation} along the lines sketched in Section \ref{sec:DispersionRelation} of the present paper. We refer the reader to \cite{Caron-Huot:2020adz} for a pedagogical treatment. In particular, Appendix B therein shows how the decomposition \eqref{master} works for certain simple correlation functions.}

We have written the coefficients of the expansion in terms of the action of the {\it dual} basis on the four-point function ${\cal G}$. The dual basis consists of the linear
 functionals $\{ \alpha^{s}_{n, \ell} \, , \beta^{s}_{n, \ell},\alpha^{t}_{n, \ell} \, , \beta^{t}_{n, \ell}  \}$, see \eqref{eq:dualityC1}, \eqref{eq:dualityC2} for their defining properties. The reason to be interested in these functionals is that they imply powerful sum rules on the OPE data. Indeed, if $\omega$ is any of the dual basis functionals, we can apply it to the crossing equation\footnote{Here we are assuming $\cG(z,\zb)$ is super-bounded. More generally, when $\cG(z,\zb)$ is merely bounded, a little more work is needed to obtain valid sum rules from the dual basis functionals.}
 \eqref{OPE} and find
\be
 \sum\limits_{\cO}f_{12\cO}f_{34\cO}\,\omega[G^{s}_{\Delta_{\cO},J_\mathcal{O}}]=
\sum\limits_{\mathcal{P}}f_{23\mathcal{P}}f_{41\mathcal{P}}\,\omega[G^{t}_{\Delta_{\mathcal{P},J_\mathcal{P}}}]\,.
\ee
Such sum rules are particularly powerful in holographic CFTs because the dual basis functionals automatically suppress the contribution of double-trace operators, meaning that the sum rules directly constrain the single-trace data. However, we do not explore this idea in the present paper. Indeed, our main purpose here is to develop the formalism while leaving most interesting physical applications for future study.\footnote{Since the original publication of the present work on the arXiv, the functionals constructed here have been further explored in several articles. In particular, \cite{Caron-Huot:2020adz} studies the formal properties of these functionals in detail, while \cite{Caron-Huot:2021enk} discusses their applications in holographic CFTs.}
 
Our main technical achievement is a general algorithm for the explicit construction of the dual basis functionals, which we illustrate in detail in many low-lying examples. In fact, we describe two independent methods to obtain the functionals. In the first method, we represent the functionals as double contour integrals in $w$ and $\wb$, of the schematic form
\begin{equation}
\omega [ {\cal G} ] = \int_{C_-} \frac{dw}{2 \pi i}  \int_{C_+} \frac{d \bar w}{2 \pi i} \; \mathcal{H}(w, \bar w) {\cal G} (w, \bar w) \,.
\end{equation}
For a given choice of dual basis element, the kernel $\mathcal{H}(w, \bar w)$ is fixed by imposing the correct structure of  zeros on the double-trace conformal blocks. Consider for example $\beta^{s}_{ N , L }$.
For each integer $\ell$, we regard   $\beta^{s}_{N, L} [G^{s}_{\Delta, \ell}]$ as a function of real $\Delta$, and impose that it has double zeros at $\Delta = 2 \Df + 2n + \ell$, {except} 
for $n =N$ and $\ell = L$, where it must have a simple zero. We also impose that  $\beta^{s}_{N, L} [G^{t}_{\Delta, \ell}]$ has double zeros for all $\Delta = 2 \Df + 2n + \ell$. These conditions ensure that    $\beta^{s}_{N, L}$ is the functional dual to the primal basis vector $\partial_{\Delta}G^{s}_{\Delta_{N,L},L}$.  It turns out that an ansatz for $\mathcal{H}(w, \bar w)$ as a meromorphic function does the job.  The algorithm is general. We work out explicitly the whole infinite family of $\beta$ functionals with $N=0$ and general $L$, and several other low-lying examples of both $\alpha$ and $\beta$. The expressions for the kernels turn out to be surprisingly simple. 

The second method for obtaining the functionals extends to higher-dimensional CFTs the logic of the ``Polyakov bootstrap'', which has already been applied to the description of analytic functionals in the CFT$_1$ and BCFT cases.  A physical four-point function admits the usual convergent OPE expansion in either the s- or the t-channel of \eqref{OPE}. We will explain that a superbounded ${\cal G}$ admits an alternative  expansion, where we sum over {\it both} the s- and t-channel spectra, and with  OPE coefficients as in (\ref{OPE}),
\begin{equation} \label{PolyakovSum}
{\cal G} =  \sum\limits_{\cO}f_{12\cO}f_{34\cO} P^{s}_{\Delta_{\cO},J_\mathcal{O}}(z,\bar{z}) + 
\sum\limits_{\mathcal{P}}f_{23\mathcal{P}}f_{41\mathcal{P}}P^{t}_{\Delta_{\mathcal{P}},J_\mathcal{P}}(z,\bar{z})\,.
\end{equation}
Here $P^{s}_{\Delta,J}$ and $P^{t}_{\Delta,J}$ are what we call the the s- and t-channel ``Polyakov-Regge'' blocks, respectively.\footnote{We use the novel terminology   ``Polyakov-Regge'' blocks to distinguish
our functions from the ``Polyakov blocks'' discussed in recent literature, e.g., \cite{Gopakumar:2016wkt,Gopakumar:2016cpb,Gopakumar:2018xqi}, which are supposed to be 
completely crossing symmetric  (invariant under crossing of all three channels,  $s \leftrightarrow t \leftrightarrow u$). As explained in \cite{Mazac:2018qmi}, Polyakov blocks with spin $ J > 0$ simply {\it do not exist} in higher-dimensional CFTs,  if one insists on good Regge behavior. The Polyakov-Regge blocks introduced here circumvent this no-go theorem by violating full crossing symmetry while maintaining good Regge behaviour.}
 $P^{s}_{\Delta,J}$  is defined as the unique superbounded function
with the same $s$-channel double discontinuity as the conformal block $G^{s}_{\Delta,J}$, and vanishing $t$-channel double-discontinuity,
\be
\dDisc_s P^{s}_{\Delta,J} = \dDisc_s G^{s}_{\Delta,J} \, , \qquad \dDisc_t P^{s}_{\Delta,J} = 0 \,.
\ee
The t-channel Polyakov-Regge block $P^{t}_{\Delta,J}$ is defined in the obvious way, with s $\leftrightarrow$ t. 
Compatibility of  (\ref{OPE})  and (\ref{PolyakovSum}) implies
 the following s- and t-channel  expansions for $P^{s}_{\Delta,J}(z,\zb)$,
\bea
P^{s}_{\Delta,J}(z,\zb) & = & G^{s}_{\Delta,J}(z,\zb)-
\sum\limits_{n,\ell}\!\left\{\alpha^{s}_{n,\ell}[G^{s}_{\Delta,J}]G^{s}_{\Delta_{n,\ell},\ell}(z,\zb) +
\beta^{s}_{n,\ell}[G^{s}_{\Delta,J}]\partial_{\Delta}G^{s}_{\Delta_{n,\ell},\ell}(z,\zb)\right\}  \nonumber \\
& = & \sum\limits_{n,\ell}\left\{\alpha^{t}_{n,\ell}[G^{s}_{\Delta,J}]G^{t}_{\Delta_{n,\ell},\ell}(z,\zb) +
\beta^{t}_{n,\ell}[G^{s}_{\Delta,J}]\partial_{\Delta}G^{t}_{\Delta_{n,\ell},\ell}(z,\zb)\right\}\,.
\label{eq:polyakovSIntro}
\eea
In other terms,  the Polyakov-Regge blocks encode the action of our functionals on general s- and t-channel conformal block. 
These definitions may appear at first rather convoluted. What makes them useful is the fact that we can essentially identify  Polyakov-Regge blocks with exchange Witten diagrams. More precisely,
\be
P^{s}_{\Delta,J}(z,\zb) = A^{-1}\mathcal{W}^{s}_{\Delta,J}(z,\zb) + \mathcal{C}(z,\zb)\, .
\ee
Here $\mathcal{W}^{s}_{\Delta,J}$ is the s- or t-channel Witten diagram in $AdS_{d+1}$ with bulk-to-bulk propagator of quantum numbers $(\Delta, J)$, $A$ a normalization factor and $\mathcal{C}$ a ``Regge improvement'' term which consists of a finite sum of contact diagrams. The improvement term  is uniquely fixed by requiring  superboundedness of $P^{s}_{\Delta,J}$.  As there are well-developed techniques
to compute and expand Witten diagrams, this logic gives an efficient way to determine the explicit action of the functionals on conformal blocks.

We have so far restricted to the space ${\cal U}$ of superbounded functions. Physical four-point functions typically belong to the larger space ${\cal V} \supset {\cal U}$ of just bounded functions,
and so we are ultimately interested in constructing well-defined functionals on ${\cal V}$. Clearly  ${\cal V}^* \subset {\cal U}^*$. 
It is easy to see that the set (\ref{master}) of double-trace blocks and their derivatives, which we claim is  a basis for  ${\cal U}$, is overcomplete in  ${\cal V}$. 
There are linear relations arising from the existence of bounded contact Witten diagrams, which can be separately expanded in either s- {\it or} the t-channel double-trace blocks and their derivatives.  A  functional in ${\cal V}^*$ must annihilate such all such contact Witten diagrams.
 As it turns out, it is possible to obtain well-defined functionals  in  ${\cal V}^*$   by taking {\it finite} linear combinations of the basis  $\{ \alpha^{s, t}_{n, \ell} \, , \beta^{s, t}_{n, \ell}  \}$ of  ${\cal U}^*$. 

We must admit that some aspects of our story are still somewhat heuristic. In particular, we do not have a proof of completeness of the primal basis  (\ref{master}) of double-trace blocks and their derivatives in the space ${\cal U}$ of superbounded functions. In fact we lack a precise understanding of the {\it topology} that we should impose on ${\cal U}$ to give full mathematical justification to this statement. Nevertheless, we emphasize that while our derivation may be heuristic, the end product is a set of
 fully valid functionals. Indeed, we can directly check that their action
 commutes with the conformal block expansion of the four-point function in the sense emphasized in \cite{Rychkov:2017tpc}. Acting with our functionals on the crossing equation yields rigorous, non-perturbative sum rules. Making sense of the notion of completness of the resulting set of sum rules is an interesting, but secondary question.

Finally, there appears to be a rather close connection between our logic and the conformal dispersion relation recently discovered by Carmi and Caron-Huot \cite{DispersionRelation}. Their dispersion relation provides a natural decomposition of a four-point function $\cG$ into $\cG^{t}+\cG^{s}$ such that $\dDisc_s[\cG^{t}] = \dDisc_t[\cG^{s}]=0$. This is the same as our decomposition \eqref{master}, where the two terms correspond to the two curly brackets. From the practical point of view, the relationship yields another systematic way of obtaining all dual basis functionals, from the expansion of the inversion kernel of Carmi and Caron-Huot.

\medskip

The rest of the paper is organized as follows.  In Section \ref{sec:SingleVariable}, we use CFT$_1$ as a warm-up example to illustrate the main ideas of this paper. We present a double-trace basis for the single-cross ratio four-point function, and construct its dual basis. We emphasize the role of the dispersion relation and its connection to the analytic functionals. We begin the discussion of the higher-dimensional case in Section \ref{sec:Spaces}, where we delineate the space of correlation functions. We propose a basis of the function space and its dual basis in Section \ref{sec:bases}, and we also define and discuss the properties of the Polyakov-Regge blocks. In Section \ref{sec:ExplicitResults}, we give the first method to explicitly construct the functionals by using integration kernels. In Section \ref{ssec:witten}, we present the second method which obtains the functionals from the conformal block decomposition of the Polyakov-Regge blocks. We discuss the connection of our results to the dispersion relation of Carmi and Caron-Huot in Section \ref{sec:DispersionRelation}. Finally, we conclude in Section \ref{sec:Discussion} by outlining a few future directions.


\section{Warm-up: Single Variable}\label{sec:SingleVariable}

\subsection{The space of functions}
We will start by illustrating the basic logic of our paper on a simpler example. Consider the correlation function $\cG(z)$ of four conformal primaries in a 1D CFT
\be
\langle\phi_1(x_1)\phi_2(x_2)\phi_3(x_3)\phi_4(x_4)\rangle = (|x_{13}||x_{24}|)^{-2\Df}\,\mathcal{G}(z)\,.
\ee
Here $x_i$ are coordinates on a line and the cross-ratio $z$ is defined as follows
\be
z = \frac{x_{12}x_{34}}{x_{13}x_{24}}\,,
\ee
where $x_{ij} = x_i-x_j$. For simplicity, we will take all four external operators $\phi_i(x)$ to have the same scaling dimension $\Df$, but we do not in general assume that they are  identical operators. The configurations with ordering $x_1<x_2<x_3<x_4$ map to $z\in(0,1)$. In unitary theories, the correlation function $\cG(z)$ can be analytically continued from $z\in(0,1)$ to the complex plane and this continuation is holomorphic away from branch points at $z=0$ and $z=1$. From now on, $\cG(z)$ will refer to this holomorphic function, with branch cuts lying at $(-\infty,0]$ and $[1,\infty)$. $\cG(z)$ can be expanded using the s-channel and t-channel OPEs
\be
\cG(z) = \sum\limits_{\cO}f_{12\cO}f_{34\cO}G^{s}_{\Delta_{\cO}}(z)=
\sum\limits_{\mathcal{P}}f_{23\mathcal{P}}f_{41\mathcal{P}}G^{t}_{\Delta_{\mathcal{P}}}(z)\,,
\ee
where $G^{s}_{\Delta}(z)$ and $G^{t}_{\Delta}(z)$ are the s-channel and t-channel $sl(2)$ blocks
\ba
G^{s}_{\Delta}(z) &= z^{\Delta-2\Df}{}_2F_1(\Delta,\Delta;2\Delta;z)\\
G^{t}_{\Delta}(z) &= (1-z)^{\Delta-2\Df}{}_2F_1(\Delta,3\Delta;2\Delta;1-z)\,.
\ea
A standard argument shows that that in unitary theories $\cG(z)$ is bounded as $|z|\rightarrow \infty$. The argument goes as follows. For all $z\in\mathcal{R} = \mathbb{C}\backslash((-\infty,0]\cup[1,\infty))$, we have
\ba
|\cG(z)| &\leq \sum\limits_{\cO}|f_{12\cO}||f_{34\cO}||G^{s}_{\Delta_{\cO}}(z)| \leq\\
&\leq \sqrt{\sum\limits_{\cO}|f_{12\cO}|^2|G^{s}_{\Delta_{\cO}}(z)|}\sqrt{\sum\limits_{\cO}|f_{34\cO}|^2|G^{s}_{\Delta_{\cO}}(z)|}\,,
\ea
where the first inequality is a consequence of the convergent OPE and the second inequality is Cauchy-Schwarz. So it is enough to show that the arguments of the square roots are both bounded as $|z|\rightarrow\infty$. For $z\in(0,1)$, the arguments are equal to physical four-point functions corresponding to $\langle\phi_1\phi_1\phi_2\phi_2\rangle$ and $\langle\phi_3\phi_3\phi_4\phi_4\rangle$. As $z\rightarrow 1$, these correlators are bounded by the contribution of identity in the t-channel, i.e. by $(1-z)^{-2\Df}$. We can relate the limit $|z|\rightarrow \infty$ to $z\rightarrow 1$ by switching to the $\rho$ variable \cite{Hogervorst:2013sma}
\be
\rho(z) = \frac{z}{\left(1+\sqrt{1-z}\right)^2}\,,
\label{eq:rho}
\ee
which maps $\mathbb{C}\backslash[1,\infty)$ to the open unit disk. $z\rightarrow1$ maps to $\rho(z)\rightarrow1$ and $z\rightarrow\pm i\infty$ to $\rho(z)\rightarrow -1$. The s-channel conformal blocks $z^{2\Df}G^{s}_{\Delta}(z)$ have an expansion into powers of $\rho(z)$ with positive coefficients. It follows that we can use the bound as $z\rightarrow1$ to bound the behaviour as $|z|\rightarrow\infty$. The result is that both square roots in the above inequality are bounded by a constant in this limit, which completes the argument.

This leads us to define a vector space $\mathcal{V}_1$ consisting of {\it all complex functions which are holomorphic in $\mathcal{R}$ and which are bounded by a constant at infinity.} We have explained that all four-point functions in unitary theories are inside $\mathcal{V}_1$. Our goal is to find a useful basis for $\mathcal{V}_1$.

\subsection{Dispersion relation and a function basis}

\begin{figure}[htbp]
\begin{center}
\includegraphics[width=0.5\linewidth]{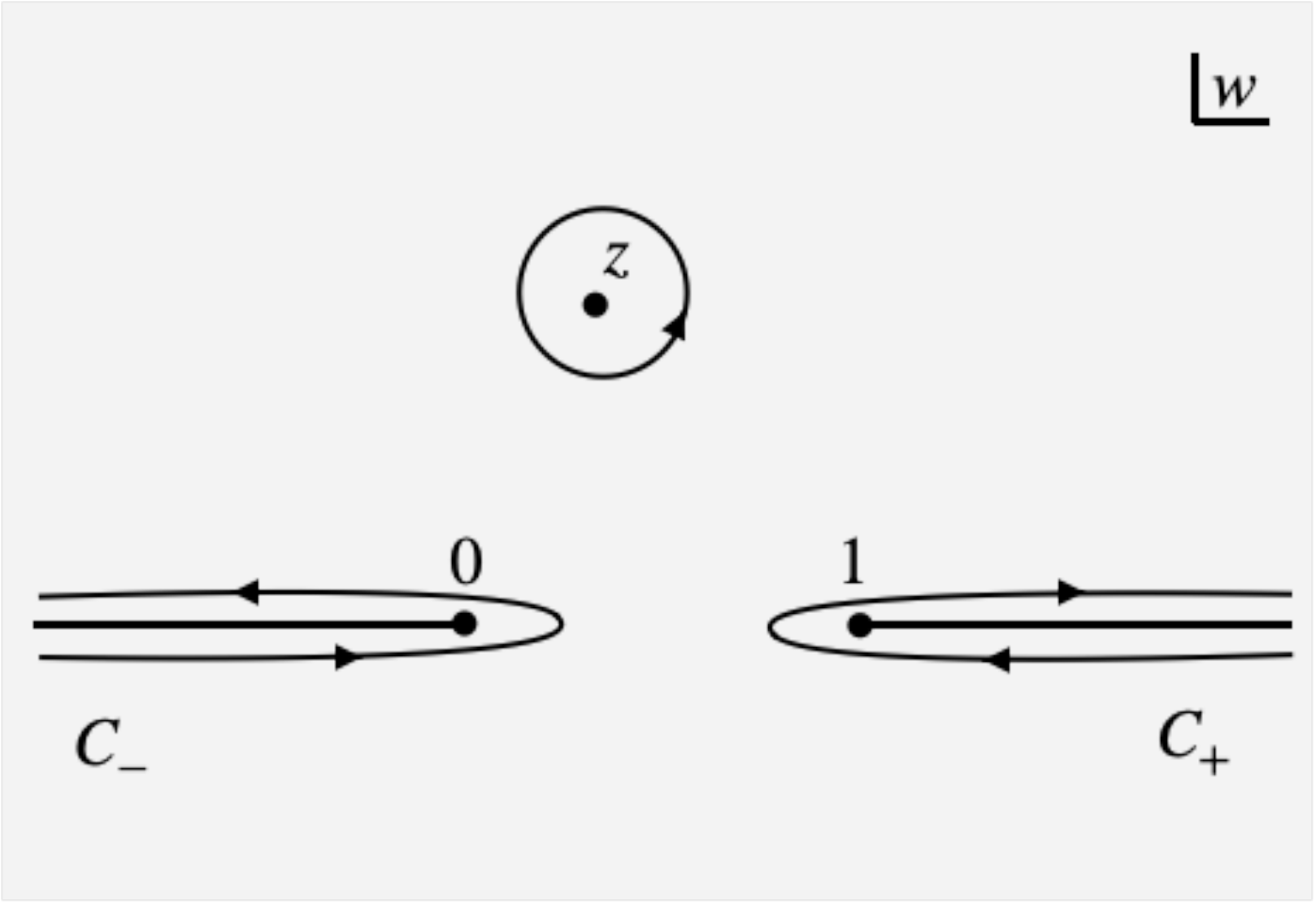}
\caption{An illustration of the contours $C_\pm$.}
\label{contourCpm}
\end{center}
\end{figure}

The first step in doing so is to write a dispersion relation for $\cG(z)$. We start from Cauchy's integral formula
\be
\cG(z) = \oint\frac{dw}{2\pi i}\frac{\cG(w)}{w-z}\,,
\ee
where the contour encircles the point $w=z$. The dispersion relation is obtained by deforming the contour so that it wraps the two branch cuts. In order to be able to drop the contribution from infinity, we need to assume that $\cG(z)$ decays at infinity. Thus let us define $\mathcal{U}_1$ to be the space consisting of {\it functions $\cG(z)\in\mathcal{V}_1$  which additionally satisfy $\cG(z) = O(|z|^{-\epsilon})$ as $|z|\rightarrow\infty$}, for some $\epsilon>0$. Note that there is generally no reason for physical four-point functions to satisfy this more stringent condition.\footnote{See \cite{Caron-Huot:2020adz} for examples where the condition is satisfied, as well as for a discussion of subtractions, which are needed to proceed when the condition is not satisfied.} Assuming $\cG(z)\in\mathcal{U}_1$, we find the dispersion relation
\be
\cG(z) = \cG^{t}(z) + \cG^{s}(z)\,,
\label{eq:dispersion1}
\ee
where
\be
\cG^{t}(z) = \int\limits_{C_+}\frac{dw}{2\pi i}\frac{\cG(w)}{w-z}\,,\qquad
\cG^{s}(z) = -\int\limits_{C_-}\frac{dw}{2\pi i}\frac{\cG(w)}{w-z}\,.
\label{eq:dispersion2}
\ee
Here $C_+$ and $C_-$ are contours wrapping the two branch cuts as shown in Figure \ref{contourCpm}. $C_+$ passes in between $z$ and the right branch cut and $C_-$ between $z$ and the left branch cut. We can rewrite the contour integrals as integrals over the discontinuities of $\cG(z)$
\be
\cG^{t}(z) = \int\limits_{1}^{\infty}\frac{dw}{2\pi i}\frac{\Disc_t[\cG(w)]}{w-z}\,,\qquad
\cG^{s}(z) = \int\limits_{-\infty}^{0}\frac{dw}{2\pi i}\frac{\Disc_s[\cG(w)]}{w-z}\,,
\label{eq:dispersion3}
\ee
where
\ba
\Disc_{s}[\cG(w)] &= \cG(w+i0^+)-\cG(w-i0^+)\quad\textrm{for }w\in(-\infty,0)\\
\Disc_{t}[\cG(w)] &= \cG(w+i0^+)-\cG(w-i0^+)\quad\textrm{for }w\in(1,\infty)\,.
\ea
The subscript $s$ or $t$ merely emphasizes around which OPE singularity is the discontinuity taken. In the cases where the integrals \eqref{eq:dispersion3} do not converge (such as when $\cG(z)$ has a strong enough singularity at $z=0$ or $z=1$), we need to use \eqref{eq:dispersion2}.

We can see from \eqref{eq:dispersion2} that $\cG^{t}(z)$ is holomorphic away from $z\in[1,\infty)$ and $\cG^{s}(z)$ is holomorphic away from $z\in(-\infty,0]$. Equivalently, we can express this as
\be
\Disc_s[\cG^{t}(z)] = 0\,,\qquad \Disc_t[\cG^{s}(z)] = 0\,.
\ee
This suggests that $\cG^{t}(z)$ can be expanded in s-channel conformal blocks with double-trace scaling dimensions $\Delta_n = 2\Df+n$, where $n=0,1,\ldots$. Indeed, it is precisely for these values of $\Delta$ that $G^{s}_{\Delta}(z)$ is holomorphic at $z=0$. To see the expansion more explicitly, let us first expand the integrand of $\cG^{t}(z)$ in \eqref{eq:dispersion2} around $z=0$ in the double-trace blocks
\be
\frac{1}{w-z} = \sum\limits_{n=0}^{\infty}H_n(w) G^{s}_{\Delta_n}(z)\,.
\label{eq:exp1}
\ee
To determine $H_n(w)$, we use the following duality relation for conformal blocks
\be
\oint\limits_{|z|=\epsilon}\frac{dz}{2\pi i}z^{-2}k_{x+n}(z)k_{1-x-m}(z) = \delta_{nm}\,,
\label{eq:dualityk}
\ee
where $n,m\in\mathbb{N}$ and
\be
k_h(z) = z^h{}_2F_1(h,h;2h;z)\,.
\ee
Hence
\be
H_n(w) = \oint\limits_{|z|=\epsilon}\frac{dz}{2\pi i}\frac{z^{2\Df-2}}{w-z}k_{1-2\Df-n}(z)\,.
\ee
From here, it is not hard to determine
\be
H_n(w) = 
\frac{(-1)^n (2\Df)^2_n}{n!(4\Df+n-1)_n}
w^{-1}{}_3F_2\left(1,-n,4\Df+n-1;2\Df,2\Df;w^{-1}\right)\,.
\label{eq:hExplicit}
\ee
When does the expansion \eqref{eq:exp1} converge? One can check that at large $n$, the terms in the series go as a constant times
\be
\left[\frac{\rho(z)}{\rho(w)}\right]^n\,,
\ee
where $\rho(z)$ is defined in \eqref{eq:rho}. Thefore, for any $z\in\mathbb{C}\backslash[1,\infty)$, we can place the contour $C_+$ in a region where $|\rho(w)|>|\rho(z)|$ so that the sum \eqref{eq:exp1} converges uniformly in $w$ and we can interchange it with the integral. We find
\be
\cG^{t}(z) = \sum\limits_{n=0}^{\infty} a^{s}_n G^{s}_{\Delta_n}(z)\,,
\ee
where
\be
a^{s}_n = \int\limits_{C_+}\frac{dw}{2\pi i} H_n(w) \cG(w)\,.
\label{eq:omegaS1D}
\ee
This shows that $\cG^{t}(z)$ can be expanded in the s-channel double-trace conformal blocks. Similarly, we can write
\be
\frac{1}{z-w} = -\frac{1}{(1-z)-(1-w)} = 
-\sum\limits_{n=0}^{\infty}H_n(1-w) G^{t}_{\Delta_n}(z)
\ee
to get the expansion
\be
\cG^{s}(z) = \sum\limits_{n=0}^{\infty} a^{t}_n G^{t}_{\Delta_n}(z)\,,
\ee
where
\be
a^{t}_n =-\int\limits_{C_-}\frac{dw}{2\pi i} H_n(1-w) \cG(w)\,.
\label{eq:omegaT1D}
\ee
Note that it follows from \eqref{eq:dispersion2} that if $\cG(z)\in\mathcal{U}_1$, then also $\cG^{t}(z), \cG^{s}(z)\in\mathcal{U}_1$.

We have explained that any function $\cG(z)\in\mathcal{U}_1$ can be written as a sum of two infinite sums as follows
\be
\cG(z) = \sum\limits_{n=0}^{\infty}a^{s}_nG^{s}_{\Delta_n}(z) + \sum\limits_{n=0}^{\infty}a^{t}_nG^{t}_{\Delta_n}(z)
\label{eq:expansion1}
\ee
for an appropriate choice of coefficients $a^{s}_n$ and $a^{t}_n$, where each sum is in $\mathcal{U}_1$. Clearly, the procedure fixes the coefficients uniquely. It is instructive to prove the uniqueness of $a^{s}_n$ and $a^{t}_n$ in a different way. Suppose
\be
\sum\limits_{n=0}^{\infty}a^{s}_nG^{s}_{\Delta_n}(z) + \sum\limits_{n=0}^{\infty}a^{t}_nG^{t}_{\Delta_n}(z) =
\sum\limits_{n=0}^{\infty}b^{s}_nG^{s}_{\Delta_n}(z) + \sum\limits_{n=0}^{\infty}b^{t}_nG^{t}_{\Delta_n}(z)\,.
\ee
Moving all the s-channel blocks to the LHS and all t-channel blocks to the RHS, we get
\be
\sum\limits_{n=0}^{\infty}a^{s}_nG^{s}_{\Delta_n}(z)-\sum\limits_{n=0}^{\infty}b^{s}_nG^{s}_{\Delta_n}(z) = \sum\limits_{n=0}^{\infty}b^{t}_nG^{t}_{\Delta_n}(z) - \sum\limits_{n=0}^{\infty}a^{t}_nG^{t}_{\Delta_n}(z)\,.
\ee
The LHS is holomorphic away from $[1,\infty)$ and the RHS is holomorphic away from $(-\infty,0]$. Therefore, either side is an entire function. Since this function is bounded by $|z|^{-\epsilon}$ as $|z|\rightarrow\infty$, it must vanish identically. In other words
\be
a^{s}_n = b^{s}_n\,,\qquad a^{t}_n = b^{t}_n\,.
\ee
In summary, the set $G^{s}_{\Delta_n}(z)$, $G^{t}_{\Delta_n}(z)$ with $n=0,1,\ldots$ forms a basis for the function space $\mathcal{U}_1$.

When working with infinite-dimensional vector spaces such as $\mathcal{U}_1$, we need to be careful with defining the notion of a basis. The most direct analogue of a basis of finite-dimensional vector space is the so-called Hamel basis. This is a linearly-independent set of vectors such that every element of the vector space can be written as a \emph{finite} linear combination of the basis vectors. The set of all $G^{s}_{\Delta_n}(z)$, $G^{t}_{\Delta_n}(z)$ do \emph{not} form a Hamel basis for $\mathcal{U}_1$ because there are elements of $\mathcal{U}_1$ which can only be written as infinite sums of $G^{s}_{\Delta_n}(z)$ and $G^{t}_{\Delta_n}(z)$. An example is $\cG(z) = 1/z^{2\Df}$, which is an infinite sum over all $G^{t}_{\Delta_n}(z)$. It is clearly useful to define another notion of a basis, where we are allowed to take infinite linear combinations of the basis vectors. However, in order to make sense of infinite sums of vectors, we first need to equip the vector space with a topology, since without a topology there is no sense in which an infinite sequence of vectors converges. Given a topology, we define a Schauder basis as a linearly independent set such that every vector is a convergent sum of vectors in the basis. What we have shown in this subsection is that $G^{s}_{\Delta_n}(z)$, $G^{t}_{\Delta_n}(z)$ form a Schauder basis for $\mathcal{U}_1$, where we take the topology on $\mathcal{U}_1$ to be such that the notion of convergence coincides with uniform convergence on compact subsets of the doubly-cut complex plane of Figure \ref{contourCpm}.

\subsection{The dual basis: the functionals}

We will be interested in the dual basis. The dual basis consists of linear functionals $\omega^{s}_n$, $\omega^{t}_n$ belonging to the continuous dual space $\mathcal{U}_1^*$ and satisfying
\be
\begin{array}{l l}
\omega^{s}_{n}[G^{s}_{\Delta_m}] = \delta_{nm} &\quad\omega^{s}_{n}[G^{t}_{\Delta_m}] = 0\vspace{.1in}\\
\omega^{t}_{n}[G^{s}_{\Delta_m}] = 0 &\quad\omega^{t}_{n}[G^{t}_{\Delta_m}] = \delta_{mn}\,.
\end{array}
\ee
This means we can use $\omega^{s}_n$ and $\omega^{t}_n$ to extract the coefficients $a^{s}_n$ and $a^{t}_n$ in \eqref{eq:expansion1} and write
\be
\cG(z) = \sum\limits_{n=0}^{\infty}\omega_n^{s}[\cG(z)]G^{s}_{\Delta_n}(z) + \sum\limits_{n=0}^{\infty}\omega_n^{t}[\cG(z)]G^{t}_{\Delta_n}(z)\,.
\label{eq:expansion2}
\ee
Looking at \eqref{eq:omegaS1D} and \eqref{eq:omegaT1D}, we find the following explicit formulas for the action of $\omega_n^{s}$ and $\omega_n^{t}$
\ba
\omega_n^{s}[\cG(z)] &= \!\!\int\limits_{\frac{1}{2}-i\infty}^{\frac{1}{2}+i\infty}\!\!\frac{dz}{2\pi i}\,H_n(z) \cG(z)\\
\omega_n^{t}[\cG(z)] &= \!\!\int\limits_{\frac{1}{2}-i\infty}^{\frac{1}{2}+i\infty}\!\!\frac{dz}{2\pi i}\,H_n(1-z) \cG(z) = \omega_n^{s}[\cG(1-z)]\,,
\label{eq:omegaBoth1D}
\ea
with $H_n(z)$ given by \eqref{eq:hExplicit}, where we deformed the contours to a more symmetric configuration. Since $H_n(z)$ is meromorphic with a pole only at $z=0$, it is obvious that $\omega_n^{s}$ indeed annihilates all $G^{t}_{\Delta_m}(z)$ and $\omega_n^{t}$ annihilates all $G^{s}_{\Delta_m}(z)$. Indeed, in these cases one can always deform the entire contour away to infinity.

Let us consider a very simple application of these functionals. We would like to find all four-point functions $\mathcal{G}(z)\in\mathcal{V}_1$ which satisfy crossing symmetry $\cG(z)=\cG(1-z)$ and whose OPE contains only the identity and double-trace conformal blocks
\be
\cG(z) = G^{s}_0(z) + \sum\limits_{n=0}^{\infty}a_n G^{s}_{\Delta_n}(z) = 
G^{t}_0(z) + \sum\limits_{n=0}^{\infty}a_n G^{t}_{\Delta_n}(z)\,.
\label{eq:crossing1D}
\ee
For simplicity, let us first assume that $\cG(z)\in\mathcal{U}_1$. Then we can apply $\omega^{s}_n$ to \eqref{eq:crossing1D} to learn
\be
a_n = \omega^{s}_n[G^{t}_0(z) - G^{s}_0(z)]\,.
\ee
In other words, the solution in $\mathcal{U}_1$ is unique. It is not hard to evaluate $a_n$ explicitly starting from \eqref{eq:hExplicit}
\be
a_n =
\frac{(2\Df)^2_n}{n! (4\Df+n-1)_n}\,.
\ee
This corresponds to
\be
\cG(z) = z^{-2\Df}+(1-z)^{-2\Df}\,,
\ee
which is just the four-point function $\langle\phi\bar{\phi}\phi\bar{\phi}\rangle$ in the mean field theory of a complex field $\phi$. We see that for $\Df>0$, we have indeed $\cG(z)\in\mathcal{U}_1$. To solve the same problem in $\mathcal{V}_1$, we need to face the fact that the basis functionals $\omega^{s}_n$ do not belong to $\mathcal{V}_1^*$. The reason is that the kernel $H_n(z)$ goes as $z^{-1}$ at large $z$. The integral in \eqref{eq:omegaS1D} therefore does not converge when acting on general functions in $\mathcal{V}_1$ and furthermore $\omega^{s}_n$ may not always be exchanged with infinite sums of functions converging to a function in $\mathcal{V}_1$. To construct functionals in $\mathcal{V}_1^*$, we need to take linear combinations of basis functionals $\omega^{s}_n$ such that their kernel is $O(z^{-2})$ as $z\rightarrow\infty$. Looking at \eqref{eq:hExplicit}, it is easy to see the following functionals do the job
\be
\widetilde{\omega}^{s}_n = \omega^{s}_n - 
\frac{(-1)^n (2\Df)^2_n}{n!(4\Df+n-1)_n}\omega^{s}_0\,.
\label{eq:subtraction1}
\ee
Applying $\widetilde{\omega}^{s}_n$ to \eqref{eq:crossing1D}, we learn the most general solution in $\mathcal{V}_1$ satisfies
\be
a_n - \frac{(-1)^n (2\Df)^2_n}{n!(4\Df+n-1)_n} a_0 = [1-(-1)^n]\frac{(2\Df)^2_n}{n! (4\Df+n-1)_n}\,.
\ee
This equation has a one-parameter family of solutions
\be
a_n = [1+(-1)^n\lambda]\frac{(2\Df)^2_n}{n! (4\Df+n-1)_n}\,,
\ee
which corresponds to the one-parameter family of four-point functions
\be
\cG(z) = z^{-2\Df}+(1-z)^{-2\Df}+\lambda\,.
\ee
For $\lambda=1$, we find the four-point function $\langle\phi\phi\phi\phi\rangle$ in the mean field theory of a real field $\phi$. Note that we could use any $\omega^{s}_{k}$ with $k\neq n$ instead of $\omega^{s}_{0}$ to perform the subtraction in \eqref{eq:subtraction1}. This would lead to an identical conclusion about the space of solutions.

Readers who are familiar with the construction of 1D functionals in \cite{Mazac:2018ycv} may notice that the above story is slightly different. The basis of \cite{Mazac:2018ycv} contains the conformal blocks as well as their derivatives, but with the conformal dimensions $\Delta_n$ restricted to $n$ even or odd. By contrast, the basis in this section does not involve the derivative of conformal blocks, and $n$ takes all values of $ \mathbb{N}$ for $\Delta_n$. The cardinality of the two basis is the same, as the derivative conformal blocks are traded for conformal blocks whose dimensions have the opposite parity. This comment however can be safely ignored for readers who have not been exposed to the previous 1D work.


\section{The Function Space of Four-Point Correlators}\label{sec:Spaces}
\subsection{Preliminaries}
We will now move on to the actual goal of this paper, which is the construction of a double-trace primal and dual basis for conformal four-point functions in $d\geq 2$. Let us consider the four-point function of scalar primary operators $\phi_i$ (with $i=1,\ldots,4$) in a $d$-dimensional unitary CFT. As in the previous section, we will assume the $\phi_i$s have the same scaling dimension $\Df$ for simplicity. Later, we will specialize further to the case when all $\phi_i$ are identical. The four-point function takes the form
\be
\langle\phi_1(x_1)\phi_2(x_2)\phi_3(x_3)\phi_4(x_4)\rangle = (|x_{13}||x_{24}|)^{-2\Df}\,
\mathcal{G}(z,\bar{z})\,,
\label{eq:4ptFunction}
\ee
where $x_i\in\mathbb{R}^{d}$ and $z$, $\zb$ are the cross-ratios
\be
z \zb = \frac{x^2_{12}x^2_{34}}{x^2_{13}x^2_{24}}\qquad
(1-z)(1-\zb) = \frac{x^2_{14}x^2_{23}}{x^2_{13}x^2_{24}}\,,
\ee
where $x_{ij} = x_i-x_j$. In the Euclidean signature, $z$ and $\zb$ are complex conjugate. We will assume $\cG(z,\zb)=\cG(\zb,z)$, which is automatic in $d\geq3$ and requires symmetry under parity in $d=2$. $\cG(z,\zb)$ can be expanded in the s- and t-channel OPE as follows
\be
\cG(z,\zb) = \sum\limits_{\cO}f_{12\cO}f_{34\cO}\,G^{s}_{\Delta_{\cO},J_{\cO}}(z,\zb)=
\sum\limits_{\mathcal{P}}f_{23\mathcal{P}}f_{41\mathcal{P}}\,G^{t}_{\Delta_{\mathcal{P}},J_{\mathcal{P}}}(z,\zb)\,,
\ee
where $\cO$ and $\mathcal{P}$ are conformal primaries appearing respectively in the s- and t-channel OPE. $G^{s}_{\Delta,J}(z,\zb)$ and $G^{t}_{\Delta,J}(z,\zb)$ are the s- and t-channel conformal block for a primary of dimension $\Delta$ and spin $J$. The two are related by crossing
\be
G^{t}_{\Delta,J}(z,\zb) = G^{s}_{\Delta,J}(1-z,1-\zb)\,.
\ee
Note that as a result of the convention \eqref{eq:4ptFunction}, our s-channel conformal blocks include the slightly unconventional prefactor $(z\zb)^{-\Df}$. The conformal blocks are normalized as follows
\be
G^{s}_{\Delta,J}(z,\zb) \sim z^{\frac{\Delta-J}{2}-\Df}\zb^{\frac{\Delta+J}{2}-\Df}\quad\textrm{for}\quad 0<z\ll\zb\ll1\,.
\ee
If $\phi_1 = \phi_2$ or $\phi_3=\phi_4$, then only primaries with $J$ even appear in the s-channel OPE and the four-point function satisfies
\be
\cG(z,\zb) =\left[(1-z)(1-\zb)\right]^{-\Df}\cG\left(\mbox{$\frac{z}{z-1}$},\mbox{$\frac{\zb}{\zb-1}$}\right)\,.
\ee
Similarly, if $\phi_1 = \phi_4$ or $\phi_2=\phi_3$, then only even spin appears in the t-channel and we have
\be
\cG(z,\zb) =\left(z\zb\right)^{-\Df}\cG\left(\mbox{$\frac{1}{z}$},\mbox{$\frac{1}{\zb}$}\right)\,.
\ee
Finally, if $\phi_1 = \phi_3$ or $\phi_2=\phi_4$, then only even spin appears in the u-channel and we have
\be
\cG(z,\zb) = \cG\left(1-z,1-\zb\right)\,.
\ee

Next, let us discuss the analytic continuation of $\cG(z,\zb)$ to independent complex $z$ and $\zb$. A standard argument involving positivity of expansion of $G_{\Delta,J}(z,\zb)$ in powers of the $\rho$ coordinate \eqref{eq:rho} shows \cite{Hartman:2015lfa}  that in unitary theories $\cG(z,\zb)$ can be analytically continued to a function holomorphic for $z,\zb\in\mathcal{R} = \mathbb{C}\backslash((-\infty,0]\cup[1,\infty))$.\footnote{A function of two complex variables is holomorphic if it is holomorphic as a function of each variable with the other variable fixed.} 
Note that this is the domain where both the s- and t-channel OPEs converge -- our analysis singles out these two channels.
The same argument also shows that this function is bounded by a constant as $z,\zb\rightarrow \pm i\infty$. To see this in another way, note that the limit $z\rightarrow i\infty$, $\zb\rightarrow -i\infty$ is controlled by the u-channel OPE and thus by the dimension of the lightest primary in the u-channel OPE. In the worst-case scenario, this primary is the identity operator, which leads to $\cG(z,\zb) \sim 1$ as $z\rightarrow i\infty$, $\zb\rightarrow -i\infty$. The same asymptotics holds whenever the u-channel OPE converges, e.g. when $z\rightarrow i\infty$ and $\zb$ is fixed in the lower half-plane. When $z,\zb$ both lie in the upper (or lower) half-plane, the u-channel OPE does not converge and unitarity is needed to show that $\cG(z,\zb)$ stays bounded. Note that the limit $z,\zb\rightarrow i\infty$ with $z/\zb$ fixed is the u-channel Regge limit. Boundedness in this limit is just the statement that the u-channel Regge intercept $J^{(u)}_0$ is less than or equal to one. More generally, if $z=r e^{i\theta_1}$ and $\zb=r e^{i\theta_2}$ with $\theta_{1,2}\in(0,\pi)$, then $\cG(z,\zb)\sim f(\theta_1-\theta_2)\,r^{J^{(u)}_0-1}$ as $r\rightarrow\infty$ with $\theta_{1,2}$ fixed.
 See section 2.1 of reference \cite{Mazac:2018ycv} for a more detailed discussion of the various kinematical limits occuring for $(z,\zb)\in\mathcal{R}\times\mathcal{R}$.

\subsection{Spaces $\mathcal{U}$ and $\mathcal{V}$}
These observations lead us to the following definition. Let $\mathcal{V}$ be the space of holomorphic functions $\mathcal{R}\times\mathcal{R}\rightarrow\mathbb{C}$ which are symmetric in the two variables and bounded by a constant away from $z,\zb=0,1$. More precisely $\cG(z,\zb)\in\mathcal{V}$ if $\cG(z,\zb)=\cG(\zb,z)$ and for every $\epsilon>0$ there exists $A>0$ such that for all $(z,\zb)\in\mathcal{R}\times\mathcal{R}$ satisfying $|z|>\epsilon$, $|\zb|>\epsilon$, $|1-z|>\epsilon$ and $|1-\zb|>\epsilon$, we have $|\cG(z,\zb)|<A$. We have explained that if $\cG(z,\zb)$ is a four-point function of the form \eqref{eq:4ptFunction} in a unitary theory, then $\cG(z,\zb)\in\mathcal{V}$. $\mathcal{V}$ is analogous to the space $\mathcal{V}_1$ from the previous section.

Note that physical four-point functions are single-valued in the Euclidean signature. This is an additional constraint on the monodromy of the function $\cG(z,\zb)$ around $(z,\zb)=(0,0)$ and $(z,\zb)=(1,1)$. Specifically, $\cG(z\circlearrowright 0,\zb\circlearrowleft 0) = \cG(z,\zb)$ and $\cG(z\circlearrowright 1,\zb\circlearrowleft 1) = \cG(z,\zb)$, where $\cG(z\circlearrowright w,\zb\circlearrowleft \wb)$ denotes $\cG(z,\zb)$ after an analytic continuation around $(w,\wb)$ by angles $2\pi$ in the directions shown. However, we do not impose these monodromy constraints on functions in $\mathcal{V}$. This ensures, among other things, that individual s- and t-channel conformal blocks belong to $\mathcal{V}$. Indeed $G^{s}_{\Delta,J}(z,\zb)$ and $G^{t}_{\Delta,J}(z,\zb)$ are holomorphic in $\mathcal{R}\times\mathcal{R}$ and are bounded as $z,\zb\rightarrow\infty$ as long as $\Df>\frac{d-2}{4}$, which is true for all $\Df$ at or above the scalar unitarity bound. 

We would like to find a generalization of the construction from the previous section to the present two-variable context. In other words, we want to find a Schauder basis for $\mathcal{V}$ consisting of double-trace conformal blocks in the s- and t-channel. It will turn out that this is indeed possible with the same caveat that we encountered in Section \ref{sec:SingleVariable}: In order to ensure uniqueness of the expansion coefficients, we need to work in a smaller function space $\mathcal{U}\subset\mathcal{V}$. The space $\mathcal{U}$ consists of functions in $\mathcal{V}$ which satisfy a certain stronger boundedness condition as $z,\zb\rightarrow\infty$. More precisely $\cG(z,\zb)\in\mathcal{U}$ if and only if $\cG(z,\zb)\in\mathcal{V}$ and there exist some constants $R>0$, $\epsilon>0$ and $A>0$ such that for all $(z,\zb)\in\mathcal{R}\times\mathcal{R}$ satisfying $|z|>R$, $|\zb|>R$ we have $|\cG(z,\zb)|\leq A |z|^{-\frac{1}{2}-\epsilon}|\zb|^{-\frac{1}{2}-\epsilon}$. In particular, $\cG(z,\zb)\in\mathcal{U}$ implies that the u-channel Regge intercept of $\cG(z,\zb)$ is negative. If this definition seems slightly ad hoc at the moment, we offer the following comment. If $\cG(z,\zb)$ is a physical correlator with u-channel Regge intercept $J_0^{(u)}$, we can try using the Lorentzian inversion formula to extract its u-channel OPE coefficient function $c_{u}(\Delta,J)$. The formula can be trusted only for $J>J_0^{(u)}$. Thus if $\cG(z,\zb)\in\mathcal{V}$, we can trust the formula only for $J>1$ and $\cG(z,\zb)$ is not uniquely determined by it. On the other hand, if $\cG(z,\zb)\in\mathcal{U}$, we can trust if for all $J\geq 0$ and thus $\cG(z,\zb)$ is uniquely determined.

\subsection{A first attempt at a double-trace expansion}\label{ssec:attempt}
The double-trace blocks are labelled by a pair of non-negative inetegers $n,\ell\in\mathbb{N}$, with $\ell$ being the spin. Their scaling dimension is
\be
\Delta_{n,\ell} = 2\Df + 2n+\ell\,.
\ee
The s-channel double-trace blocks are holomorphic at $(z,\zb)=(0,0)$:
\be
G^{s}_{\Delta_{n,\ell},\ell}(z,\zb) \sim z^{n}\zb^{n+\ell}\quad\textrm{for}\quad 0<z\ll\zb\ll1\,.
\ee
We see that the s-channel double-trace blocks provide a basis for functions symmetric under $z\leftrightarrow\zb$ and holomorphic in a neighbourhood of $(z,\zb)=(0,0)$. Similarly, the t-channel double-trace blocks form a basis for symmetric functions holomorphic in the neighbourhood of $(z,\zb)=(1,1)$. How should we find the expansion of a general $\cG(z,\zb)\in \mathcal{U}$ into a double-trace basis? We could try to generalize the argument of Section \ref{sec:SingleVariable} and write down a two-variable dispersion relation, starting from the Cauchy's integral formula applied to both variables
\be
\cG(z,\zb) = \!\!\!\oint\limits_{|w-z|=\epsilon}\!\!\!\frac{dw}{2\pi i}\!\!\!\oint\limits_{|\wb-\zb|=\epsilon}\!\!\!\frac{d\wb}{2\pi i}\,
\frac{\cG(w,\wb)}{(w-z)(\wb-\zb)}\,.
\label{eq:NaiveDispersion1}
\ee
After we deform both $w$ and $\wb$ contours to wrap the branch cuts, we find four terms
\ba
\cG(z,\zb) &=
\oint\limits_{C_+}\frac{dw}{2\pi i}\oint\limits_{C_+}\frac{d\wb}{2\pi i}\frac{\cG(w,\wb)}{(w-z)(\wb-\zb)}-
\oint\limits_{C_-}\frac{dw}{2\pi i}\oint\limits_{C_-}\frac{d\wb}{2\pi i}\frac{\cG(w,\wb)}{(w-z)(\wb-\zb)}-\\
&-\oint\limits_{C_+}\frac{dw}{2\pi i}\oint\limits_{C_-}\frac{d\wb}{2\pi i}\frac{\cG(w,\wb)}{(w-z)(\wb-\zb)}+
\oint\limits_{C_-}\frac{dw}{2\pi i}\oint\limits_{C_+}\frac{d\wb}{2\pi i}\frac{\cG(w,\wb)}{(w-z)(\wb-\zb)}\,,
\label{eq:NaiveDispersion2}
\ea
where $C_+$ wraps the branch cut $[1,\infty)$ while $C_-$ wraps the branch cut $(-\infty,0]$ as in Figure \ref{contourCpm}. The first term on the RHS of \eqref{eq:NaiveDispersion2} is holomorphic at $(z,\zb)=(0,0)$ and thus can be expanded in the s-channel double-trace blocks. Similarly, the second term can be expanded in the t-channel double-trace blocks. However, the third and fourth term on the RHS of \eqref{eq:NaiveDispersion2} generally do not admit either expansion. In other words, the idea of starting from \eqref{eq:NaiveDispersion1} fails to produce the kind of expansion we are interested in.

\subsection{A new look at the double discontinuity}\label{ssec:dDisc}
Although the naive dispersion relation did not lead to a desired double-trace expansion, it is nevertheless worth taking a closer look at the first two terms on the RHS of \eqref{eq:NaiveDispersion2}. These terms turn out to be closely related to the double discontinuity which plays an important role in the Lorentzian inversion formula (LIF), and will provide us with useful insight to find the correct answer in the next section.

Just like we did in Section \ref{sec:SingleVariable}, we can wrap the contours tightly around the branch cuts to find
\ba
&\oint\limits_{C_+}\frac{dw}{2\pi i}\oint\limits_{C_+}\frac{d\wb}{2\pi i}\frac{\cG(w,\wb)}{(w-z)(\wb-\zb)} =
\int\limits_{1}^{\infty}\frac{dw}{2\pi i}\int\limits_{1}^{\infty}\frac{d\wb}{2\pi i}\frac{\Disc_t\overline{\Disc}_t[\cG(w,\wb)]}{(w-z)(\wb-\zb)}
\\
&\oint\limits_{C_-}\frac{dw}{2\pi i}\oint\limits_{C_-}\frac{d\wb}{2\pi i}\frac{\cG(w,\wb)}{(w-z)(\wb-\zb)} =
\int\limits_{-\infty}^{0}\frac{dw}{2\pi i}\int\limits_{-\infty}^{0}\frac{d\wb}{2\pi i}\frac{\Disc_s\overline{\Disc}_s[\cG(w,\wb)]}{(w-z)(\wb-\zb)}\,.
\ea
We see the appearance of the ``squared discontinuity'', which is just an ordinary discontinuity in $\wb$ followed by a discontinuity in $w$ (or equivalently vice versa)
\ba
\Disc_{s,t}\overline{\Disc}_{s,t}[\cG(w,\wb)] &= 
\cG(w+i0^+,\wb+i0^+)-\cG(w-i0^+,\wb+i0^+)-\\
&-\cG(w+i0^+,\wb-i0^+)+\cG(w-i0^+,\wb-i0^+)\,.
\label{eq:SquaredDisc}
\ea
It turns out that the squared discontinuity is very closely related to the double discontinuity, which features prominently in the Lorentzian inversion formula (LIF) \cite{Caron-Huot:2017vep}. Recall that the LIF is a formula for the OPE coefficient function $c(\Delta,J)$ in one channel in terms of the double discontinuities around the OPE singularities of the other two channels. The case relevant to us is the formula giving the u-channel coefficient function $c_{u}(\Delta,J)$ in terms of the double discontinuities around the s- and t-channel singularity
\ba
c_{u}(\Delta,J) &=
\int\limits_{-\infty}^{0}\int\limits_{-\infty}^{0}\!\! dw d\wb\,g^{s}_{\Delta,J}(w,\wb) \dDisc_s[\cG(w,\wb)]+\\
&+(-1)^{J}\int\limits_{1}^{\infty}\!\int\limits_{1}^{\infty}dw d\wb\,g^{t}_{\Delta,J}(w,\wb) \dDisc_t[\cG(w,\wb)]\,,
\label{eq:LIF}
\ea
where $g^{s}_{\Delta,J}(w,\wb)$ and $g^{t}_{\Delta,J}(w,\wb)$ are certain kernels related to conformal blocks, whose precise form will not be important. The double discontinuities are defined as follows
\ba
\dDisc_s[\cG(w,\wb)] &= \cG_{\mathrm{E}}(w,\wb) - \frac{1}{2}\cG_{\mathrm{E}}(w,\wb\circlearrowright0)- \frac{1}{2}\cG_{\mathrm{E}}(w,\wb\circlearrowleft0)\quad w,\wb\in(-\infty,0)\\
\dDisc_t[\cG(w,\wb)] &= \cG_{\mathrm{E}}(w,\wb) - \frac{1}{2}\cG_{\mathrm{E}}(w,\wb\circlearrowright1)- \frac{1}{2}\cG_{\mathrm{E}}(w,\wb\circlearrowleft1)\quad w,\wb\in(1,\infty)\,.
\label{eq:dDisc}
\ea
The first term on the RHS is the Euclidean correlator $\cG(w+i 0^+,\wb-i0^+)$. The second and third term are its analytic continuation in $\wb$. It is important to note that the LIF only applies to correlators which are single-valued in the Euclidean signature because it describes an expansion of $\cG(w,\wb)$ into Euclidean conformal partial waves, which are Euclidean single-valued themselves. If $\cG(w,\wb)$ is Euclidean single-valued, we can relate the terms on the RHS of \eqref{eq:SquaredDisc} to those on the RHS of \eqref{eq:dDisc} as follows:
\ba
&\cG(w+i0^+,\wb+i0^+)=\cG_{\mathrm{E}}(w,\wb\circlearrowleft 0)\\
&\cG(w-i0^+,\wb+i0^+)=\cG_{\mathrm{E}}(w,\wb)\\
&\cG(w+i0^+,\wb-i0^+)=\cG_{\mathrm{E}}(w,\wb)\\
&\cG(w-i0^+,\wb-i0^+)=\cG_{\mathrm{E}}(w,\wb\circlearrowright0)
\ea
for $w,\wb\in(-\infty,0)$ and with the RHS of the first and fourth line exchanged when $w,\wb\in(1,\infty)$. In other words
\ba
\Disc_s\overline{\Disc}_s[\cG(w,\wb)] &= - 2\,\dDisc_s[\cG(w,\wb)]\\
\Disc_t\overline{\Disc}_t[\cG(w,\wb)] &= - 2\,\dDisc_t[\cG(w,\wb)]\,.
\ea
We found that for Euclidean single-valued functions, the squared discontinuity is the same as the double discontinuity (up to a $-2$ factor).

\section{The Primal and Dual Basis}\label{sec:bases}
\subsection{Our proposal}\label{ssec:proposal}
We can now correct our attempt at a double-trace expansion of functions in $\mathcal{U}$ by combining the intuition from Sections \ref{sec:SingleVariable} and \ref{ssec:dDisc}. In Section \ref{sec:SingleVariable}, we found that $\cG(z)\in\mathcal{U}_1$ can be written as a sum of two terms $\cG^{t}(z)+\cG^{s}(z)$ such that $\Disc_s[\cG^{t}(z)] = 0$ and $\Disc_t[\cG^{s}(z)] = 0$. Furthermore, $\cG^{t}(z)$ was uniquely determined by $\Disc_t[\cG(z)]$ and $\cG^{s}(z)$ was uniquely determined by $\Disc_s[\cG(z)]$. $\cG^{t}(z)$ was then expanded in the set of s-channel conformal blocks which satisfy $\Disc_s[G^{s}_{\Delta}(z)] = 0$ and similarly $\cG^{s}(z)$ in the t-channel blocks satisfying $\Disc_t[G^{t}_{\Delta}(z)] = 0$.

Observations of Section \ref{ssec:dDisc} suggest that in the present two-variable context the double discontinuity plays an analogous role to the one played by the single discontinuity in the single-variable case. Specifically, we conjecture that for any $\cG(z,\zb)\in\mathcal{U}$, we can write
\be
\cG(z,\zb) = \cG^{t}(z,\zb) + \cG^{s}(z,\zb)\,,
\label{eq:Gsplit}
\ee
where
\begin{enumerate}
\item $\cG^{t}(z,\zb),\,\cG^{s}(z,\zb)\in\mathcal{U}$.
\item $\cG^{t}(z,\zb)$ is Euclidean single-valued around $(z,\zb) = (0,0)$ and $\cG^{s}(z,\zb)$ is Euclidean single-valued around $(z,\zb) = (1,1)$.
\item $\dDisc_s[\cG^{t}(z,\zb)] = 0$ and $\dDisc_t[\cG^{s}(z,\zb)] = 0$.
\end{enumerate}
Furthermore, this decomposition of $\cG(z,\zb)$ is unique. One of the main goals of the rest of this paper will be to accumulate evidence for this conjecture. This proposal is partly inspired by the u-channel LIF \eqref{eq:LIF}. Indeed, suppose that $\cG(z,\zb)$ is Euclidean single-valued around both $z=\zb=0$ and $z=\zb=1$. Then the LIF applies to it and gives a natural decomposition of $\cG(z,\zb)$ into two parts, corresponding to the two terms on the RHS of \eqref{eq:LIF}. The first term on the RHS of \eqref{eq:LIF} depends only on $\dDisc_s[\cG(z,\zb)]$ and gives rise to our $\cG^{s}(z,\zb)$, while the second term depends only on $\dDisc_t[\cG(z,\zb)]$ and gives rise to our $\cG^{t}(z,\zb)$. Properties 1, 2 and 3 of $\cG^{t}(z,\zb)$ and $\cG^{s}(z,\zb)$ are not immediate consequences of the LIF but presumably can be shown using the dispersion relation of Carmi and Caron-Huot \cite{DispersionRelation}. In any case, if $\cG(z,\zb)$ is a general element of $\mathcal{U}$ which is not Euclidean single-valued, we are not allowed to use the LIF and thus a different idea is required to prove our conjectured proposal rigorously.

Assuming the conjecture is correct, we get a Schauder basis for $\mathcal{U}$ in a similar manner to what we saw in Section \ref{sec:SingleVariable}. We need to expand $\cG^{t}(z,\zb)$ in a complete set of functions which are Euclidean single-valued around $(z,\zb)=(0,0)$ and whose s-channel double discontinuity vanishes. It is natural to use s-channel conformal blocks. The s-channel blocks $G^{s}_{\Delta,\ell}(z,\zb)$ are Euclidean single-valued around $(z,\zb)=(0,0)$ for $\ell\in\mathbb{N}$. They satisfy
\be
\dDisc_s[G^{s}_{\Delta,\ell}(z,\zb)] = 2\sin^2\left[\frac{\pi}{2}(\Delta-\ell-2\Df)\right] G^{s}_{\Delta,\ell}(z,\zb)\,.
\ee
The crucial point is that since $\dDisc_s[G^{s}_{\Delta,\ell}(z,\zb)]$ has double zeros as a function of $\Delta$ at the double-trace dimensions $\Delta=\Delta_{n,\ell} = 2\Df+2n+\ell$, the basis consists of both the double-trace conformal blocks and their derivatives with respect to $\Delta$, i.e.
\be
\dDisc_s[G^{s}_{\Delta_{n,\ell},\ell}(z,\zb)] = \dDisc_s[\partial_{\Delta}G^{s}_{\Delta_{n,\ell},\ell}(z,\zb)] = 0
\ee
for all $n,\ell\in\mathbb{N}$. Indeed, we have
\ba
G^{s}_{\Delta_{n,\ell},\ell}(z,\zb) &\sim z^{n}\zb^{n+\ell}\\
\partial_{\Delta}[G^{s}_{\Delta_{n,\ell},\ell}(z,\zb)] &\sim \frac{\log(z)+\log(\zb)}{2}z^{n}\zb^{n+\ell}
\ea
as $\zb,z/\zb\rightarrow 0$ and thus all $G^{s}_{\Delta_{n,\ell},\ell}(z,\zb)$ and $\partial_{\Delta}G^{s}_{\Delta_{n,\ell},\ell}(z,\zb)$ are linearly independent. Conversely, it is not hard to show that any $\cG^{t}(z,\zb)$ satisfying conditions 2 and 3 can be written as $\cG^{t}(z,\zb) = f(z,\zb) + [\log(z)+\log(\zb)]g(z,\zb)$ for some $f(z,\zb), g(z,\zb)$ which are holomorphic at $(z,\zb)=(0,0)$. Thus $\cG^{t}(z,\zb)$ can be expanded in the s-channel double-trace blocks and their $\Delta$-derivatives.

Since identical arguments apply in the t-channel, we conclude that
\be
G^{s}_{\Delta_{n,\ell},\ell}(z,\zb)\,,\quad \partial_{\Delta}G^{s}_{\Delta_{n,\ell},\ell}(z,\zb)\,,\quad
G^{t}_{\Delta_{n,\ell},\ell}(z,\zb)\,,\quad \partial_{\Delta}G^{t}_{\Delta_{n,\ell},\ell}(z,\zb)
\label{eq:basis}
\ee
with $n,\ell\in\mathbb{N}$ is a Schauder basis for $\mathcal{U}$ in the sense that any $\cG(z,\zb)\in\mathcal{U}$ can be written as \eqref{eq:Gsplit} with $\cG^{t}(z,\zb),\cG^{s}(z,\zb)\in\mathcal{U}$ and
\ba
\cG^{t}(z,\zb) &= \sum\limits_{n,\ell}\left[
a^{s}_{n,\ell}\,G^{s}_{\Delta_{n,\ell},\ell}(z,\zb) +
b^{s}_{n,\ell}\,\partial_{\Delta}G^{s}_{\Delta_{n,\ell},\ell}(z,\zb)\right]\\
\cG^{s}(z,\zb) &= \sum\limits_{n,\ell}\left[
a^{t}_{n,\ell}\,G^{t}_{\Delta_{n,\ell},\ell}(z,\zb) +
b^{t}_{n,\ell}\,\partial_{\Delta}G^{t}_{\Delta_{n,\ell},\ell}(z,\zb)\right]\,,
\label{eq:basisExpansion}
\ea
where here and in the following the sums over $n$ and $\ell$ range over all nonnegative integers. Furthermore, given $\cG(z,\zb)$ the coefficients $a^{s}_{n,\ell}$, $b^{s}_{n,\ell}$, $a^{t}_{n,\ell}$ and $b^{t}_{n,\ell}$ are uniquely fixed.

Before we start collecting evidence for the central conjecture of this paper, let us introduce the dual basis. The dual basis is a basis for $\mathcal{U}^{*}$, i.e. for the space of continuous linear functionals on $\mathcal{U}$. It consists of functionals denoted
\be
\alpha^{s}_{n,\ell}\,,\quad \beta^{s}_{n,\ell}\,,\quad
\alpha^{t}_{n,\ell}\,,\quad \beta^{t}_{n,\ell}\,,
\label{eq:dualBasis}
\ee
where $n,\ell\in\mathbb{N}$. These functionals are dual to elements of the primal basis \eqref{eq:basis} according to
\be
\begin{array}{l}
\alpha^{s}_{n,\ell}\vspace{.2cm}\\
\beta^{s}_{n,\ell}\vspace{.2cm}\\
\alpha^{t}_{n,\ell}\vspace{.2cm}\\
\beta^{t}_{n,\ell}
\end{array}
\qquad\textrm{is dual to}\qquad
\begin{array}{l}
G^{s}_{\Delta_{n,\ell},\ell}\vspace{.17cm}\\
\partial_{\Delta}G^{s}_{\Delta_{n,\ell},\ell}\vspace{.17cm}\\
G^{t}_{\Delta_{n,\ell},\ell}\vspace{.17cm}\\
\partial_{\Delta}G^{t}_{\Delta_{n,\ell},\ell}\,.
\end{array}
\ee
This means the dual basis acts on the primal basis as follows
\be
\begin{array}{l l}
\alpha^{s}_{n,\ell}[G^{s}_{\Delta_{n',\ell'},\ell'}] = \delta_{n n'}\delta_{\ell\ell'}
&\quad\alpha^{s}_{n,\ell}[\partial_{\Delta}G^{s}_{\Delta_{n',\ell'},\ell'}] = 0\vspace{.2cm}\\
\beta^{s}_{n,\ell}[G^{s}_{\Delta_{n',\ell'},\ell'}] = 0
&\quad\beta^{s}_{n,\ell}[\partial_{\Delta}G^{s}_{\Delta_{n',\ell'},\ell'}] = \delta_{n n'}\delta_{\ell\ell'}\vspace{.4cm}\\
\alpha^{s}_{n,\ell}[G^{t}_{\Delta_{n',\ell'},\ell'}] = 0
&\quad\alpha^{s}_{n,\ell}[\partial_{\Delta}G^{t}_{\Delta_{n',\ell'},\ell'}] = 0\vspace{.2cm}\\
\beta^{s}_{n,\ell}[G^{t}_{\Delta_{n',\ell'},\ell'}] = 0
&\quad\beta^{s}_{n,\ell}[\partial_{\Delta}G^{t}_{\Delta_{n',\ell'},\ell'}] = 0
\end{array}
\label{eq:dualityC1}
\ee
and
\be
\begin{array}{l l}
\alpha^{t}_{n,\ell}[G^{s}_{\Delta_{n',\ell'},\ell'}] = 0
&\quad\alpha^{t}_{n,\ell}[\partial_{\Delta}G^{s}_{\Delta_{n',\ell'},\ell'}] = 0\vspace{.2cm}\\
\beta^{t}_{n,\ell}[G^{s}_{\Delta_{n',\ell'},\ell'}] = 0
&\quad\beta^{t}_{n,\ell}[\partial_{\Delta}G^{s}_{\Delta_{n',\ell'},\ell'}] = 0\vspace{.4cm}\\
\alpha^{t}_{n,\ell}[G^{t}_{\Delta_{n',\ell'},\ell'}] = \delta_{n n'}\delta_{\ell\ell'}
&\quad\alpha^{t}_{n,\ell}[\partial_{\Delta}G^{t}_{\Delta_{n',\ell'},\ell'}] = 0\vspace{.2cm}\\
\beta^{t}_{n,\ell}[G^{t}_{\Delta_{n',\ell'},\ell'}] = 0
&\quad\beta^{t}_{n,\ell}[\partial_{\Delta}G^{t}_{\Delta_{n',\ell'},\ell'}] = \delta_{n n'}\delta_{\ell\ell'}
\end{array}
\label{eq:dualityC2}
\ee
for all $n,n',\ell,\ell'\in\mathbb{N}$. The dual basis functionals can be used to extract the coefficients $a^{s}_{n,\ell}$, $b^{s}_{n,\ell}$, $a^{t}_{n,\ell}$ and $b^{t}_{n,\ell}$ in the expansions \eqref{eq:basisExpansion}. This means that for any $\cG(z,\zb)\in\mathcal{U}$, we can write
\ba
\cG(z,\zb) &= \sum\limits_{n,\ell}
\left\{\alpha^{s}_{n,\ell}[\cG]G^{s}_{\Delta_{n,\ell},\ell}(z,\zb) +
\beta^{s}_{n,\ell}[\cG]\partial_{\Delta}G^{s}_{\Delta_{n,\ell},\ell}(z,\zb)\right\}+\\
&\phantom{,}+\sum\limits_{n,\ell}
\left\{\alpha^{t}_{n,\ell}[\cG]G^{t}_{\Delta_{n,\ell},\ell}(z,\zb) +
\beta^{t}_{n,\ell}[\cG]\partial_{\Delta}G^{t}_{\Delta_{n,\ell},\ell}(z,\zb)\right\}\,.
\ea

The duality conditions \eqref{eq:dualityC1},\eqref{eq:dualityC2} imply that the action of s-channel functionals $\alpha^{s}_{n,\ell}$ and $\beta^{s}_{n,\ell}$ on t-channel conformal blocks $G^{t}_{\Delta,J}(z,\zb)$ has double zeros on all double-trace spectrum and their action on s-channel conformal blocks $G^{s}_{\Delta,J}(z,\zb)$ has double zeros on all but one double-trace operator. One of the main technical achievements of this paper is an explicit construction of linear functionals \eqref{eq:dualBasis}. By the virtue of their double-zero structure, these functionals imply interesting new sum rules on the CFT data in general CFTs.

\subsection{Linear independence of the primal basis}
Let us prove that our proposed basis \eqref{eq:basis} is indeed linearly independent. Suppose that
\be
\cG^{t}(z,\zb) + \cG^{s}(z,\zb) = 0\,,
\label{eq:LinDep}
\ee
where $\cG^{t}(z,\zb),\,\cG^{s}(z,\zb)\in\mathcal{U}$ and
\ba
\cG^{t}(z,\zb) &= \sum\limits_{n,\ell}\left[
a^{s}_{n,\ell}\,G^{s}_{\Delta_{n,\ell},\ell}(z,\zb) +
b^{s}_{n,\ell}\,\partial_{\Delta}G^{s}_{\Delta_{n,\ell},\ell}(z,\zb)\right]\\
\cG^{s}(z,\zb) &= \sum\limits_{n,\ell}\left[
a^{t}_{n,\ell}\,G^{t}_{\Delta_{n,\ell},\ell}(z,\zb) +
b^{t}_{n,\ell}\,\partial_{\Delta}G^{t}_{\Delta_{n,\ell},\ell}(z,\zb)\right]\,.
\ea
We want to show that necessarilly $a^{s}_{n,\ell}=b^{s}_{n,\ell}=a^{t}_{n,\ell}=b^{t}_{n,\ell}=0$. The idea will be to use the Lorentzian inversion formula (LIF) in the u-channel to show $\cG^{t}(z,\zb) = \cG^{s}(z,\zb) = 0$. We will apply the LIF to $\cG^{t}(z,\zb)$. Since the LIF only applies to functions which are Euclidean single-valued, we first need to show $\cG^{t}(z,\zb)$ has this property. By assumption, $\cG^{t}(z,\zb)$ is single-valued around $(z,\zb)=(0,0)$ and $\cG^{s}(z,\zb)$ is single-valued around $(z,\zb)=(1,1)$. Furthermore, from \eqref{eq:LinDep}, we know $\cG^{t}(z,\zb) = -\cG^{s}(z,\zb)$, so both sides are single-valued around both $(z,\zb)=(0,0)$ and $(z,\zb)=(1,1)$ and thus in the entire Euclidean plane. Since $\cG^{t}(z,\zb)\in\mathcal{U}$, its u-channel Regge intercept is negative and thus the LIF produces $c_{u}(\Delta,J)$ valid for all spins. But $\dDisc_s[\cG^{t}(z,\zb)] = 0$ and $\dDisc_t[\cG^{t}(z,\zb)] = -\dDisc_t[\cG^{s}(z,\zb)] = 0$, so the LIF implies $c(\Delta,J)$ vanishes identically. The last statement immediately implies $\cG^{t}(z,\zb) = \cG^{s}(z,\zb) = 0$, which we wanted to show. Thus the proposed basis is indeed linearly independent.

For most physical applications, we are interested in working in the bigger space $\mathcal{V}$, rather than its subspace $\mathcal{U}$. In that light, it is important to understand what happens to linear independence of the set of vectors \eqref{eq:basis} when we move to the space $\mathcal{V}$. The short answer is that they are no longer linearly independent and that their linear relations can be parametrized in terms of certain contact Witten diagrams. Assume $\cG^{t}(z,\zb)+\cG^{s}(z,\zb) = 0$ with $\cG^{t}(z,\zb),\,\cG^{s}(z,\zb)\in\mathcal{V}$. We can still apply the LIF to $\cG^{t}(z,\zb)$. However, this time the u-channel Regge intercept of $\cG^{t}(z,\zb)$ can be as high as one, so we can only trust the LIF for spin two and higher. This means $c_{u}(\Delta,J)$ is supported on $J=0,1$. In other words, the u-channel OPE of $\cG^{t}(z,\zb)$ only contains scalar and spin-one conformal blocks. The possible choices for $\cG^{t}(z,\zb)$ are best understood in Mellin space \cite{Mack:2009mi,Penedones:2010ue}. The Mellin amplitude $\mathcal{M}(s,t)$ is a function of Mellin variables $s,t$. It is also useful to introduce $u$ satisfying $s+t+u=4\Df$. We focus on polynomial Mellin amplitudes since these automatically lead to a correlator with vanishing double discontinuity in both s- and t-channel. Such amplitudes are equal to contact Witten diagrams in $\mathrm{AdS}_{d+1}$. The u-channel Regge limit corresponds to $s,t\rightarrow\infty$ with $u=4\Df-s-t$ fixed. For polynomial amplitudes, the Regge intercept in a given channel agrees with the maximal spin present in the OPE of that channel. Thus there are two infinite classes of contact Witten diagrams which give relations between the vectors \eqref{eq:basis}
\be
\mathcal{M}^{(0)}_j(s,t)=u^j\,,\qquad \mathcal{M}^{(1)}_j(s,t)=(s-t)u^j
\ee
for $j=0,1,\ldots$. The first one only contains scalars in the u-channel and double-traces up to spin $j$ in the s- and t-channel, while the second one has only spin-one double-traces in the u-channel and up to spin $j+1$ in s- and t-channel. As we have explained, each of these contact diagrams gives rise to a linear relation among \eqref{eq:basis} of the form
\ba
&\phantom{-\;}\sum\limits_{n,\ell}\left[
a^{s}_{n,\ell}\,G^{s}_{\Delta_{n,\ell},\ell}(z,\zb) +
b^{s}_{n,\ell}\,\partial_{\Delta}G^{s}_{\Delta_{n,\ell},\ell}(z,\zb)\right]=\\
&-\sum\limits_{n,\ell}\left[
a^{t}_{n,\ell}\,G^{t}_{\Delta_{n,\ell},\ell}(z,\zb) +
b^{t}_{n,\ell}\,\partial_{\Delta}G^{t}_{\Delta_{n,\ell},\ell}(z,\zb)\right]\,,
\ea
where both sides of the equality are inside the space $\mathcal{V}$.

\subsection{Polyakov-Regge blocks}\label{subsec:PolyakovRegge}
In order to establish the central claim of Section \ref{ssec:proposal}, it remains to be shown that the set of functions \eqref{eq:basis} indeed generates all of $\mathcal{U}$ in the sense detailed above. We do not have a general proof of this fact, but in this section we will demonstrate that it holds for a large set of functions -- namely all s- and t-channel conformal blocks of arbitrary dimension $\Delta$ and integer spin $J$. This will naturally lead to a relation between our formalism and the Polyakov-Mellin approach to the conformal bootstrap \cite{Gopakumar:2016wkt,Gopakumar:2016cpb,Gopakumar:2018xqi}.

Without loss of generality, let us restrict to s-channel conformal blocks $G^{s}_{\Delta,J}(z,\zb)$. We assume that $\Df>d/4$ so that $G^{s}_{\Delta,J}(z,\zb)\in\mathcal{U}$. We would like to show that it is possible to write
\ba
G^{s}_{\Delta,J}(z,\zb) &= 
\sum\limits_{n,\ell}\left\{\alpha^{s}_{n,\ell}[G^{s}_{\Delta,J}]G^{s}_{\Delta_{n,\ell},\ell}(z,\zb) +
\beta^{s}_{n,\ell}[G^{s}_{\Delta,J}]\partial_{\Delta}G^{s}_{\Delta_{n,\ell},\ell}(z,\zb)\right\}+\\
&\;+\sum\limits_{n,\ell}\left\{\alpha^{t}_{n,\ell}[G^{s}_{\Delta,J}]G^{t}_{\Delta_{n,\ell},\ell}(z,\zb) +
\beta^{t}_{n,\ell}[G^{s}_{\Delta,J}]\partial_{\Delta}G^{t}_{\Delta_{n,\ell},\ell}(z,\zb)
\right\}\,,
\label{eq:blocksFromDT}
\ea
where both curly brackets on the RHS are inside $\mathcal{U}$. We wrote the coeffients in the expansion as the dual basis functionals acting on the LHS. This expansion becomes more familiar if we move all the s-channel blocks to the LHS and leave the t-channel blocks on the RHS. Thus, let us define the following function
\be
P^{s}_{\Delta,J}(z,\zb) = G^{s}_{\Delta,J}(z,\zb)-
\sum\limits_{n,\ell}\!\left\{\alpha^{s}_{n,\ell}[G^{s}_{\Delta,J}]G^{s}_{\Delta_{n,\ell},\ell}(z,\zb) +
\beta^{s}_{n,\ell}[G^{s}_{\Delta,J}]\partial_{\Delta}G^{s}_{\Delta_{n,\ell},\ell}(z,\zb)\right\}.
\label{eq:polyakovS}
\ee
We will call $P^{s}_{\Delta,J}(z,\zb)$ the s-channel {\it Polyakov-Regge block}. The expansion \eqref{eq:blocksFromDT} is equivalent to saying that $P^{s}_{\Delta,J}(z,\zb)$ also admits the following t-channel expansion
\be
P^{s}_{\Delta,J}(z,\zb) = \sum\limits_{n,\ell}\left\{\alpha^{t}_{n,\ell}[G^{s}_{\Delta,J}]G^{t}_{\Delta_{n,\ell},\ell}(z,\zb) +
\beta^{t}_{n,\ell}[G^{s}_{\Delta,J}]\partial_{\Delta}G^{t}_{\Delta_{n,\ell},\ell}(z,\zb)\right\}\,.
\label{eq:polyakovT}
\ee
Recall that $G^{s}_{\Delta,J}(z,\zb)$ and the two curly brackets in \eqref{eq:blocksFromDT} are all required to be in $\mathcal{U}$. It follows that also $P^{s}_{\Delta,J}(z,\zb)\in\mathcal{U}$. Therefore, showing that the expansion \eqref{eq:blocksFromDT} exists is equivalent to showing that there is $P^{s}_{\Delta,J}(z,\zb)\in\mathcal{U}$ whose s- and t-channel OPEs have the form \eqref{eq:polyakovS} and \eqref{eq:polyakovT}.

This OPE structure is characteristic of exchange Witten diagrams. Consider the tree-level s-channel exchange Witten diagram $\mathcal{W}^{s}_{\Delta,J}(z,\zb)$ of a bulk field of dimension $\Delta$ and spin $J$ in $\mathrm{AdS}_{d+1}$. The s-channel OPE of $\mathcal{W}^{s}_{\Delta,J}(z,\zb)$ contains the single-trace conformal block $G^{s}_{\Delta,J}(z,\zb)$ dressed by an infinite set of double-trace conformal blocks $G^{s}_{\Delta_{n,\ell},\ell}(z,\zb)$ and $\partial_{\Delta}G^{s}_{\Delta_{n,\ell},\ell}(z,\zb)$ with spins $\ell\leq J$. The t-channel OPE of $\mathcal{W}^{s}_{\Delta,J}(z,\zb)$ contains only double-trace conformal blocks $G^{t}_{\Delta_{n,\ell},\ell}(z,\zb)$ and $\partial_{\Delta}G^{t}_{\Delta_{n,\ell},\ell}(z,\zb)$. The physical interpretation of the double-traces contributions is as the tree-level corrections to the mean-field OPE coefficients and scaling dimensions induces by the exchange. Let us normalize $\mathcal{W}^{s}_{\Delta,J}(z,\zb)$ so that $G^{s}_{\Delta,J}(z,\zb)$ appears in the s-channel with unit coefficient.

Equations \eqref{eq:polyakovS} and \eqref{eq:polyakovT} show that $P^{s}_{\Delta,J}(z,\zb)$ has the right OPE structure to be the s-channel exchange Witten diagram. More precisely, we should have
\be
P^{s}_{\Delta,J}(z,\zb) = A^{-1}\mathcal{W}^{s}_{\Delta,J}(z,\zb) + \mathcal{C}(z,\zb)\,,
\ee
where $A^{-1}$ is an overall normalization and $\mathcal{C}(z,\zb)$ is a {\it finite} linear combination of four-point contact diagrams. Each contact diagram contains only double-trace conformal blocks (and their $\Delta$-derivatives) in each channel, and therefore preserving the OPE structure.\footnote{In fact, $\mathcal{W}^{s}_{\Delta,J}(z,\zb)$ is itself ambiguous since the three-point coupling between two scalars and a massive spinning field is ambiguous. In the context of tree level exchange diagrams, this ambiguity becomes a contact term and therefore can be absorbed in $\mathcal{C}(z,\zb)$.} By adding such contact diagrams we can correct the u-channel Regge behavior of $\mathcal{W}^{s}_{\Delta,J}(z,\zb)$ (which is generically outside of $\mathcal{U}$), such that $P^{s}_{\Delta,J}(z,\zb)=A^{-1}\mathcal{W}^{s}_{\Delta,J}(z,\zb) + \mathcal{C}(z,\zb)$ is inside $\mathcal{U}$. Therefore, showing that the expansion \eqref{eq:blocksFromDT} exists is equivalent to showing that for every exchange diagram $\mathcal{W}^{s}_{\Delta,J}(z,\zb)$, one can find such a finite linear combination of contact diagrams $\mathcal{C}(z,\zb)$.

A particularly transparent way to see that this is indeed always possible is to work in Mellin space, as we will show in detail in Section \ref{ssec:witten}. Alternatively, we can construct $P^{s}_{\Delta,J}(z,\zb)$ by using the Lorentzian inversion formula as follows. Let us denote the u-channel OPE coefficient function $c_u(\Delta,J)$ of $P^{s}_{\Delta',J'}(z,\zb)$ by $\mathcal{I}^{s}_{u}(\Delta,J;\Delta',J'|\Df)$. The LIF \eqref{eq:LIF} states that
\ba
\mathcal{I}^{s}_{u}(\Delta,J;\Delta',J'|\Df) &=
\int\limits_{-\infty}^{0}\int\limits_{-\infty}^{0}\!\! dw d\wb\,g^{s}_{\Delta,J}(w,\wb) \dDisc_s[P^{s}_{\Delta',J'}(w,\wb)]+\\
&+(-1)^{J}\int\limits_{1}^{\infty}\!\int\limits_{1}^{\infty}dw d\wb\,g^{t}_{\Delta,J}(w,\wb) \dDisc_t[P^{s}_{\Delta',J'}(w,\wb)]\,.
\ea
From \eqref{eq:polyakovS}, \eqref{eq:polyakovT}, we get
\ba
\dDisc_s[P^{s}_{\Delta',J'}(w,\wb)] &= \dDisc_s[G^{s}_{\Delta',J'}(w,\wb)]\\
\dDisc_t[P^{s}_{\Delta',J'}(w,\wb)] &=0\,.
\ea
Thus\footnote{Note that $\mathcal{I}^{s}_{u}(\Delta,J;\Delta',J'|\Df)$ is closely related to the 6j-symbol of the conformal group \cite{Liu:2018jhs}.}
\be
\mathcal{I}^{s}_{u}(\Delta,J;\Delta',J'|\Df) =
\int\limits_{-\infty}^{0}\int\limits_{-\infty}^{0}\!\! dw d\wb\,g^{s}_{\Delta,J}(w,\wb) \dDisc_s[G^{s}_{\Delta',J'}(w,\wb)]\,.
\label{eq:LIFforPR}
\ee
Since $P^{s}_{\Delta',J'}(z,\zb)\in\mathcal{U}$, this formula is valid for all $J\geq 0$ and therefore it uniquely fixes $P^{s}_{\Delta',J'}(z,\zb)$. Note that the formula is analytic in $J$ for $\mathrm{Re}[J]\geq 0$. This leads to an equivalent characterization of $P^{s}_{\Delta',J'}(z,\zb)$: it is a sum of the s-channel exchange and contact diagrams, where the contact diagrams are uniquely fixed by requiring that the u-channel OPE data are analytic in spin all the way down to (and including) spin zero.\footnote{Strictly speaking, the argument does not show $P^{s}_{\Delta,J}(z,\zb)$ with the right properties exists but only that if it does, it is unique. In Section \ref{ssec:witten}, we will give a Mellin-space formula for $P^{s}_{\Delta,J}(z,\zb)$ which should dispel any lingering doubts about its existence.}

The expansion of t-channel conformal blocks in the primal basis \eqref{eq:basis} is exactly equivalent to the expansion of s-channel blocks. We write
\ba
G^{t}_{\Delta,J}(z,\zb) &= 
\sum\limits_{n,\ell}\left\{\alpha^{s}_{n,\ell}[G^{t}_{\Delta,J}]G^{s}_{\Delta_{n,\ell},\ell}(z,\zb) +
\beta^{s}_{n,\ell}[G^{t}_{\Delta,J}]\partial_{\Delta}G^{s}_{\Delta_{n,\ell},\ell}(z,\zb)\right\}+\\
&\;+\sum\limits_{n,\ell}\left\{\alpha^{t}_{n,\ell}[G^{t}_{\Delta,J}]G^{t}_{\Delta_{n,\ell},\ell}(z,\zb) +
\beta^{t}_{n,\ell}[G^{t}_{\Delta,J}]\partial_{\Delta}G^{t}_{\Delta_{n,\ell},\ell}(z,\zb)
\right\}\,,
\ea
and define the t-channel Polyakov-Regge blocks as
\ba
P^{t}_{\Delta,J}(z,\zb) &= G^{t}_{\Delta,J}(z,\zb)\,-\\
&\;\;\;\;-\sum\limits_{n,\ell}\left\{\alpha^{t}_{n,\ell}[G^{t}_{\Delta,J}]G^{t}_{\Delta_{n,\ell},\ell}(z,\zb) +
\beta^{t}_{n,\ell}[G^{t}_{\Delta,J}]\partial_{\Delta}G^{t}_{\Delta_{n,\ell},\ell}(z,\zb)\right\} =\\
& =\sum\limits_{n,\ell}\left\{\alpha^{s}_{n,\ell}[G^{t}_{\Delta,J}]G^{s}_{\Delta_{n,\ell},\ell}(z,\zb) +
\beta^{s}_{n,\ell}[G^{t}_{\Delta,J}]\partial_{\Delta}G^{s}_{\Delta_{n,\ell},\ell}(z,\zb)\right\}\,.
\ea
Note that thanks to the symmetry between the s- and t-channel, we must have
\ba
\alpha^{t}_{n,\ell}[\cG(z,\zb)] &= \alpha^{s}_{n,\ell}[\cG(1-z,1-\zb)]\\
\beta^{t}_{n,\ell}[\cG(z,\zb)] &= \beta^{s}_{n,\ell}[\cG(1-z,1-\zb)]\,.
\label{eq:DualBasisSymmetry}
\ea
Since we also have $G^{t}_{\Delta,J}(z,\zb) = G^{s}_{\Delta,J}(1-z,1-\zb)$, it follows that
\be
P^{t}_{\Delta,J}(z,\zb) = P^{s}_{\Delta,J}(1-z,1-\zb)\,.
\ee

For practical applications of our formalism, we would like to know the functional actions on general conformal blocks, i.e. find formulas for the functions
\be
\alpha^{s}_{n,\ell}[G^{s}_{\Delta,J}]\,,\quad \beta^{s}_{n,\ell}[G^{s}_{\Delta,J}]\,,\quad
\alpha^{t}_{n,\ell}[G^{s}_{\Delta,J}]\,,\quad \beta^{t}_{n,\ell}[G^{s}_{\Delta,J}]\,.
\label{eq:functionalActions}
\ee
While the LIF \eqref{eq:LIFforPR} fixes these functions uniquely, it does so in a rather non-explicit manner. Indeed, in order to go from \eqref{eq:LIFforPR} to \eqref{eq:functionalActions}, we would first need to resum the entire u-channel OPE corresponding to $\mathcal{I}^{s}_{u}(\Delta,J;\Delta',J'|\Df)$, and then re-expand the result in s- and t-channel. While this may be possible with enough hard work, we can circumvent it in two different ways. The first is an explicit construction of the dual basis functionals in terms of contour integrals in $z$ and $\zb$, and the second one is a construction of the Polyakov-Regge blocks in Mellin space. We will explain both methods in Section \ref{sec:ExplicitResults} and Section \ref{ssec:witten}.

To conclude this subsection, let us explain how the Polyakov-Regge blocks give rise to an alternative way to expand the four-point correlator. Instead of writing the four-point function as a sum of conformal blocks in either the s- or t-channel
\begin{equation}\label{GinCB}
\mathcal{G}=\sum_\mathcal{O} f_{12\mathcal{O}} f_{34\mathcal{O}} G^{s}_{\Delta_\mathcal{O},J_\mathcal{O}}= \sum_\mathcal{P} f_{23\mathcal{P}} f_{41\mathcal{P}} G^{t}_{\Delta_\mathcal{P},J_\mathcal{P}}\;,
\end{equation}
we can write it as a sum over {\it both} the s- and t-channel Polyakov-Regge blocks with the {\it same} OPE coefficients
\begin{equation}\label{GinP}
\mathcal{G}=\sum_\mathcal{O} f_{12\mathcal{O}} f_{34\mathcal{O}} P^{s}_{\Delta_\mathcal{O},J_\mathcal{O}}+ \sum_\mathcal{P} f_{23\mathcal{P}} f_{41\mathcal{P}} P^{t}_{\Delta_\mathcal{P},J_\mathcal{P}}\;.
\end{equation}
To show that (\ref{GinP}) is equivalent to (\ref{GinCB}), let us use the definitions (\ref{eq:polyakovS}), (\ref{eq:polyakovT})  for the Polyakov-Regge blocks and expand both $P^{s}_{\Delta_\mathcal{O},J_\mathcal{O}}$ and $P^{t}_{\Delta_\mathcal{P},J_\mathcal{P}}$ into s-channel conformal blocks
\begin{equation}
\begin{split}
\mathcal{G}={}&\sum_\mathcal{O} f_{12\mathcal{O}} f_{34\mathcal{O}} \left[G^{s}_{\Delta_\mathcal{O},J_\mathcal{O}}-
\sum\limits_{n,\ell}\!\left\{\alpha^{s}_{n,\ell}[G^{s}_{\Delta_\mathcal{O},J_\mathcal{O}}]G^{s}_{\Delta_{n,\ell},\ell} +
\beta^{s}_{n,\ell}[G^{s}_{\Delta_\mathcal{O},J_\mathcal{O}}]\partial_{\Delta}G^{s}_{\Delta_{n,\ell},\ell}\right\}\right]\\
{}&+\sum_\mathcal{P} f_{23\mathcal{P}} f_{41\mathcal{P}} \left[
\sum\limits_{n,\ell}\!\left\{\alpha^{s}_{n,\ell}[G^{t}_{\Delta_\mathcal{P},J_\mathcal{P}}]G^{s}_{\Delta_{n,\ell},\ell} +
\beta^{s}_{n,\ell}[G^{t}_{\Delta_\mathcal{P},J_\mathcal{P}}]\partial_{\Delta}G^{s}_{\Delta_{n,\ell},\ell}\right\}\right]\;.
\end{split}
\end{equation}
The sum over the single-trace conformal blocks reproduces (\ref{GinCB}) in the s-channel, and all we need to do is to show the rest vanish in the sum. This follows simply from the crossing equation. Under the assumption that we can exchange the order of functional action and summation (this holds, e.g., when $\mathcal{G}\in\mathcal{U}$), we find the extra terms can be written as
\begin{equation}
\begin{split}
{}&\sum\limits_{n,\ell}\!\alpha^{s}_{n,\ell}\left[\sum_\mathcal{O}f_{12\mathcal{O}} f_{34\mathcal{O}} G^{s}_{\Delta_\mathcal{O},J_\mathcal{O}}-\sum_\mathcal{P}f_{23\mathcal{P}} f_{41\mathcal{P}} G^{t}_{\Delta_\mathcal{P},J_\mathcal{P}}\right]G^{s}_{\Delta_{n,\ell},\ell}\\
+{}& \sum\limits_{n,\ell}\!\beta^{s}_{n,\ell}\left[\sum_\mathcal{O}f_{12\mathcal{O}} f_{34\mathcal{O}} G^{s}_{\Delta_\mathcal{O},J_\mathcal{O}}-\sum_\mathcal{P}f_{23\mathcal{P}} f_{41\mathcal{P}} G^{t}_{\Delta_\mathcal{P},J_\mathcal{P}}\right]\partial_\Delta G^{s}_{\Delta_{n,\ell},\ell}\;,
\end{split}
\end{equation}
which is zero thanks to the crossing equation. Similarly, if we use the t-channel expansion of the Polyakov-Regge blocks, we reproduce the t-channel conformal block decomposition. We could also reverse the logic and start from the Polyakov-Regge block decomposition (\ref{GinP}). The condition that no double-trace conformal blocks and their derivatives appear in the s- or t-channel conformal block decomposition gives rise to the following sum rules
\begin{equation}
\begin{split}
{}&\sum_\mathcal{O}f_{12\mathcal{O}} f_{34\mathcal{O}}\, \alpha^{(s,t)}_{n,\ell}[G^{s}_{\Delta_\mathcal{O},J_\mathcal{O}}]-\sum_\mathcal{P}f_{23\mathcal{P}} f_{41\mathcal{P}} \alpha^{(s,t)}_{n,\ell}[G^{t}_{\Delta_\mathcal{P},J_\mathcal{P}}]=0\;,\\
{}&\sum_\mathcal{O}f_{12\mathcal{O}} f_{34\mathcal{O}}\, \beta^{(s,t)}_{n,\ell}[G^{s}_{\Delta_\mathcal{O},J_\mathcal{O}}]-\sum_\mathcal{P}f_{23\mathcal{P}} f_{41\mathcal{P}} \beta^{(s,t)}_{n,\ell}[G^{t}_{\Delta_\mathcal{P},J_\mathcal{P}}]=0\;,
\end{split}
\end{equation}
for the CFT data. These conditions are reminiscent of the sum rules arising from the Polyakov-Mellin bootstrap  \cite{Gopakumar:2016wkt,Gopakumar:2016cpb,Gopakumar:2018xqi}.

\subsection{General external dimensions}\label{Pgenerald}
Let us briefly comment on how the discussion so far should be modified when the external operators $\phi_1,\phi_2,\phi_3,\phi_4$ in \eqref{eq:4ptFunction} have general scaling dimensions $\Delta_1,\Delta_2,\Delta_3,\Delta_4$. A physical four-point function in a unitary theory is then still holomorphic in $\mathcal{R}\times\mathcal{R}$ and satisfies a boundedness condition at infinity. Consequently, the vector spaces $\mathcal{U}$ and $\mathcal{V}$ generalize naturally. The main novelty is that for each channel, there are now two independent sets of double-trace conformal blocks. The s-channel double-traces are $G^{s}_{\Delta_1+\Delta_2+2n+\ell,\ell}$ and $G^{s}_{\Delta_3+\Delta_4+2n+\ell,\ell}$, while the t-channel double-traces are $G^{t}_{\Delta_2+\Delta_3+2n+\ell,\ell}$ and $G^{t}_{\Delta_1+\Delta_4+2n+\ell,\ell}$. We expect that the set of all s- and t-channel double-traces blocks
\be
G^{s}_{\Delta_1+\Delta_2+2n+\ell,\ell}\,,\quad G^{s}_{\Delta_3+\Delta_4+2n+\ell,\ell}\,\quad
G^{t}_{\Delta_2+\Delta_3+2n+\ell,\ell}\,,\quad G^{t}_{\Delta_1+\Delta_4+2n+\ell,\ell}
\ee
forms a Schauder basis for $\mathcal{U}$. Indeed, the correct generalization of the double discontinuity to non-equal external dimensions has simple zeros at these double-traces \cite{Caron-Huot:2017vep}
\ba
\dDisc_s[G^{s}_{\Delta,J}(z,\zb)] &= 2\sin\left[\scriptstyle{\frac{\pi}{2}(\Delta-J-\Delta_1-\Delta_2)}\right]\sin\left[\scriptstyle{\frac{\pi}{2}(\Delta-J-\Delta_3-\Delta_4)}\right] G^{s}_{\Delta,J}(z,\zb)\\
\dDisc_t[G^{t}_{\Delta,J}(z,\zb)] &= 2\sin\left[\scriptstyle{\frac{\pi}{2}(\Delta-J-\Delta_1-\Delta_4)}\right]\sin\left[\scriptstyle{\frac{\pi}{2}(\Delta-J-\Delta_2-\Delta_3)}\right] G^{t}_{\Delta,J}(z,\zb)\,.
\ea
Relatedly, exchange and contact Witten diagrams contain precisely this set of double-trace conformal blocks in the s- and t-channel OPEs. Much of the preceding arguments thus can be repeated without change. When $\Delta_1+\Delta_2 = \Delta_3 + \Delta_4$, the s-channel double-trace dimensions become degenerate\footnote{Or more generally when $\Delta_1+\Delta_2-\Delta_3-\Delta_4\in2\mathbb{Z}$.} and we need to include the derivatives $\partial_{\Delta}G^{s}_{\Delta_1+\Delta_2+2n+\ell,\ell}$.\footnote{It can also be understood from general $1/N$ analysis that the degenerate double-trace operators now acquire anomalous dimensions.} Similarly, when $\Delta_2+\Delta_3 = \Delta_1+\Delta_4$, we should include $\partial_{\Delta}G^{t}_{\Delta_2+\Delta_3+2n+\ell,\ell}$. In the rest of this paper, we will focus on the case of equal external dimensions and leave further generalization to non-equal dimensions for future work.

\section{An Explicit Construction of the Dual Basis}\label{sec:ExplicitResults}
\subsection{Functionals with double zeros on double-trace dimensions}
The aim of this section is to construct elements of the dual basis
\be
\alpha^{s}_{n,\ell}\,,\quad\beta^{s}_{n,\ell}\,,\quad\alpha^{t}_{n,\ell}\,,\quad\beta^{t}_{n,\ell}\quad\textrm{where}\quad n,\ell\in\mathbb{N}
\label{eq:DualBasisR}
\ee
as explicit linear functionals acting on functions of two complex variables $w,\wb$. This is important in particular for finding the action of the dual basis functionals on general s- and t-channel conformal blocks. These actions are themselves needed to derive sum rules on OPE  data following from the existence of the dual basis functionals. An alternative method to extract the functional actions on general conformal blocks involves decomposing exchange Witten diagrams in double-trace operators. Later in the next section, we will explain this second method in detail and show that it agrees with the explicit construction of the basis functionals.

Each element of the dual basis \eqref{eq:DualBasisR} is uniquely fixed by imposing that it acts as required on the primal basis, i.e. by equations \eqref{eq:dualityC1}, \eqref{eq:dualityC2}. In addition, we need to impose that each basis functional is inside $\mathcal{U}^*$, i.e. that it is continuous as a map from the topological vector space $\mathcal{U}$ to $\mathbb{C}$. Continuity guarantees that the functional can be swapped with infinite sums of functions in $\mathcal{U}$ which converge in $\mathcal{U}$. We will see that in practice, this property says that the action of the functionals in $\mathcal{U}$ must be ``suppressed near infinity'' in the $w,\wb$ space. Note that symmetry under $\mathrm{s}\leftrightarrow\mathrm{t}$, i.e. equation \eqref{eq:DualBasisSymmetry}, implies that  $\alpha^{t}_{n,\ell}$ and $\beta^{t}_{n,\ell}$ come for free once $\alpha^{s}_{n,\ell}$ and $\beta^{s}_{n,\ell}$ have been constructed. We will thus focus on finding $\alpha^{s}_{n,\ell}$ and $\beta^{s}_{n,\ell}$ in what follows.

The third and the fourth line of \eqref{eq:dualityC1} are equivalent to saying that when acting on $G^{t}_{\Delta,J}(w,\wb)$, all $\alpha^{s}_{n,\ell}$ and $\beta^{s}_{n,\ell}$ have double-zeros as a function of $\Delta$ on the double-trace dimensions. Similarly, the first and second line of \eqref{eq:dualityC1} imply that when any $\alpha^{s}_{n,\ell}$ or $\beta^{s}_{n,\ell}$ acts on $G^{s}_{\Delta,J}(w,\wb)$, it has double zeros on all the double-traces except for the one of spin $\ell$ and dimension $\Delta_{n,\ell}$. Thus in order to make progress, we need to learn how to construct linear functionals which have double zeros on almost all s- and t-channel double-trace conformal blocks. There is a very simple generalization of the single-variable functionals we saw in Section \ref{sec:SingleVariable} which does exactly that. The functionals of Section \ref{sec:SingleVariable} were defined as contour integrals against a suitable holomorphic kernel, where the contour starts at $-i\infty$, ends at $i\infty$ and passes in between the branch points at $w=0$ and $w=1$, see \eqref{eq:omegaBoth1D}. This guarantees that the functional has \emph{simple} zeros at most double-trace conformal blocks in both channels. In order to get \emph{double} zeros in the two-variable case, we simply consider a double contour integral in $w$ and $\wb$, where each contour is the same as in the single-variable case. As a toy example, consider arguably the simplest such functional
\be
\omega[\cG(w,\wb)] =
\!\int\limits_{\frac{1}{2}-i\infty}^{\frac{1}{2}+i\infty}\!\!\frac{dw}{2\pi i}\!\int\limits_{\frac{1}{2}-i\infty}^{\frac{1}{2}+i\infty}\!\!\frac{d\wb}{2\pi i}\,\cG(w,\wb)\,.
\ee
We will soon see that $\omega$ is not inside $\mathcal{U}^*$ so it will not quite do for our purposes, but it illustrates how we will obtain functionals with the right structure of double zeros. We claim that $\omega$ has double zeros on \emph{all} double-trace conformal blocks in both channels. The change of variables $(w,\wb)\mapsto(1-w,1-\wb)$ shows that $\omega[\cG(w,\wb)] = \omega[\cG(1-w,1-\wb)]$ so it is sufficient to exhibit the double zeros on the s-channel double traces. It is useful to manifest the analytic structure of conformal blocks at $w=\wb=0$ by writing
\be
G^{s}_{\Delta,J}(w,\wb) = (w \wb)^{\frac{\Delta-J-2\Df}{2}}\widetilde{G}_{\Delta,J}(w,\wb)\,,
\ee
where $\widetilde{G}_{\Delta,J}(w,\wb)$ is holomorphic at $w=\wb=0$ when $J\in\mathbb{N}$. The action of $\omega$ looks as follows
\be
\omega[G^{s}_{\Delta,J}] =
\!\int\limits_{\frac{1}{2}-i\infty}^{\frac{1}{2}+i\infty}\!\!\frac{dw}{2\pi i}\!\int\limits_{\frac{1}{2}-i\infty}^{\frac{1}{2}+i\infty}\!\!\frac{d\wb}{2\pi i}\,(w \wb)^{\frac{\Delta-J-2\Df}{2}}\widetilde{G}_{\Delta,J}(w,\wb)\,.
\ee
To exhibit the double zeros, let us wrap both $w$ and $\wb$ contour onto the power-law branch cut on the left.\footnote{The contribution from infinity vanishes for sufficiently large $\Df$.} We pick up the squared discontinuity $\Disc_s\overline{\Disc}_s$ of the conformal blocks. Each discontinuity contributes a factor $\sin\left[\frac{\pi}{2}(\Delta-J-2\Df)\right]$ and find
\be
\omega[G^{s}_{\Delta,J}] = \frac{1}{\pi^2}\sin^2\!\left[\frac{\pi}{2}(\Delta-J-2\Df)\right]
\int\limits_{-\infty\;\;}^{0}\!\!\!\int\limits_{-\infty\;\;}^{0}\!\!\!\!dw\,d\wb
((-w)(-\wb))^{\frac{\Delta-J-2\Df}{2}}\widetilde{G}_{\Delta,J}(w,\wb)\,.
\label{eq:omegaActionDef}
\ee
The integral converges for all $\Delta>J+2\Df-2$. This manifests the promised double zeros on all the double-trace conformal blocks. More generally, we could consider functionals of the form
\be
\omega[\cG(w,\wb)] =
\!\int\limits_{\frac{1}{2}-i\infty}^{\frac{1}{2}+i\infty}\!\!\frac{dw}{2\pi i}\!\int\limits_{\frac{1}{2}-i\infty}^{\frac{1}{2}+i\infty}\!\!\frac{d\wb}{2\pi i}\,\mathcal{H}(w,\wb)\,\cG(w,\wb)\,,
\label{eq:FunctionalGeneralN}
\ee
where $\mathcal{H}(w,\wb)$ is an arbitrary polynomial. The contour deformation still works (for sufficiently large $\Df$), showing that all of these functionals have double zeros on all double-trace blocks in both channels.

While they have a promising structure of double zeros, none of the functionals constructed so far belong to the dual space $\mathcal{U}^*$. Indeed, it is easy to find functions $\mathcal{G}(w,\wb)\in\mathcal{U}$ such that $\omega[\cG(w,\wb)]$ is not well-defined as a result of the $w$ and $\wb$ integrals not being convergent at infinity. Recall that for $\cG(w,\wb)$ to belong to $\mathcal{U}$, it should be bounded by a constant multiple of $|w|^{-1/2-\epsilon}|\wb|^{-1/2-\epsilon}$ (with $\epsilon>0$) away from $w,\wb=0,1$, and in particular near infinity. So for example $\cG(w,\wb)=[w \wb(1-w)(1-\wb)]^{t}$ belongs to $\mathcal{U}$ for $\mathrm{Re}(t)<-1/4$, but the integral in \eqref{eq:FunctionalGeneralN} certainly diverges for $\mathrm{Re}(t)\geq -1/2$, as long as $\mathcal{H}(w,\wb)$ is any nonzero polynomial. Furthermore, assuming correctness of our proposal that
\be
G^{s}_{\Delta_{n,\ell},\ell}(w,\wb)\,,\quad \partial_{\Delta}G^{s}_{\Delta_{n,\ell},\ell}(w,\wb)\,,\quad
G^{t}_{\Delta_{n,\ell},\ell}(w,\wb)\,,\quad \partial_{\Delta}G^{t}_{\Delta_{n,\ell},\ell}(w,\wb)
\ee
form a Schauder basis for $\mathcal{U}$, there can clearly be no non-trivial functional in $\mathcal{U}^*$ with double-zeros on all double-trace conformal blocks in both channels since any functional with this property would then necessarilly vanish on all functions in $\mathcal{U}$.

A natural adjustment we can make to construct functionals which do belong to $\mathcal{U}^*$ is to consider functionals of the form \eqref{eq:FunctionalGeneralN}, where $\mathcal{H}(w,\wb)$ decays sufficiently fast at infinity. The precise condition which ensures that the action of such $\omega$ on arbitary elements of $\mathcal{U}$ is well-defined is
\be
|\mathcal{H}(w,\wb)| \leq A |w|^{-\frac{1}{2}}|\wb|^{-\frac{1}{2}}
\label{eq:Hcondition}
\ee
for some $A>0$ everywhere along the contour of integration.\footnote{The same condition also ensures that $\omega$ defined by \eqref{eq:FunctionalGeneralN} commutes with infinite sums of functions which converge in $\mathcal{U}$. To show that, we need to swap the integrals in \eqref{eq:FunctionalGeneralN} with the sum. If the contour of integration were localized in a bounded subset of $\mathcal{R}\times\mathcal{R}$, we could swap the integral and sum thanks to uniform convergence. Condition \eqref{eq:Hcondition} ensures that the part of the contour near infinity can be neglected and the argument using a bounded set still holds. See \cite{Rychkov:2017tpc} for more details.} Evidently, there is no nonvanishing $\mathcal{H}(w,\wb)$ which satisfies \eqref{eq:Hcondition} and is holomorphic everywhere in $\mathbb{C}^2$. That is good news because if $\mathcal{H}(w,\wb)$ was holomorphic everyhwere, we could use the same contour deformation as above to show the functional kills the entire double-trace basis, a contradiction.

Nevertheless, to construct $\alpha^{s}_{n,\ell}$ and $\beta^{s}_{n,\ell}$, we would still like to keep $\mathcal{H}(w,\wb)$ holomorphic almost everywhere so that the contour-deformation argument almost works and $\omega$ thus kills almost all functions in the primal basis. A simple idea is to try $\mathcal{H}(w,\wb)=1/(w\wb)$. When acting on the t-channel blocks, we still find double zeros on all the double-traces since we can deform both $w$ and $\wb$ contour to the right and encounter no new singularity. When acting on the s-channel conformal blocks, the contour deformation gives
\be
\omega[G^{s}_{\Delta,J}] = \frac{1}{\pi^2}\sin^2\!\left[\frac{\pi}{2}(\Delta-J-2\Df)\right]
\int\limits_{-\infty\;\;}^{0}\!\!\!\int\limits_{-\infty\;\;}^{0}\!\!\!\!dw\,d\wb
((-w)(-\wb))^{\frac{\Delta-J-2\Df-2}{2}}\widetilde{G}_{\Delta,J}(w,\wb)\,.
\ee
Thanks to the enhanced singularity at $w=\wb=0$, the integral develops a simple pole in $\Delta$ at $2\Df+J$ for $J>0$ and a double pole at $2\Df$ for $J=0$. This pole cancels with the double zero of the $\sin^2\left[\frac{\pi}{2}(\Delta-J-2\Df)\right]$ prefactor to leave a simple zero at $\Delta=2\Df+J$ for $J>0$ and a non-zero value at $\Delta=2\Df$ for $J=0$. In other words, this functional looks like a linear combination of $\alpha^{s}_{0,0}$ and $\beta^{s}_{0,\ell}$ for all $\ell\in\mathbb{N}$.

This is progress but still not quite what we want since we would like to construct individual $\alpha^{s}_{n,\ell}$ and $\beta^{s}_{n,\ell}$. It is crucial that in order to do that, we will have to use $\mathcal{H}(w,\wb)$ which do not factorize into a function of $w$ and $\wb$. This is because if $\mathcal{H}(w,\wb)$ factorizes, we would always find that the functional is a linear combination of infinitely many dual basis functional, just like we saw in the simple example $\mathcal{H}(w,\wb)=1/(w\wb)$. We will see momentarily that essentially the simplest choice of $\mathcal{H}(w,\wb)$ which does not factorize gives us exactly what we want.

\subsection{The functional $\beta^{s}_{0,0}$}\label{ssec:beta00}
\begin{figure}[htbp]
\begin{center}
\includegraphics[width=0.5\linewidth]{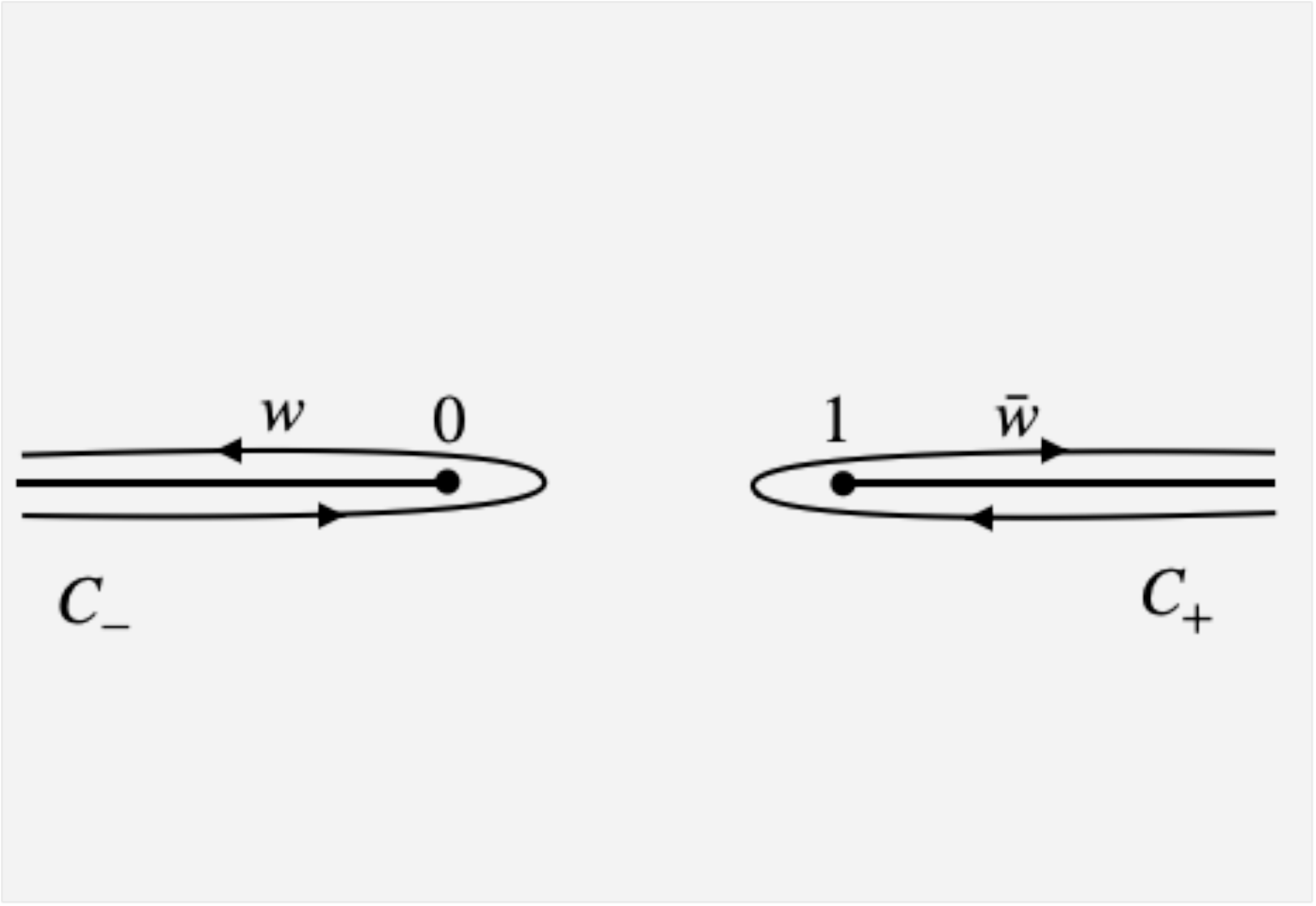}
\caption{Contour prescription for the general functional \eqref{eq:FunctionalGeneral}. The $w$ variable is integrated over contour $C_-$ and $\wb$ over $C_+$. All the dual basis functionals $\alpha^{s}_{n,\ell}$, $\beta^{s}_{n,\ell}$, $\alpha^{t}_{n,\ell}$ and $\beta^{t}_{n,\ell}$ take this form for suitable integral kernel $\mathcal{H}(w,\wb)$.}
\label{wwbcontour}
\end{center}
\end{figure}

A natural class of $\mathcal{H}(w,\wb)$ which do not factorize is meromorphic functions whose only pole is at $w=\wb$. We will soon see that all $\beta^{s}_{n,\ell}$ arise from such $\mathcal{H}(w,\wb)$. The presence of the pole at $w=\wb$ means that we need to be more careful in defining the functionals since \eqref{eq:FunctionalGeneralN} would have a singularity when the $w$ and $\wb$ integrations collide. The correct way to make sure the integral is well-defined is to choose the $w$ and $\wb$ contour so that they do not intersect. Thus, let us consider the following class of functionals
\be
\omega[\cG(w,\wb)] = \int\limits_{C_-}\!\!\frac{dw}{2\pi i} \int\limits_{C_+}\!\!\frac{d\wb}{2\pi i} \,\mathcal{H}(w,\wb) \cG(w,\wb)\,,
\label{eq:FunctionalGeneral}
\ee
where $C_-$ and $C_+$ start at complex infinity in the lower half-plane, end at complex infinity in the upper half-plane and pass in between $0$ and $1$. Furthermore, $C_-$ always stays to the left of $C_+$ so that they never intersect. It is convenient to choose $C_-$ to wrap the branch cut $(-\infty,0]$ and $C_+$ to wrap the right branch cut $[1,\infty)$, as in Figure \ref{wwbcontour}.
If $\mathcal{H}(w,\wb)$ has no singularities the interior of $\mathcal{R}\times\mathcal{R}$, then \eqref{eq:FunctionalGeneralN} and \eqref{eq:FunctionalGeneral} are clearly equivalent. All the dual basis functionals will take the form \eqref{eq:FunctionalGeneral} for suitable $\mathcal{H}(w,\wb)$.

We claim that $\beta^{s}_{0,0}$ is given by \eqref{eq:FunctionalGeneral} with
\be
\mathcal{H}(w,\wb) = \frac{\sigma_0 + \sigma_1(w+\wb)}{(\wb-w)^2}
\ee
for suitable real constants $\sigma_0$, $\sigma_1$ to be determined shortly. This $\mathcal{H}(w,\wb)$ is the most general meromorphic function of $w$ and $\wb$ which satisfies $\mathcal{H}(w,\wb)=\mathcal{H}(\wb,w)$, the growth condition \eqref{eq:Hcondition} and whose only singularity is a pole at $w=\wb$ of degree at most two. Let us check that the resulting functional has the correct structure of zeros on the double-traces, which will also fix the constants $\sigma_0$, $\sigma_1$. Consider the action of $\beta^{s}_{0,0}$ on the s-channel conformal blocks
\be
\beta^{s}_{0,0}[G^{s}_{\Delta,J}] =
\int\limits_{C_-}\!\!\frac{dw}{2\pi i} \int\limits_{C_+}\!\!\frac{d\wb}{2\pi i} \,\frac{\sigma_0 + \sigma_1(w+\wb)}{(\wb-w)^2}\,(w \wb)^{\frac{\Delta-J-2\Df}{2}}\widetilde{G}_{\Delta,J}(w,\wb)\,.
\ee
To simplify the integral, we will wrap the $C_-$ contour in $w$ variable onto the left branch cut, picking up the discontinuity of the integrand. Since $\mathcal{H}(w,\wb)$ has no discontinuity there, the entire discontinuity comes from the conformal block. We find
\ba
\beta^{s}_{0,0}[G^{s}_{\Delta,J}] &=-\frac{\sin\left[\frac{\pi}{2}(\Delta-J-2\Df)\right]}{\pi}\times\\
&\times\int\limits_{-\infty}^{0}\!\! dw \int\limits_{C_+}\!\!\frac{d\wb}{2\pi i} \,\frac{\sigma_0 + \sigma_1(w+\wb)}{(\wb-w)^2}\,((-w)( \wb))^{\frac{\Delta-J-2\Df}{2}}\widetilde{G}_{\Delta,J}(w,\wb)\,.
\label{eq:beta001}
\ea
Thanks to the sine prefactor, we can immediately conclude that $\beta^{s}_{0,0}[G^{s}_{\Delta,J}]$ vanishes on all s-channel double-traces
\be
\beta^{s}_{0,0}[G^{s}_{\Delta_{n,\ell},\ell}] = 0\,.
\ee
We would like to show that these zeros are in fact double zeros, with the exception of $n=\ell=0$, i.e. we want to show $\beta^{s}_{0,0}[\partial_{\Delta}G^{s}_{\Delta_{n,\ell},\ell}] = \delta_{n0}\delta_{\ell,0}$. To evaluate $\beta^{s}_{0,0}[\partial_{\Delta}G^{s}_{\Delta_{n,\ell},\ell}]$, we take the derivative with respect to $\Delta$ of the RHS of \eqref{eq:beta001} and set $\Delta=2\Df+2n+\ell$ and $J=\ell$. Only the term where the derivative acts on the sine survives, and we find
\be
\beta^{s}_{0,0}[\partial_{\Delta}G^{s}_{\Delta_{n,\ell},\ell}] =-\frac{1}{2}
\int\limits_{-\infty}^{0}\!\! dw \int\limits_{C_+}\!\!\frac{d\wb}{2\pi i} \,\frac{\sigma_0 + \sigma_1(w+\wb)}{(\wb-w)^2}\,G^{s}_{\Delta_{n,\ell},\ell}(w,\wb)\,.
\label{eq:beta002}
\ee
We can now deform the $C_+$ contour in variable $\wb$ to the left. The s-channel double-trace blocks $G^{s}_{\Delta_{n,\ell},\ell}(w,\wb)$ are holomorphic for all $w,\wb\notin[1,\infty)$, so the only singularity we encounter is the double pole at $\wb=w$. In other words, the $\wb$ integral localizes to a derivative $\partial_{\wb}$ evaluated at $\wb=w$
\be
\beta^{s}_{0,0}[\partial_{\Delta}G^{s}_{\Delta_{n,\ell},\ell}] =-\frac{1}{2}
\int\limits_{-\infty}^{0}\!\! dw \,\left.\partial_{\wb}\right|_{\wb=w}\!\left[(\sigma_0 + \sigma_1(w+\wb))\,G^{s}_{\Delta_{n,\ell},\ell}(w,\wb)\right]\,.
\label{eq:beta003}
\ee
Now, notice that any function satifying $f(w,\wb)=f(\wb,w)$ satisfies the identity
\be
\left.\partial_{\wb}\right|_{\wb=w} f(w,\wb) = \frac{1}{2}\partial_w f(w,w)\,.
\ee
Since the square bracket on the RHS of \eqref{eq:beta003} is symmetric under $w\leftrightarrow\wb$, we can use this identity to get
\be
\beta^{s}_{0,0}[\partial_{\Delta}G^{s}_{\Delta_{n,\ell},\ell}] =-\frac{1}{4}
\int\limits_{-\infty}^{0}\!\! dw \,\partial_{w}\!\left[(\sigma_0 + 2\,\sigma_1 w)\,G^{s}_{\Delta_{n,\ell},\ell}(w,w)\right]\,.
\label{eq:beta004}
\ee
We found the integrand is a total derivative of a function which vanishes at the lower end-point of integration (due to $G^{s}_{\Delta_{n,\ell},\ell}(w,w)$ being suppressed). This means the entire integral localizes to $w=\wb=0$
\be
\beta^{s}_{0,0}[\partial_{\Delta}G^{s}_{\Delta_{n,\ell},\ell}] =-\frac{\sigma_0}{4}
G^{s}_{\Delta_{n,\ell},\ell}(0,0)\,.
\label{eq:beta005}
\ee
Recall that $G^{s}_{\Delta_{n,\ell},\ell}(w,\wb)\sim w^{n}\wb^{n+\ell}$ in the limit $|w|\ll|\wb|\ll 1$. It follows that the RHS of \eqref{eq:beta005} is indeed nonvanishing only for $n=\ell=0$, i.e.
\be
\beta^{s}_{0,0}[\partial_{\Delta}G^{s}_{\Delta_{n,\ell},\ell}] =-\frac{\sigma_0}{4}\delta_{n0}\delta_{\ell 0}
\ee
and we should take $\sigma_0 = -4$.

Similarly, we need to study $\beta^{s}_{0,0}[G^{t}_{\Delta,J}]$ and make sure $\beta^{s}_{0,0}[G^{t}_{\Delta_{n,\ell},\ell}]=\beta^{s}_{0,0}[\partial_{\Delta}G^{t}_{\Delta_{n,\ell},\ell}]=0$ for all $n,\ell\in\mathbb{N}$. We have
\ba
\int\limits_{C_-}\!\!\frac{dw}{2\pi i} \int\limits_{C_+}\!\!\frac{d\wb}{2\pi i} \,\mathcal{H}(w,\wb) \cG(1-w,1-\wb) &= 
\int\limits_{C_+}\!\!\frac{dw}{2\pi i} \int\limits_{C_-}\!\!\frac{d\wb}{2\pi i} \,\mathcal{H}(1-w,1-\wb) \cG(w,\wb) =\\
& =\int\limits_{C_-}\!\!\frac{dw}{2\pi i} \int\limits_{C_+}\!\!\frac{d\wb}{2\pi i} \,\mathcal{H}(1-w,1-\wb) \cG(w,\wb)\,,
\ea
where we used $\cG(w,\wb)=\cG(\wb,w)$ and $\mathcal{H}(w,\wb)=\mathcal{H}(\wb,w)$. Therefore
\be
\beta^{s}_{0,0}[G^{t}_{\Delta,J}] =
\int\limits_{C_-}\!\!\frac{dw}{2\pi i} \int\limits_{C_+}\!\!\frac{d\wb}{2\pi i} \,\frac{\sigma_0+2\sigma_1 - \sigma_1(w+\wb)}{(\wb-w)^2}\,G^{s}_{\Delta,J}(w,\wb)\,.
\ee
The same manipulations as above apply and guarantee that $\beta^{s}_{0,0}[G^{t}_{\Delta_{n,\ell},\ell}]=\beta^{s}_{0,0}[\partial_{\Delta}G^{t}_{\Delta_{n,\ell},\ell}]=0$ provided $\sigma_0+2\sigma_1=0$, from which we find
\be
\mathcal{H}(w,\wb) = \frac{2(w+\wb-2)}{(\wb-w)^2}\,.
\ee
In particular, $\beta^{s}_{0,0}$ is independent of $\Df$ and spacetime dimension! This is because it relies only on the leading behaviour of the conformal blocks in the $w,\wb\rightarrow 0$ limit, which is universal.

\subsection{General elements of the dual basis}\label{ssec:GeneratingKernels}
We will now generalize the construction of $\beta^{s}_{0,0}$ to all the other elements of the dual basis. They all take the form
\ba
\alpha^{s}_{n,\ell}[\cG(w,\wb)] &= \int\limits_{C_-}\!\!\frac{dw}{2\pi i} \int\limits_{C_+}\!\!\frac{d\wb}{2\pi i} \,\mathcal{A}^{s}_{n,\ell}(w,\wb) \cG(w,\wb)\\
\beta^{s}_{n,\ell}[\cG(w,\wb)] &= \int\limits_{C_-}\!\!\frac{dw}{2\pi i} \int\limits_{C_+}\!\!\frac{d\wb}{2\pi i} \,\mathcal{B}^{s}_{n,\ell}(w,\wb) \cG(w,\wb)\\
\alpha^{t}_{n,\ell}[\cG(w,\wb)] &= \int\limits_{C_-}\!\!\frac{dw}{2\pi i} \int\limits_{C_+}\!\!\frac{d\wb}{2\pi i} \,\mathcal{A}^{t}_{n,\ell}(w,\wb) \cG(w,\wb)\\
\beta^{t}_{n,\ell}[\cG(w,\wb)] &= \int\limits_{C_-}\!\!\frac{dw}{2\pi i} \int\limits_{C_+}\!\!\frac{d\wb}{2\pi i} \,\mathcal{B}^{t}_{n,\ell}(w,\wb) \cG(w,\wb)\,,
\label{eq:basisKersGeneral}
\ea
where the contours are shown in Figure \ref{wwbcontour} and $\mathcal{A}^{s}_{n,\ell}(w,\wb)$, $\mathcal{B}^{s}_{n,\ell}(w,\wb)$, $\mathcal{A}^{t}_{n,\ell}(w,\wb)$ and $\mathcal{B}^{t}_{n,\ell}(w,\wb)$ are suitable integral kernels to be determined. All the kernels are symmetric under $w\leftrightarrow\wb$. The t-channel kernels are obtained from the s-channel kernels by the crossing transformation
\ba
\mathcal{A}^{t}_{n,\ell}(w,\wb) &= \mathcal{A}^{s}_{n,\ell}(1-w,1-\wb)\\
\mathcal{B}^{t}_{n,\ell}(w,\wb) &= \mathcal{B}^{s}_{n,\ell}(1-w,1-\wb)\,,
\ea
so we will focus on finding the s-channel kernels.

A fixed dual basis functional generally depends on $\Df$ and the dimension of spacetime $d$. This is because the primal basis functions also depend on $\Df$ and $d$. For example, for $\ell=0$ we have the following series expansions around $z=\zb=0$
\ba
&G^{s}_{\Delta_{n,0},0}(z,\zb) = z^n\zb^n\left[1+\frac{\Df+n}{2}(z+\zb)+\frac{\left(\Delta _{\phi }+n\right) \left(\Delta _{\phi }+n+1\right){}^2}{4 \left(2 \Delta _{\phi }+2 n+1\right)}(z^2+\zb^2)+\right.\\
&\qquad\qquad\qquad\qquad\quad\,\,\left.+\frac{\left(\Delta _{\phi }+n\right){}^3 \left(4 \Delta _{\phi }+4 n+4-d\right)}{2 \left(2 \Delta _{\phi }+2 n+1\right) \left(4 \Delta _{\phi }+4 n+2-d\right)}z \zb + \ldots\right]\\
&\partial_{\Delta}G^{s}_{\Delta_{n,0},0}(z,\zb) =\frac{\log(z\zb)}{2}G^{s}_{\Delta_{n,0},0}(z,\zb)+z^n\zb^n\left[\frac{1}{4}(z+\zb)+\ldots\right]\,,
\ea
where the coefficients of higher-order terms are rational functions of $\Df$ and $d$ of increasing complexity. It will be useful to deal with this detailed structure of conformal blocks once and for all by defining the generating functional
\be
\Omega_{z,\zb} = \sum\limits_{n,\ell}\left[G^{s}_{\Delta_{n,\ell},\ell}(z,\zb)\alpha^{s}_{n,\ell} +
\partial_{\Delta}G^{s}_{\Delta_{n,\ell},\ell}(z,\zb)\beta^{s}_{n,\ell}\right]\,.
\label{eq:GenFun}
\ee
$\Omega_{z,\zb}$ is a family of functionals parametrized by $z$ and $\zb$. It can be thought of as a series in $z$ and $\zb$ around $z=\zb=0$ whose coefficients are functionals in $\mathcal{U}^*$. The terms in the series are of the form $z^i\zb^j$ and $\log(z\zb)z^i\zb^j$ with $i,j\in\mathbb{N}$. We can re-expand $\Omega_{z,\zb}$ in pure-power terms
\be
\Omega_{z,\zb} = \sum\limits_{i,j=0}^{\infty}\left[\widehat{\alpha}^{s}_{i,j}+\widehat{\beta}^{s}_{i,j}\,\frac{\log(z\zb)}{2}\right]\!z^i\zb^j\,,
\ee
where $\widehat{\alpha}^{s}_{i,j}$ and $\widehat{\beta}^{s}_{i,j}$ are functionals in $\mathcal{U}^*$. We have $\widehat{\alpha}^{s}_{i,j} = \widehat{\alpha}^{s}_{j,i}$ and $\widehat{\beta}^{s}_{i,j} = \widehat{\beta}^{s}_{j,i}$ thanks to the symmetry $\Omega_{z,\zb} = \Omega_{\zb,z}$. Note that thanks to the power-series structure of conformal blocks, $\widehat{\beta}^{s}_{i,j}$ is a finite linear combination of $\beta^{s}_{n,\ell}$ functionals and $\widehat{\alpha}^{s}_{i,j}$ is a finite linear combination of $\alpha^{s}_{n,\ell}$ and $\beta^{s}_{n,\ell}$. This map is clearly a bijection so we can also write $\alpha^{s}_{n,\ell}$  (or $\beta^{s}_{n,\ell}$) as a finite linear combinations of $\widehat{\alpha}^{s}_{i,j}$ and $\widehat{\beta}^{s}_{i,j}$ (or only $\widehat{\beta}^{s}_{i,j}$), with $i\in\{0,\ldots,n\}$ and $j\in\{i,\ldots, 2n+\ell -i\}$. The precise linear combinations can be easily found from the series expansion of general conformal blocks.

The point of this definition is that $\Omega_{z,\zb}$ and hence also $\widehat{\alpha}^{s}_{i,j}$ and $\widehat{\beta}^{s}_{i,j}$ are completely universal, i.e. independent of $\Df$ and $d$. In other words, the entire $\Df$ and $d$ dependence of $\alpha^{s}_{n,\ell}$ and $\beta^{s}_{n,\ell}$ is in the coefficients of their expansion using $\widehat{\alpha}^{s}_{i,j}$ and $\widehat{\beta}^{s}_{i,j}$. We can see that $\widehat{\alpha}^{s}_{i,j}$ and $\widehat{\beta}^{s}_{i,j}$ are universal by the following argument. Consider $\cG(z,\zb)\in\mathcal{U}$ which is Euclidean single-valued around $z=\zb=0$ and satisfies $\dDisc_s[\cG(z,\zb)] = 0$. We can then expand
\be
\cG(z,\zb) = \sum\limits_{i,j=0}^{\infty}\left[\widehat{a}_{i,j}+\widehat{b}_{i,j}\,\frac{\log(z\zb)}{2}\right]\!z^i\zb^j
\label{eq:cGExpansionMon}
\ee
for some coefficients $\widehat{a}_{i,j},\widehat{b}_{i,j}\in\mathbb{C}$. $\widehat{\alpha}^{s}_{i,j}$ and $\widehat{\beta}^{s}_{i,j}$ are precisely the linear functionals which extract these coefficients, i.e.
\be
\widehat{a}_{i,j} = \widehat{\alpha}^{s}_{i,j}[\cG(z,\zb)]\,,\qquad\widehat{b}_{i,j} = \widehat{\beta}^{s}_{i,j}[\cG(z,\zb)]\,.
\ee
Clearly, the expansion \eqref{eq:cGExpansionMon} is independent of $\Df$ and $d$ and so also the corresponding functionals have to be. An equivalent way of viewing this is the following. We start from the completeness relation for any $\cG(z,\zb)\in\mathcal{U}$ in the form
\ba
\cG(z,\zb) &= \sum\limits_{n,\ell}
\left\{G^{s}_{\Delta_{n,\ell},\ell}(z,\zb)\alpha^{s}_{n,\ell}[\cG(w,\wb)] +
\partial_{\Delta}G^{s}_{\Delta_{n,\ell},\ell}(z,\zb)\beta^{s}_{n,\ell}[\cG(w,\wb)]\right\}+\\
&\phantom{,}+\sum\limits_{n,\ell}
\left\{G^{t}_{\Delta_{n,\ell},\ell}(z,\zb)\alpha^{t}_{n,\ell}[\cG(w,\wb)] +
\partial_{\Delta}G^{t}_{\Delta_{n,\ell},\ell}(z,\zb)\beta^{t}_{n,\ell}[\cG(w,\wb)]\right\}\,.
\ea
This can be rewritten using the generating functional as
\be
\cG(z,\zb) = \Omega_{z,\zb}[\cG(w,\wb)] + \Omega_{1-z,1-\zb}[\cG(1-w,1-\wb)]\,.
\label{eq:dispersionOurs1}
\ee
Since $\Omega_{z,\zb}$ is independent of $\Df$ and $d$, this is a universal identity between functions in $\mathcal{U}$.

Since the basis functionals are defined by contour integrals \eqref{eq:basisKersGeneral}, we should define the generating kernel
\be
\mathcal{K}(z,\zb;w,\wb) =
\sum\limits_{n,\ell}\left[G^{s}_{\Delta_{n,\ell},\ell}(z,\zb)\mathcal{A}^{s}_{n,\ell}(w,\wb) +
\partial_{\Delta}G^{s}_{\Delta_{n,\ell},\ell}(z,\zb)\mathcal{B}^{s}_{n,\ell}(w,\wb)\right]\,,
\label{eq:GeneratingKernel}
\ee
so that the basis kernels, which are functions of $w$ and $\wb$, can be extracted by expanding $\mathcal{K}(z,\zb;w,\wb)$ in $z$ and $\zb$ in s-channel double-trace conformal blocks. We will give a conjectural closed formula for $\mathcal{K}(z,\zb;w,\wb)$ in Section \ref{sec:DispersionRelation}, building on the work of Carmi and Caron-Huot \cite{DispersionRelation}. It follows from the above that $\mathcal{K}(z,\zb;w,\wb)$ is completely universal, i.e. independent of $\Df$ and $d$. We can expand it in $z,\zb$ as
\be
\mathcal{K}(z,\zb;w,\wb) = \sum\limits_{i,j=0}^{\infty}\left[\widehat{\mathcal{A}}^{s}_{i,j}(w,\wb)+\widehat{\mathcal{B}}^{s}_{i,j}(w,\wb)\,\frac{\log(z\zb)}{2}\right]\!z^i\zb^j\,,
\ee
where $\widehat{\mathcal{A}}^{s}_{i,j}(w,\wb)$ and $\widehat{\mathcal{B}}^{s}_{i,j}(w,\wb)$ are kernels defining the functionals $\widehat{\alpha}^{s}_{i,j}$ and $\widehat{\beta}^{s}_{i,j}$ via
\ba
\widehat{\alpha}^{s}_{i,j}[\cG(w,\wb)] &= \int\limits_{C_-}\!\!\frac{dw}{2\pi i} \int\limits_{C_+}\!\!\frac{d\wb}{2\pi i} \,\widehat{\mathcal{A}}^{s}_{i,j}(w,\wb)\cG(w,\wb)\\
\widehat{\beta}^{s}_{i,j}[\cG(w,\wb)] &= \int\limits_{C_-}\!\!\frac{dw}{2\pi i} \int\limits_{C_+}\!\!\frac{d\wb}{2\pi i} \,\widehat{\mathcal{B}}^{s}_{i,j}(w,\wb)\cG(w,\wb)\,.
\ea
The kernels $\widehat{\mathcal{A}}^{s}_{i,j}(w,\wb)$ and $\widehat{\mathcal{B}}^{s}_{i,j}(w,\wb)$ are independent of $\Df$ and $d$. We will present an algorithm for finding closed formulas for these kernels in the next subsection. The kernels $\mathcal{A}^{s}_{n,\ell}(w,\wb)$ and $\mathcal{B}^{s}_{n,\ell}(w,\wb)$ can then be obtained as finite linear combinations of $\widehat{\mathcal{A}}^{s}_{i,j}(w,\wb)$ and $\widehat{\mathcal{B}}^{s}_{i,j}(w,\wb)$ as explained above.

\subsection{Constructing the $\beta$ kernels}\label{ssec:generalBeta}
We will start by finding $\widehat{\mathcal{B}}^{s}_{i,j}(w,\wb)$. Recall that
\be
\widehat{\mathcal{B}}^{s}_{0,0}(w,\wb) = \mathcal{B}^{s}_{0,0}(w,\wb) = \frac{2(w+\wb-2)}{(\wb-w)^2}\,.
\label{eq:BHat001}
\ee
The general kernel $\widehat{\mathcal{B}}^{s}_{i,j}(w,\wb)$ looks as follows
\be
\widehat{\mathcal{B}}^{s}_{i,j}(w,\wb) = \frac{p_{i,j}(w,\wb)}{(\wb-w)^{2(i+j+1)}}\,,
\label{eq:BHatAnsatz}
\ee
where $p_{i,j}(w,\wb)$ is a symmetric polynomial of total degree $2i+2j+1$, i.e. containing only terms $w^m\wb^n$ with $m+n\leq 2i+2j+1$. The upper bound on the degree ensures $\widehat{\mathcal{B}}^{s}_{i,j}(w,\wb)$ is suppressed at infinity so that the resulting functional is inside $\mathcal{U}^*$. Let us explain how the polynomial $p_{i,j}(w,\wb)$ is fixed by requiring that $\widehat{\beta}^{s}_{i,j}$ has the correct action. Let $\cG(w,\wb)\in\mathcal{U}$ be Euclidean single-valued satisfying $\dDisc_s[\cG] = 0$ so that
\be
\cG(w,\wb) = \cG_{0}(w,\wb) + \cG_{1}(w,\wb)\frac{\log(w\wb)}{2}\,,
\ee
where $\cG_{0}(w,\wb)=\cG_{0}(\wb,w)$ and $\cG_{1}(w,\wb)=\cG_{1}(\wb,w)$ are holomorphic at $w=\wb=0$. $\widehat{\beta}^{s}_{i,j}$ is defined by the following actions
\ba
&\widehat{\beta}^{s}_{i,j}[\cG(w,\wb)] = \frac{1}{i!\,j!}\left.[\partial^{i}_w\partial^{j}_{\wb}\cG_{1}(w,\wb)]\right|_{w=\wb=0}\\
&\widehat{\beta}^{s}_{i,j}[\cG(1-w,1-\wb)] = 0\,.
\label{eq:betaHatConstraints}
\ea
Let us take the ansatz \eqref{eq:BHatAnsatz} and impose these equations, starting from the first line. The following steps are a generalization of the manipulations familiar from Section \ref{ssec:beta00}. We have
\be
\widehat{\beta}^{s}_{i,j}[\cG(w,\wb)] = \int\limits_{C_-}\!\!\frac{dw}{2\pi i} \int\limits_{C_+}\!\!\frac{d\wb}{2\pi i} \,
\frac{p_{i,j}(w,\wb)}{(\wb-w)^{2(i+j+1)}}
\left[\cG_{0}(w,\wb) + \cG_{1}(w,\wb)\frac{\log(w\wb)}{2}\right]\,.
\ee
Let us wrap the $w$ contour onto the left branch cut, picking up the discontinuity, which comes only from the $\log(w)$ factor inside the square bracket.
\be
\widehat{\beta}^{s}_{i,j}[\cG(w,\wb)] = -\frac{1}{2}\int\limits_{-\infty}^{0}\!\!dw \int\limits_{C_+}\!\!\frac{d\wb}{2\pi i} \,
\frac{p_{i,j}(w,\wb)}{(\wb-w)^{2(i+j+1)}}\cG_{1}(w,\wb)\,.
\ee
Now, let us deform the $\wb$ contour to the left, where the only singularity we encounter is the pole at $\wb=w$. We pick up the residue and find
\be
\widehat{\beta}^{s}_{i,j}[\cG(w,\wb)] = -\frac{1}{2\,(2i+2j+1)!}\int\limits_{-\infty}^{0}\!\!\!dw\,
\left.\partial^{2i+2j+1}_{\wb}\right|_{\wb=w}\left[p_{i,j}(w,\wb)\cG_{1}(w,\wb)\right]\,.
\label{eq:betaGeneralStep1}
\ee
We would like to write the integrand as a total derivative with respect to $w$ of a function of $w$, so that the integral localizes to a boundary term at $w=\wb=0$. This is possible since any holomorphic function $f(w,\wb)$ which satisfies $f(w,\wb)=f(\wb,w)$ satisfies the following identity for any $n\in\mathbb{N}$
\be
\left.\partial^{2n+1}_{\wb}\right|_{\wb=w}f(w,\wb) = \partial_{w}\left[\sum\limits_{m=0}^{n}\frac{(-1)^{m}}{1+\delta_{m,n}}f^{(m,2n-m)}(w,w)\right]\,,
\ee
where $f^{(i,j)}(w,\wb) = \partial^i_w\partial^j_{\wb}f(w,\wb)$. After using this identity in \eqref{eq:betaGeneralStep1}, we get
\be
\widehat{\beta}^{s}_{i,j}[\cG(w,\wb)] = \frac{-1}{\scriptstyle{2\,(2i+2j+1)!}}
\sum\limits_{m=0}^{i+j}\frac{(-1)^{m}}{1+\delta_{m,i+j}}\left.\partial^{m}_{w}\partial^{2i+2j-m}_{\wb}[p_{i,j}(w,\wb)\cG_{1}(w,\wb)]\right|_{w=\wb=0}\,.
\ee
When we evaluate the derivatives acting on the square bracket and set $w=\wb=0$, we find a linear combination of derivatives $\left.\partial^k_w\partial^l_{\wb}\cG_{1}(w,\wb)\right|_{w=\wb=0}$. Which linear combination we get depends on the polynomial $p_{i,j}(w,\wb)$. Therefore, the first line of \eqref{eq:betaHatConstraints} constrains the coefficients in $p_{ij}(w,\wb)$
\ba
&\frac{-1}{2\,(2i+2j+1)!}\sum\limits_{m=0}^{i+j}\frac{(-1)^{m}}{1+\delta_{m,i+j}}\left.\partial^{m}_{w}\partial^{2i+2j-m}_{\wb}[p_{i,j}(w,\wb)\cG_{1}(w,\wb)]\right|_{w=\wb=0} = \\
&\qquad\qquad\qquad\qquad\qquad\qquad\qquad\qquad\quad\;=\frac{1}{i!\,j!}\left.[\partial^{i}_w\partial^{j}_{\wb}\cG_{1}(w,\wb)]\right|_{w=\wb=0}\,.
\label{eq:const1}
\ea
We can impose the second line of \eqref{eq:betaHatConstraints} by identical manipulations, with replacement $p_{i,j}(w,\wb)\mapsto p_{i,j}(1-w,1-\wb)$
\be
\sum\limits_{m=0}^{i+j}\frac{(-1)^{m}}{1+\delta_{m,i+j}}\left.\partial^{m}_{w}\partial^{2i+2j-m}_{\wb}[p_{i,j}(1-w,1-\wb)\cG_{1}(w,\wb)]\right|_{w=\wb=0} = 0\,,
\label{eq:const2}
\ee
which is another constraint on the polynomial $p_{i,j}(w,\wb)$. Just like we saw in the simplest example $i=j=0$ in section \ref{ssec:beta00}, these two constraints fix the polynomial and thus the kernel $\widehat{\mathcal{B}}^{s}_{i,j}(w,\wb)$ uniquely. For several low-lying example, we found
\ba
\widehat{\mathcal{B}}^{s}_{0,0}(w,\wb) &= \frac{2 (w+\wb -2)}{(w-\wb )^2}\\
\widehat{\mathcal{B}}^{s}_{0,1}(w,\wb) &= -\frac{4 \left(3 w^2 \wb -2 w^2+3 w \wb ^2-8 w \wb +3 w-2 \wb ^2+3 \wb\right)}{(w-\wb )^4}\\
\widehat{\mathcal{B}}^{s}_{1,1}(w,\wb) &= 12(w-\wb)^{-6}\left(2 w^4 \wb -w^4+18 w^3 \wb ^2-26 w^3 \wb +6 w^3+18 w^2 \wb ^3-\right.\\
&\qquad\qquad\qquad\quad-66 w^2 \wb ^2+54 w^2 \wb -6 w^2+2 w \wb ^4-26 w \wb ^3+54 w \wb ^2-\\
&\qquad\qquad\qquad\quad\left.-28 w \wb -\wb ^4+6 \wb ^3-6 \wb ^2\right)\,.
\ea
Experimentally, we also found a general formula for $i=0$ and arbitrary $j\in\mathbb{N}$
\ba
\widehat{\mathcal{B}}^{s}_{0,j}(w,\wb) = &(\wb-w)^{-2j-2}\times\\
&\times\sum_{m,n=0}^{j+1}
\frac{2 (-1)^{m+n-j}(j+1)(m+n-2-2 j)(j+1)!(m+n)!}{m!\,n!\,(j+1-m)!(j+1-n)!(m+n-j)!}
w^m\wb^n\,,
\ea
where as usual, the factorial $(m+n-j)!$ in the denominator is assumed infinite whenever $m+n-j<0$. This means we can now also write a closed formula for the $\mathcal{B}^{s}_{n,\ell}(w,\wb)$ kernels on the leading Regge trajectory
\be
\mathcal{B}^{s}_{0,\ell}(w,\wb) =
\sum\limits_{j=0}^{\ell}\frac{(-1)^{\ell-j}(\Df+j)^2_{\ell-j}}{(\ell-j)! (2\Df+j+\ell -1)_{\ell-j}}\widehat{\mathcal{B}}^{s}_{0,j}(w,\wb)\,.
\ee
Note that there is no $d$ dependence since the $z\rightarrow 0$, fixed $\zb$ limit of conformal blocks is independent of $d$.

\subsection{Constructing the $\alpha$ kernels}\label{ssec:generalAlpha}
To have a complete picture, it remains to construct the kernels $\widehat{\mathcal{A}}^{s}_{i,j}(w,\wb)$ of the universal $\widehat{\alpha}^{s}_{i,j}$ functionals. We claim that they take the following form
\be
\widehat{\mathcal{A}}^{s}_{i,j}(w,\wb) = \frac{1}{2}\log\!\left[\frac{w(1-w)\wb(1-\wb)}{(\wb-w)^4}\right]\widehat{\mathcal{B}}^{s}_{i,j}(w,\wb) + \widehat{\mathcal{C}}^{s}_{i,j}(w,\wb)\,,
\label{eq:AHatGeneral}
\ee
where $\widehat{\mathcal{C}}^{s}_{i,j}(w,\wb)$ is meromorphic. In fact, $\widehat{\mathcal{C}}^{s}_{i,j}(w,\wb)$ can be decomposed into $\widehat{\mathcal{B}}^{s}_{m,n}(w,\wb)$ and $\widehat{\mathcal{B}}^{t}_{m,n}(w,\wb)$ kernels\footnote{Recall that $\widehat{\mathcal{B}}^{t}_{m,n}(w,\wb) = \widehat{\mathcal{B}}^{s}_{m,n}(1-w,1-\wb)$.} with $m+n\leq i+j$. For example,
\ba
\widehat{\mathcal{A}}^{s}_{0,0}(w,\wb) &= 
\frac{(w+\wb -2)}{(w-\wb )^2}\left\{\log\!\left[\frac{w(1-w)\wb(1-\wb)}{(\wb-w)^4}\right]+4\right\} =\\
&=\frac{1}{2}\log\!\left[\frac{w(1-w)\wb(1-\wb)}{(\wb-w)^4}\right]\!\widehat{\mathcal{B}}^{s}_{0,0}(w,\wb) +
2\,\widehat{\mathcal{B}}^{s}_{0,0}(w,\wb)\,.
\label{eq:AHat001}
\ea
The most notable new ingredient is the appearance of the logarithm. Note that when $w\in C_-$ and $\wb\in C_+$, then
\be
\mathrm{Re}\!\left[\frac{w(1-w)\wb(1-\wb)}{(\wb-w)^4}\right]>0\,.
\ee
so the logarithm is well-defined. To understand better where the structure of the kernels comes from, let us first spell out the definining properties of the $\widehat{\alpha}^{s}_{i,j}$ functionals. Let $\cG(w,\wb)\in\mathcal{U}$ be Euclidean single-valued satisfying $\dDisc_s[\cG] = 0$ so that
\be
\cG(w,\wb) = \cG_{0}(w,\wb) + \cG_{1}(w,\wb)\frac{\log(w\wb)}{2}\,,
\ee
where $\cG_{0}(w,\wb)=\cG_{0}(\wb,w)$ and $\cG_{1}(w,\wb)=\cG_{1}(\wb,w)$ are holomorphic at $w=\wb=0$. $\widehat{\alpha}^{s}_{i,j}$ is required to have the actions
\ba
&\widehat{\alpha}^{s}_{i,j}[\cG(w,\wb)] = \frac{1}{i!\,j!}\left.[\partial^{i}_w\partial^{j}_{\wb}\cG_{0}(w,\wb)]\right|_{w=\wb=0}\\
&\widehat{\alpha}^{s}_{i,j}[\cG(1-w,1-\wb)] = 0\,,
\label{eq:alphaHatConstraints}
\ea
which should be compared with \eqref{eq:betaHatConstraints}. We can proceed using a similar contour-deformation argument as in Section \ref{ssec:generalBeta}. The first difference is that when we wrap the $w$ contour onto the left branch cut, we now also pick the discontinuity of $\widehat{\mathcal{A}}^{s}_{i,j}(w,\wb)$ due to $\log(w)$. Since $\log(w)$ is multiplied by $\widehat{\mathcal{B}}^{s}_{i,j}(w,\wb)$, we can conclude that
\ba
&\widehat{\alpha}^{s}_{i,j}[\cG(w,\wb)] = \frac{1}{i!\,j!}\left.[\partial^{i}_w\partial^{j}_{\wb}\cG_{0}(w,\wb)]\right|_{w=\wb=0} + \omega^{s}_{i,j}[\cG_{1}(w,\wb)]\\
&\widehat{\alpha}^{s}_{i,j}[\cG(1-w,1-\wb)] = \omega^{s}_{i,j}[\cG_{1}(1-w,1-\wb)]\,,
\ea
where $\omega^{s}_{i,j}$ is some linear functional. The remainder term $\widehat{\mathcal{C}}^{s}_{i,j}(w,\wb)$ is present in \eqref{eq:AHatGeneral} to guarantee that in fact $\omega^{s}_{i,j}$ vanishes identically and thus $\widehat{\alpha}^{s}_{i,j}$ acts as required.\footnote{We currently do not have a full explanation for why the argument of the logarithm must be $\frac{w(1-w)\wb(1-\wb)}{(\wb-w)^4}$. We originally found these functionals in a very different way, similar to the method used in \cite{Mazac:2016qev}, which automatically produced a factor of $\log\!\left[\frac{w(1-w)\wb(1-\wb)}{(\wb-w)^4}\right]$. It would be interesting to understand it more directly using a contour deformation.}

Let us conclude with several explicit examples. We introduce the shorthand notation
\be
\mathcal{L}(w,\wb) = \log\!\left[\frac{w(1-w)\wb(1-\wb)}{(\wb-w)^4}\right]\,.
\ee
We found the following closed formulas
\ba
\widehat{\mathcal{A}}^{s}_{0,0}(w,\wb) &= \frac{\mathcal{L}(w,\wb)}{2}\,\widehat{\mathcal{B}}^{s}_{0,0}(w,\wb) +
2\,\widehat{\mathcal{B}}^{s}_{0,0}(w,\wb)\\
\widehat{\mathcal{A}}^{s}_{0,1}(w,\wb) &= \frac{\mathcal{L}(w,\wb)}{2}\,\widehat{\mathcal{B}}^{s}_{0,1}(w,\wb)
-\frac{1}{6}\,\widehat{\mathcal{B}}^{s}_{0,0}(w,\wb)
+\frac{8}{3}\,\widehat{\mathcal{B}}^{s}_{0,1}(w,\wb)
+\frac{1}{3}\,\widehat{\mathcal{B}}^{t}_{0,0}(w,\wb)\\
\widehat{\mathcal{A}}^{s}_{0,2}(w,\wb) &= \frac{\mathcal{L}(w,\wb)}{2}\,\widehat{\mathcal{B}}^{s}_{0,2}(w,\wb)
-\frac{1}{20}\,\widehat{\mathcal{B}}^{s}_{0,0}(w,\wb)
-\frac{2}{5}\,\widehat{\mathcal{B}}^{s}_{0,1}(w,\wb)
+\frac{52}{15}\,\widehat{\mathcal{B}}^{s}_{0,2}(w,\wb)-\\
&\qquad\qquad\qquad\qquad\quad\;\;-\frac{1}{10}\,\widehat{\mathcal{B}}^{s}_{1,1}(w,\wb)
+\frac{1}{5}\,\widehat{\mathcal{B}}^{t}_{0,0}(w,\wb)
+\frac{3}{10}\,\widehat{\mathcal{B}}^{t}_{0,1}(w,\wb)\\
\widehat{\mathcal{A}}^{s}_{1,1}(w,\wb) &= \frac{\mathcal{L}(w,\wb)}{2}\,\widehat{\mathcal{B}}^{s}_{1,1}(w,\wb)
-\frac{1}{5}\,\widehat{\mathcal{B}}^{s}_{0,0}(w,\wb)
-\frac{13}{5}\,\widehat{\mathcal{B}}^{s}_{0,1}(w,\wb)
+\frac{26}{5}\,\widehat{\mathcal{B}}^{s}_{0,2}(w,\wb)+\\
&\qquad\qquad\qquad\qquad\quad\;\;+\frac{19}{15}\,\widehat{\mathcal{B}}^{s}_{1,1}(w,\wb)
-\frac{1}{5}\,\widehat{\mathcal{B}}^{t}_{0,0}(w,\wb)
+\frac{6}{5}\,\widehat{\mathcal{B}}^{t}_{0,1}(w,\wb)\,.
\ea

\section{Functional Actions from Witten Diagrams}\label{ssec:witten}
\subsection{Regge-improved Witten diagrams and Polyakov-Regge blocks}
The actions of functionals appear in the definition (\ref{eq:polyakovS}), (\ref{eq:polyakovT}) of Polyakov-Regge blocks  as the decomposition coefficients of double-trace conformal blocks. On the other hand,  as we briefly commented on in Section \ref{subsec:PolyakovRegge} Polyakov-Regge blocks can be identified as appropriate combinations of exchange and contact Witten diagrams in AdS space. In this section we make this connection more precise. This gives us an independently method to obtain the actions of the functionals in $\mathcal{U}^*$, by relating them to the conformal block decomposition coefficients of Witten diagrams.

Let us begin by recalling that the conformal block decomposition of the s- and t-channel Polyakov-Regge blocks with conformal dimension $\Delta$ and spin $J$ have the same structure as the exchange Witten diagrams for a field with the same quantum numbers. Unlike the Polyakov-Regge blocks, however, the exchange Witten diagrams generically do not live in the space of functions $\mathcal{U}$ as they have worse behavior in the u-channel Regge limit. Nevertheless, as we will show below, it is always possible to improve the Regge behavior of the exchange Witten diagrams by adding a {\it finite} number of contact Witten diagrams with up to $2(J-1)$ derivatives in the quartic vertices. The existence of such Regge-improved exchange Witten diagrams is particularly evident in Mellin space \cite{Mack:2009mi,Penedones:2010ue}. In Mellin space, we write the correlator as
\begin{equation}
\mathcal{G}(U,V)=\int_{-i\infty}^{i\infty}\frac{dsdt}{(4\pi i)^2} U^{\frac{s}{2}-\Delta_\phi}V^{\frac{t}{2}-\Delta_\phi}\mathcal{M}(s,t)\Gamma^2\left(\frac{2\Delta_\phi-s}{2}\right)\Gamma^2\left(\frac{2\Delta_\phi-t}{2}\right)\Gamma^2\left(\frac{s+t-2\Delta_\phi}{2}\right)\;.
\end{equation}
The requirement that $\mathcal{G}\in \mathcal{U}$ translates into the u-channel Regge behavior of the Mellin amplitude \cite{Costa:2012cb} that 
\begin{equation}
\mathcal{M}(s,t)\sim s^{-\epsilon}\;,\quad\quad s\to\infty\;,u\;\text{fixed}
\end{equation}
where $u=4\Delta_\phi-s-t$. Generally, the s, t-channel exchange Mellin amplitudes have the following structure 
\begin{equation}
\mathcal{M}^{s}_{\Delta,J}(s,t)=\sum_{m=0}^\infty \frac{P_{J,m}(t,u)}{s-(\Delta-J)-2m}+Q_{J-1}(t,u)\;,
\end{equation}
\begin{equation}
\mathcal{M}^{t}_{\Delta,J}(s,t)=\sum_{m=0}^\infty \frac{P_{J,m}(s,u)}{t-(\Delta-J)-2m}+Q_{J-1}(s,u)\;.
\end{equation}
Here $P_{J,m}(t,u)$ are degree-$J$ polynomials in $t$ and $u$, and is symmetric or anti-symmetry depending on whether $J$ is even or odd. Similarly, $Q_{J-1}(t,u)$ is a degree-$(J-1)$ polynomial, and has the same symmetric properties as $P_{J,m}(t,u)$. We should note that the above exchange Mellin amplitudes are computed with respect to a specific choice of the cubic vertices. Different choices of the cubic vertices amount to the ambiguities of adding contact Witten diagrams with up to $2(J-1)$ derivatives \cite{Costa:2014kfa}. In Mellin space, these ambiguities correspond to the possibility of adding to the exchange Mellin amplitude a polynomial of degree $J-1$.
Let us now focus on the u-channel Regge limit. Both $\mathcal{M}^{s}_{\Delta,J}$ and $\mathcal{M}^{t}_{\Delta,J}$ have growth $s^{J-1}$ for generically chosen cubic vertices. In the scalar $J=0$ case, the Mellin amplitudes already exhibit super-bounded behavior, and therefore the corresponding exchange diagrams in position space live in $\mathcal{U}$. For spin $J\geq 1$, it is clearly possible to add a finite collection of contact Witten diagrams with up to $2(J-1)$ derivatives such that the improved Mellin amplitude behaves as $s^{-1}$ in the u-channel Regge limit. The improved exchange Witten diagrams are then identified with the Polyakov-Regge blocks, after normalizing the single-trace conformal block to have unit coefficient.

This gives a concrete prescription for constructing Polyakov-Regge blocks from Witten diagrams. By further computing the conformal block decomposition coefficients, we can extract the action of the functionals. Below we will demonstrate this procedure for two explicit examples with $J=0$ and $J=1$, and obtain functional actions on conformal blocks with spins 0 and 1.

\subsection{Spin $J=0$}
Let us first consider the scalar case where the exchanged field has $J=0$. In position space, the exchange Witten diagrams are defined as 
\begin{equation}
W^{s}_{\Delta,J=0}=\int \frac{d^{d+1}z}{z_0^{d+1}}\frac{d^{d+1}w}{w_0^{d+1}}G_{B\partial}^{\Delta_\phi}(x_1,z)G_{B\partial}^{\Delta_\phi}(x_2,z)G_{BB}^{\Delta}(z,w)G_{B\partial}^{\Delta_\phi}(x_3,w)G_{B\partial}^{\Delta_\phi}(x_4,w)\;,
\end{equation}
\begin{equation}
W^{t}_{\Delta,J=0}=\int \frac{d^{d+1}z}{z_0^{d+1}}\frac{d^{d+1}w}{w_0^{d+1}}G_{B\partial}^{\Delta_\phi}(x_1,z)G_{B\partial}^{\Delta_\phi}(x_4,z)G_{BB}^{\Delta}(z,w)G_{B\partial}^{\Delta_\phi}(x_3,w)
G_{B\partial}^{\Delta_\phi}(x_3,w)\end{equation}
where $G_{B\partial}^{\Delta_\phi}(x_i,z)$ are the bulk-to-boundary propagators
\begin{equation}
G_{B\partial}^{\Delta_\phi}(x_i,z)=\left(\frac{z_0}{z_0^2+(\vec{z}-\vec{x}_i)^2}\right)^{\Delta_\phi}\;,
\end{equation}
and $G_{BB}^\Delta(z,w)$ is the bulk-to-bulk propagator which satisfies the equation of motion
\begin{equation}
(\square-\Delta(\Delta-d))G_{BB}^\Delta(z,w)=\delta^{d+1}(z-w)\;.
\end{equation}
The Mellin amplitudes of the scalar exchange Witten diagrams read
\begin{equation}
\mathcal{M}_{\Delta,J=0}^{s}=\sum_{m=0}^\infty \frac{a_m}{s-\Delta-2m}\;,
\end{equation}
\begin{equation}
\mathcal{M}_{\Delta,J=0}^{t}=\sum_{m=0}^\infty \frac{a_m}{t-\Delta-2m}
\end{equation}
where 
\begin{equation}
a_m=-\frac{ \pi ^{d/2} \Gamma \left(\frac{\Delta-d}{2}+\Delta_\phi \right)^2\Gamma \left(m+\frac{\Delta}{2}-\Delta_\phi +1\right)^2}{8 \Gamma (\Delta_\phi )^4 \Gamma (m+1) \Gamma \left(\frac{\Delta}{2}-\Delta_\phi +1\right)^2 \Gamma \left(-\frac{d}{2}+m+\Delta+1\right)}\;.
\end{equation}
In the u-channel Regge limit,
\begin{equation}
\mathcal{M}_{\Delta,J=0}^{s}\,,\mathcal{M}_{\Delta,J=0}^{t}\sim s^{-1}\;,
\end{equation}
therefore we have
\begin{equation}
\mathcal{W}^{s}_{\Delta,J=0}\in \mathcal{U}\;,\quad \mathcal{W}^{t}_{\Delta,J=0}\in \mathcal{U}
\end{equation}
where we defined
\begin{equation}
W^{s,t}_{\Delta,J=0}=\frac{1}{(x_{12}^{2}x_{34}^2)^{\Delta_\phi}}\mathcal{W}^{s,t}_{\Delta,J=0}\;.
\end{equation}
The scalar exchange Witten diagrams admit the following conformal block decomposition
\begin{equation}
\begin{split}
\mathcal{W}^{s}_{\Delta,J=0}={}&A G^{s}_{\Delta,0}+\sum_{n=0}^{\infty} A_{n,0} G^{s}_{\Delta_{n,0},0}+\sum_{n=0}^\infty D_{n,0}\partial G^{s}_{\Delta_{n,0},0}\;,\\ 
={}& \sum_{\ell=0}^\infty \sum_{n=0}^\infty B_{n,\ell} G^{t}_{\Delta_{n,\ell},\ell}+\sum_{\ell=0}^\infty \sum_{n=0}^\infty C_{n,\ell} \partial G^{t}_{\Delta_{n,\ell},\ell}\;,
\end{split}
\end{equation}
\begin{equation}
\begin{split}
\mathcal{W}^{t}_{\Delta,J=0}={}&A G^{t}_{\Delta,0}+\sum_{n=0}^{\infty} A_{n,0} G^{t}_{\Delta_{n,0},0}+\sum_{n=0}^\infty D_{n,0}\partial G^{t}_{\Delta_{n,0},0}\;,\\ 
={}& \sum_{\ell=0}^\infty \sum_{n=0}^\infty B_{n,\ell} G^{s}_{\Delta_{n,\ell},\ell}+\sum_{\ell=0}^\infty \sum_{n=0}^\infty C_{n,\ell} \partial G^{s}_{\Delta_{n,\ell},\ell}\;.
\end{split}
\end{equation}
We can identify the $J=0$ Polyakov-Regge blocks as
\begin{equation}
P_{\Delta,J=0}^{s}=\frac{1}{A} \mathcal{W}^{s}_{\Delta,J=0}\;,\quad P_{\Delta,J=0}^{t}=\frac{1}{A} \mathcal{W}^{t}_{\Delta,J=0}\;,
\end{equation}
which gives
\begin{equation}
\begin{split}
{}&\alpha^{s}_{n,\ell}[G^{s}_{\Delta,0}]=-\frac{A_{n,0}}{A}\delta_{\ell,0}\;,\quad \beta^{s}_{n,\ell}[G^{s}_{\Delta,0}]=-\frac{D_{n,0}}{A}\delta_{\ell,0}\;,\\
{}&\alpha^{t}_{n,\ell}[G^{s}_{\Delta,0}]=\frac{B_{n,\ell}}{A}\;,\quad \beta^{t}_{n,\ell}[G^{s}_{\Delta,0}]=\frac{C_{n,\ell}}{A}\;,
\end{split}
\end{equation}
and similarly the action of functionals on the t-channel conformal block $G^{t}_{\Delta,0}$. Here and below, we will focus on the action of $\beta$ functionals which admit simpler expressions. These functional actions are related to the anomalous dimensions. From the decomposition coefficients in the direct channel, we find
\begin{equation}\small
\begin{split}
\beta^{s}_{n,\ell}[G^{s}_{\Delta,0}]={}&\frac{4 \Gamma (\Delta) \Gamma \left(-\frac{d}{2}+\Delta+1\right)}{(n!)^2 \Gamma \left(\frac{\Delta}{2}\right)^4 \Gamma \left(\Delta_\phi -\frac{\Delta}{2}\right)^2 (-\Delta+2 \Delta_\phi +2 n) \Gamma \left(\frac{\Delta-d}{2}+\Delta_\phi \right)^2 (-d+\Delta+2 \Delta_\phi +2 n)}\\
{}&\times \frac{\Gamma (n+\Delta_\phi )^4 \Gamma \left(-\frac{d}{2}+n+2 \Delta_\phi \right)^2}{\Gamma (2 (n+\Delta_\phi )) \Gamma \left(2 (n+\Delta_\phi )-\frac{d}{2}\right)}\delta_{\ell,0}\;.
\end{split}
\end{equation}
In the crossed channel, from the anomalous dimensions of the leading double-trace operators ($n=0$), we have 
\begin{equation}\small
\begin{split}
{}&\beta^{t}_{0,\ell}[G^{s}_{\Delta,0}]=-\frac{\Gamma (\Delta) \Gamma (\Delta_\phi )^2 \left((\Delta_\phi )_{\ell }\right){}^2 \Gamma (\ell +\Delta_\phi )^2 \Gamma \left(-\frac{d}{2}+\ell +2 \Delta_\phi \right) \Gamma \left(-\frac{d}{2}+\ell +\Delta+2 \Delta_\phi \right)}{\ell ! \Gamma \left(\frac{\Delta}{2}\right)^2 \Gamma \left(\Delta_\phi -\frac{\Delta}{2}\right)^2 (\ell +2 \Delta_\phi -1)_{\ell } \Gamma \left(\ell +\frac{\Delta}{2}+\Delta_\phi \right)^2 \Gamma \left(-\frac{d}{2}+\ell +\frac{\Delta}{2}+2 \Delta_\phi \right)^2}\\
{}&\times {}_7F_6 \left(\begin{array}{c}\frac{\Delta}{2},\frac{\Delta}{2},\frac{-d+\Delta+2 \Delta_\phi }{2},\frac{-d+\Delta+2 \Delta_\phi }{2},\frac{-d+2 \Delta+4 \Delta_\phi +2 \ell +2}{4} ,2 \Delta_\phi +\ell -1,\frac{-d+2 \Delta+4 \Delta_\phi +2 \ell -2}{2} \\\frac{-d+2 \Delta+2}{2},\frac{-d+2 \Delta+4 \Delta_\phi +2 \ell -2}{4},\frac{\Delta+2 \Delta_\phi +2 \ell }{2} ,\frac{\Delta+2 \Delta_\phi +2 \ell }{2} ,\frac{-d+\Delta+4 \Delta_\phi +2 \ell }{2},\frac{-d+\Delta+4 \Delta_\phi +2 \ell }{2} \end{array};1\right)\;.
\end{split}
\label{eq:7F6}
\end{equation}
Here we have used the result of \cite{Sleight:2018ryu} (see also \cite{Sleight:2018epi}) for $D_{0,\ell}$ which holds down to $\ell=0$ when $J=0$.\footnote{Note that the particular ${}_7F_6$ appearing in \eqref{eq:7F6} can be written as a combination of two ${}_4F_3$ hypergeometrics, see equation (2.28) of reference \cite{Sleight:2018ryu}.} To obtain $\beta^{t}_{n,\ell}[G^{s}_{\Delta,0}]$ for $n>0$, we need the decomposition coefficients $D_{n,\ell}$ for the sub-leading double-trace operators. These coefficients uniquely determined by the the recursion relations discovered in \cite{Zhou:2018sfz}, and therefore all $\beta^{t}_{n>0,\ell}[G^{s}_{\Delta,0}]$ functional actions are recursively fixed. 

Let us consider the special case with $d=2$, $\Delta_\phi=1$. We have
\begin{equation}
\beta^{t}_{0,0}[G^{s}_{\Delta,0}]=\frac{\sin ^2\left(\frac{\pi  \Delta}{2}\right) \Gamma (\Delta)^2 \psi ^{(2)}\left(\frac{\Delta}{2}\right)}{\pi ^2 \Gamma \left(\frac{\Delta}{2}\right)^4}\;,
\end{equation}
which reproduces the result obtained by using kernels.

\subsection{Spin $J=1$}
We now look at the case with $J=1$. The vector exchange Witten diagrams are defined as
\begin{equation}
W_{\Delta,J=1}^{s}=\int \frac{d^{d+1}z}{z_0^{d+1}}\frac{d^{d+1}w}{w_0^{d+1}} J_\mu (x_1,x_2;z) G^{\mu\nu,\Delta}_{BB}(z,w) J_\nu (x_3,x_4,w)\;,
\end{equation}
\begin{equation}
W_{\Delta,J=1}^{t}=\int \frac{d^{d+1}z}{z_0^{d+1}}\frac{d^{d+1}w}{w_0^{d+1}} J_\mu (x_1,x_4;z) G^{\mu\nu,\Delta}_{BB}(z,w) J_\nu (x_2,x_3,w)
\end{equation}
where we have coupled the vector field minimally to the conserved current 
\begin{equation}
J_\mu (x_i,x_j;z)=\frac{1}{2}\left(\partial_\mu G^{\Delta_\phi}_{B\partial}(x_i,z)G^{\Delta_\phi}_{B\partial}(x_j,z)-G^{\Delta_\phi}_{B\partial}(x_i,z) \partial_\mu G^{\Delta_\phi}_{B\partial}(x_j,z)\right)\;,
\end{equation}
and $G^{\mu\nu,\Delta}_{BB}(z,w)$ is the bulk-to-bulk propagator for spin-1 particle. The Mellin amplitudes have the following form 
\begin{equation}
\mathcal{M}^{s}_{\Delta,J=1}(s,t)=\sum_{m=0}^\infty \frac{b_m(t-u)}{s-(\Delta-1)-2m}\;,
\end{equation}
\begin{equation}
\mathcal{M}^{t}_{\Delta,J=1}(s,t)=\sum_{m=0}^\infty \frac{b_m(s-u)}{t-(\Delta-1)-2m}
\end{equation}
where 
\begin{equation}
b_m=\frac{\pi ^{d/2} \Gamma \left(\frac{1}{2} (-d+\Delta+1)+\Delta_\phi \right)^2 \Gamma \left(m+\frac{\Delta}{2}-\Delta_\phi +\frac{1}{2}\right)^2}{16 \Gamma (\Delta_\phi )^4\Gamma (m+1) \Gamma \left(\frac{1}{2} (\Delta-2 \Delta_\phi +1)\right)^2 \Gamma \left(-\frac{d}{2}+m+\Delta+1\right)}\;.
\end{equation}
In the u-channel Regge limit,
\begin{equation}
\mathcal{M}^{s}_{\Delta,J=1}\sim s^0\;, \quad \mathcal{M}^{t}_{\Delta,J=1}\sim s^0\;.
\end{equation}
We can add a zero-derivative contact term to cancel the leading $\mathcal{O}(s^0)$ contributions, such that both Mellin amplitudes have improved Regge behavior $\mathcal{O}(s^{-1})$
\begin{equation}
W_{\Delta,J=1}^{s,imp}=W_{\Delta,J=1}^{s}-\frac{1}{4}D_{\Delta_\phi \Delta_\phi \Delta_\phi \Delta_\phi}\;,
\end{equation}
\begin{equation}
W_{\Delta,J=1}^{t,imp}=W_{\Delta,J=1}^{t}+\frac{1}{4}D_{\Delta_\phi \Delta_\phi \Delta_\phi \Delta_\phi}\;.
\end{equation}
The improved exchange Witten diagrams can then be identified with the Polyakov-Regge blocks after a rescaling 
\begin{equation}
P_{\Delta,J=1}^{s}=\frac{1}{\tilde{A}}\mathcal{W}_{\Delta,J=1}^{s,imp}\;,\quad P_{\Delta,J=1}^{t}=\frac{1}{\tilde{A}}\mathcal{W}_{\Delta,J=1}^{t,imp}
\end{equation}
where 
\begin{equation}
W^{s,t}_{\Delta,J=1}=\frac{1}{(x_{12}^{2}x_{34}^2)^{\Delta_\phi}}\mathcal{W}^{s,t}_{\Delta,J=1}\;,
\end{equation}
and $\tilde{A}$ is the coefficient of the single-trace conformal block in the direct channel.

From the conformal block decomposition coefficients of $\mathcal{W}_{\Delta,J=1}^{s,imp}$ and $\mathcal{W}_{\Delta,J=1}^{t,imp}$, we can extract the action of functionals on conformal blocks with $J=1$. In the direct channel, the improved vector exchange Witten diagrams contain double-trace operators with only spin $\ell=0,1$. Therefore $\alpha^{s}_{n,\ell}[G^{s}_{\Delta,1}]$, $\beta^{s}_{n,\ell}[G^{s}_{\Delta,1}]$, $\alpha^{t}_{n,\ell}[G^{t}_{\Delta,1}]$, $\beta^{t}_{n,\ell}[G^{t}_{\Delta,1}]$ are only nonzero when $\ell=0,1$. For the $\beta$ functions, the explicit results read
\begin{equation}\small
\begin{split}
\beta^{s}_{n,0}[G^{s}_{\Delta,1}]={}&\frac{2^{\Delta}  \Gamma \left(\frac{\Delta+2}{2}\right) \Gamma \left(-\frac{d}{2}+\Delta+1\right) \cos ^2\left(\frac{\pi  (\Delta-2 \Delta_\phi )}{2}\right) \Gamma \left(\frac{\Delta-2 \Delta_\phi +1}{2}\right)^2 }{\pi ^{5/2} (n!)^2 \Gamma \left(\frac{\Delta-1}{2}\right) \Gamma \left(\frac{\Delta+1}{2}\right)^2 \Gamma (2 (n+\Delta_\phi )) }\\
{}&\times \frac{\Gamma (n+\Delta_\phi )^4 \Gamma \left(-\frac{d}{2}+n+2 \Delta_\phi \right)^2}{\Gamma \left(\frac{1}{2} (-d+\Delta+1)+\Delta_\phi \right)^2 \Gamma \left(2 (n+\Delta_\phi )-\frac{d}{2}\right)}\;,
\end{split}
\end{equation}
\begin{equation}\small
\begin{split}
\beta^{s}_{n,1}[G^{s}_{\Delta,1}]={}&\frac{(-1)^{n+1} 2^{\Delta-2 \Delta_\phi -2 n+1}\Gamma \left(\frac{\Delta+2}{2}\right)\Gamma \left(-\frac{d}{2}+n+2 \Delta_\phi +1\right) \Gamma \left(\frac{d-4 (n+\Delta_\phi )}{2}\right) }{\Gamma \left(\frac{\Delta-1}{2}\right) \Gamma \left(\frac{\Delta+1}{2}\right)^2 \Gamma \left(n+\Delta_\phi +\frac{3}{2}\right) \Gamma \left(\frac{d-2n-4\Delta_\phi }{2}\right)\Gamma \left(\frac{1+2\Delta_\phi -\Delta}{2}\right)^2 \Gamma \left(\frac{-d+\Delta+1+2\Delta_\phi}{2} \right)^2}\\
{}&\times \frac{\Gamma \left(-\frac{d}{2}+\Delta+1\right) \Gamma (n+\Delta_\phi +1)^3}{(\Delta_\phi +n) \Gamma (n+1)^2 (\Delta-2 \Delta_\phi -2 n-1) (-d+\Delta+2 \Delta_\phi +2 n+1)}\;.
\end{split}
\end{equation}
To obtain the $\beta$ functionals on the crossed channel conformal blocks, we need to use the crossed channel decomposition of the exchange Witten diagrams. These coefficients are related to the anomalous dimensions computed in \cite{Sleight:2018ryu}, and their result is valid down to $\ell=1$. We find for $\ell\geq 1$
\begin{equation}\label{crossedanom}
\beta^{t}_{0,\ell}[G^{s}_{\Delta,1}]=n_{\Delta,1}\sum_{j=0}^2\beta_{\ell,j}\phi_{\ell-j}\left(\frac{d+2}{4},\frac{d+2}{4},\Delta_\phi-\frac{d-2}{4},\Delta_\phi-\frac{d-2}{4},\frac{2\Delta-d}{4},\frac{2\Delta-d}{4}\right)
\end{equation}
where 
\begin{equation}
n_{\Delta,1}=\frac{\pi ^{-\frac{d}{2}-1} 4^{\Delta} \Delta \Gamma \left(\frac{\Delta}{2}\right)^2 \Gamma \left(-\frac{d}{2}+\Delta+1\right)}{(d-\Delta-1) \Gamma (\Delta) \Gamma \left(\frac{\Delta+1}{2}\right)^2 \Gamma \left(-\frac{\Delta}{2}+\Delta_\phi +\frac{1}{2}\right)^2 \Gamma \left(\frac{1}{2} (-d+\Delta+1)+\Delta_\phi \right)^2}\;,
\end{equation}
\begin{equation}
\beta_{\ell,0}=-\frac{\pi ^{d/2} (d+2 \ell ) (\Delta_\phi +\ell ) (2 \Delta_\phi +\ell -1) (2 \Delta_\phi +\ell )}{4 \Delta_\phi ^2 \Gamma \left(\frac{d}{2}+1\right) (2 \Delta_\phi +2 \ell -1)}\;,
\end{equation}
\begin{equation}
\beta_{\ell,1}=\frac{\pi ^{d/2} \ell  (-d+2 \Delta_\phi +2) (2 \Delta_\phi +\ell -1)}{4 \Delta_\phi ^2 \Gamma \left(\frac{d}{2}+1\right)}\;,
\end{equation}
\begin{equation}
\beta_{\ell,2}=\frac{\pi ^{d/2} (\ell -1) \ell  (\Delta_\phi +\ell -1) (-d+4 \Delta_\phi +2 \ell -2)}{4 \Delta_\phi ^2 \Gamma \left(\frac{d}{2}+1\right) (2 \Delta_\phi +2 \ell -1)}\;,
\end{equation}
and
\begin{equation}
\begin{split}
\phi_\ell(a_i)={}&\Gamma(a_1+a_2)\Gamma(a_1+a_3)\Gamma(a_1+a_4)\Gamma(a_2+a_5)\Gamma(a_3+a_5)\Gamma(a_4+a_5)\\
{}&\times \frac{\Gamma(a_2+a_3+\ell)\Gamma(a_2+a_4+\ell)\Gamma(a_3+a_4+\ell)\Gamma(a_6-a_5+\ell+1)}{\Gamma(1-a_5+a_6)}\psi(a;b,c,d,e,f)\;,\\
{}& a=a_1+a_2+a_3+a_4+2a_5+\ell-1\;,\\
{}& b=a_1+a_5\;,\quad c=a_2+a_5\;,\quad d=a_3+a_5\;, e=a_4+a_5\;,\\
{}& f=a_1+a_2+a_3+a_4-a_5-a_6+\ell-1\;,
\end{split}
\end{equation}
is defined in terms of 
\begin{equation}\small
\begin{split}
\psi(a;b,c,d,e,f)={}&\frac{\Gamma(a+1)}{\Gamma(1+a-b)\Gamma(1+a-c)\Gamma(1+a-d)\Gamma(1+a-e)\Gamma(2+2a-b-c-d-e-f)}\\
{}&\times \frac{1}{\Gamma(1+a-f)} {}_7F_6\left(\begin{array}{c} a, 1+\frac{a}{2}, b, c, d, e, f\\ \frac{a}{2}, 1+a-b, 1+a-c, 1+a-d, 1+a-e, 1+a-f \end{array};1\right)\;.
\end{split}
\end{equation}
To further obtain the $\ell=0$ functional $\beta^{t}_{0,0}[G^{s}_{\Delta,1}]$, we can use the recursion relations of \cite{Zhou:2018sfz}. More specifically, the functional action can be written as 
\begin{equation}
 \beta^{t}_{0,0}[G^{s}_{\Delta,1}]=Y+Z
 \end{equation}
where $Z$ comes from the improvement term
\begin{equation}\small
Z=\frac{2^{\Delta-2 \Delta_\phi } \Gamma \left(\frac{\Delta}{2}+1\right) \Gamma (\Delta_\phi )^3 \Gamma \left(-\frac{d}{2}+\Delta+1\right) \Gamma \left(2 \Delta_\phi -\frac{d}{2}\right) (\cos (\pi  (\Delta-2 \Delta_\phi ))+1) \Gamma \left(\frac{\Delta-2 \Delta_\phi +1}{2}\right)^2}{\pi ^2 \Gamma \left(\frac{\Delta-1}{2}\right) \Gamma \left(\frac{\Delta+1}{2}\right)^2 \Gamma \left(\Delta_\phi +\frac{1}{2}\right) \Gamma \left(\frac{1}{2} (-d+\Delta+1)+\Delta_\phi \right)^2}\;,
\end{equation}
and $Y$ comes from $W_{\Delta,J=1}^{s}$ and satisfy the recursion relation\footnote{This recursion relation follows from the equation of motion identity
\begin{equation}
\mathbf{EOM}^{t}[W^{s}_{\Delta,J=1}]=D_{\Delta_\phi \Delta_\phi \Delta_\phi \Delta_\phi}\;,
\end{equation}
and the recursive properties of conformal blocks under the equation of motion operator in the crossed channel.  See Section 2.2  and Section 4 of \cite{Zhou:2018sfz} for more details.} 
\begin{equation}
\begin{split}
{}&(d (-\Delta+\Delta_\phi +1)+\Delta^2-1) Y\\
{}&+(-2d) \beta^{t}_{0,1}[G^{s}_{\Delta,1}]=-\frac{(\Delta-1) \Delta_\phi  2^{\Delta-2 \Delta_\phi } \Gamma \left(\frac{\Delta+2}{2}\right) \Gamma (\Delta_\phi )^3 \Gamma \left(\frac{2+2\Delta-d}{2}\right) \Gamma \left(-\frac{2+4\Delta_\phi-d}{2}\right)}{\pi ^2 \Gamma \left(\frac{\Delta+1}{2}\right)^3 \Gamma \left(\Delta_\phi +\frac{1}{2}\right) }\\
{}&\quad\quad\quad\quad\quad\quad\quad\quad\quad \times  \frac{(\cos (\pi  (\Delta-2 \Delta_\phi ))+1) \Gamma \left(\frac{\Delta-2 \Delta_\phi +1}{2}\right)^2}{\Gamma \left(\frac{-d+\Delta+2 \Delta_\phi +1}{2} \right)^2}\;.
\end{split}
\end{equation}
In obtaining the above $n=0$ functionals, we have used the anomalous dimensions for double-trace operators with the leading twist. By inputing these anomalous dimensions into the recursion relations of \cite{Zhou:2018sfz}, we can efficiently generate the anomalous dimensions of all the double-trace operators with sub-leading twists. We can then assemble them to obtain the functional actions $\beta^{t}_{n,\ell}[G^{s}_{\Delta,1}]$ for $n>0$.

As a special example, let us consider the special case with $d=2$ and $\Delta_\phi=1$. We have
\begin{equation}
\beta^{s}_{0,0}[G^{s}_{\Delta,1}]=-\frac{\Delta (\cos (\pi  \Delta)+1) \Gamma (\Delta-1) \Gamma (\Delta)}{\pi ^2 \Gamma \left(\frac{\Delta+1}{2}\right)^4}\;,
\end{equation}
\begin{equation}
\begin{split}
\beta^{t}_{0,0}[G^{s}_{\Delta,1}]={}&\frac{2^{\Delta-2} \Delta \Gamma (\Delta-1) \Gamma \left(\frac{\Delta}{2}\right)}{\pi ^{5/2} (\Delta-1) \Gamma \left(\frac{\Delta+1}{2}\right)^3} \bigg(2 (3 \Delta-1) (\cos (\pi  \Delta)+1)\\
{}&+(\Delta-1)^3 \cos ^2\left(\frac{\pi  \Delta}{2}\right) \psi ^{(2)}\left(\frac{\Delta-1}{2}\right)\bigg)\;.
\end{split}
\end{equation}
This agrees perfectly with the functionals constructed using kernels.

While  in principle the above procedure can be repeated to obtain the actions of functionals on conformal blocks with any spin $J$, we notice that it is quite cumbersome in practice. The major challenge is to obtain the crossed channel decomposition coefficients for the leading double-trace operators ($n=0$) with $\ell<J$. The closed form expressions of \cite{Sleight:2018epi,Sleight:2018ryu} do not apply to these coefficients, which are sensitive to the details of the contact terms. It is worth studying these low-spin operators in greater detail, and streamlining the calculation. Once the OPE coefficients of the leading double-trace operators are determined, it is straightforward to apply the recursion relations of \cite{Zhou:2018sfz} to obtain the coefficients of all the sub-leading double-trace operators. This will provide enough information to write down the basis of functional actions with arbitrary  $n$ and $\ell$.


\section{The Dispersion Relation of Carmi and Caron-Huot}\label{sec:DispersionRelation}
\subsection{Review of the dispersion relation}
The purpose of this section is to point out a close connection between our approach and the conformal dispersion relation derived recently by Carmi and Caron-Huot \cite{DispersionRelation}.\footnote{We would like to thank the authors of \cite{DispersionRelation} for sharing with us the expression of their dispersion kernel  prior to publication.} From the practical point of view, the connection will provide a closed formula for the generating kernel $\mathcal{K}(z,\zb;w,\wb)$, defined in \eqref{eq:GeneratingKernel}, and thus an independent method to obtain the kernels $\mathcal{A}^{s,t}_{n,\ell}(w,\wb)$, $\mathcal{B}^{s,t}_{n,\ell}(w,\wb)$. As in most of the paper, in this section we will assume the external operators have identical dimensions.

The dispersion relation of \cite{DispersionRelation} reconstructs a physical four-point function $\cG(z,\zb)$ from its double discontinuities. More precisely, there are three independent dispersion relations for any $\cG(z,\zb)$, each corresponding to one choice of an OPE channel (s,t or u). The language of the rest of our paper directly corresponds to the u-channel dispersion relation, but we will stick to explaining the s-channel dispersion relation for now. The s-channel dispersion relation expresses $\cG(z,\zb)$ as an integral of its double discontinuities around the t- and u-channel OPE limits. The relation takes the simplest form if we additionally assume $\cG(z,\zb)$ is super-bounded in the s-channel Regge limit, meaning the s-channel Regge intercept $J^{(s)}_0$ is negative. The dispersion relation states that
\be
\cG(z,\zb) = \cG^{t|s}(z,\zb)+\cG^{u|s}(z,\zb)\,,
\label{eq:DR1}
\ee
where
\ba
(z \zb)^{\Df}\mathcal{G}^{t|s}(z,\zb) &= \int\limits_{0}^{1}\!\frac{dw}{w^2}\int\limits_{0}^{1}\!\frac{d\wb}{\wb^2}\,K^{t|s}(z,\zb;w,\wb)\,\dDisc_t[(w \wb)^{\Df}\mathcal{G}(w,\wb)]\\
(z \zb)^{\Df}\mathcal{G}^{u|s}(z,\zb) &=\!\!\int\limits_{-\infty}^{0}\!\frac{dw}{w^2}\!\!\int\limits_{-\infty}^{0}\!\frac{d\wb}{\wb^2}\,K^{u|s}(z,\zb;w,\wb)\,\dDisc_u[(w \wb)^{\Df}\mathcal{G}(w,\wb)]\,.
\label{eq:DR2}
\ea
The notation stresses that $\cG^{t|s}(z,\zb)$ and $\cG^{u|s}(z,\zb)$ are respectively the contribution of the t- and u-channel double discontinuity to the s-channel dispersion relation. Some factors $(z\zb)^{\Df}$ are present because of our u-channel-friendly definition of $\cG(z,\zb)$, see \eqref{eq:4ptFunctionIntro}. We stress that for the relation to hold, $\cG(z,\zb)$ should be single-valued in the entire Euclidean plane. Indeed, the derivation of \eqref{eq:DR1}+\eqref{eq:DR2} in \cite{DispersionRelation} starts from the Lorentzian inversion formula, which requires $\cG(z,\zb)$ to be Euclidean single-valued. Equation \eqref{eq:DR1}+\eqref{eq:DR2} is a completely universal formula of two-variable complex analysis applying to all s-channel super-bounded and Euclidean single-valued functions $\cG(z,\zb)$. In particular, $K^{t|s}(z,\zb;w,\wb)$ and $K^{u|s}(z,\zb;w,\wb)$ are universal kernels independent of the external dimension $\Df$ and spacetime dimension $d$. The authors of \cite{DispersionRelation} found the following formulas for the kernels. Firstly, the symmetry under switching operators 1 and 2 determines $K^{u|s}(z,\zb;w,\wb)$ in terms of $K^{t|s}(z,\zb;w,\wb)$
\be
K^{u|s}(z,\zb;w,\wb) = K^{t|s}\!\left(\mbox{$\frac{z}{z-1}$},\mbox{$\frac{\zb}{\zb-1}$};\mbox{$\frac{w}{w-1}$},\mbox{$\frac{\wb}{\wb-1}$}\right)\,.
\ee
To write down $K^{t|s}(z,\zb;w,\wb)$, let us recall the $\rho$ variable \eqref{eq:rho} and let us use the more compact notation $\rho_z=\rho(z)$. $K^{t|s}(z,\zb;w,\wb)$ decomposes as a sum of two terms
\be
K^{t|s}(z,\zb;w,\wb) =
\mathcal{P}^{t|s}(z,\zb;w,\wb)\delta(\rho_z\rho_{\zb}\rho_{\wb}-\rho_w)+
\mathcal{Q}^{t|s}(z,\zb;w,\wb)\theta(\rho_z\rho_{\zb}\rho_{\wb}-\rho_w)\,,
\label{eq:kTS}
\ee
where $\delta$ is the Dirac delta distribution and $\theta$ is the Heaviside step function. Then \cite{DispersionRelation}
\ba
\mathcal{P}^{t|s}(z,\zb;w,\wb) &= \frac{16 \rho_{w} \rho_{\wb} \rho_{z} \rho_{\zb} \left(1-\rho_{w}\rho_{\wb}\right)}{\pi  (1-\rho_{\wb} \rho_{z}) (1-\rho_{\wb} \rho_{\zb})\sqrt{ \left(1-\rho_{w}^2\right)\left(1-\rho_{\wb}^2\right) \left(1-\rho_{z}^2\right) \left(1-\rho_{\zb}^2\right)}}\\
\mathcal{Q}^{t|s}(z,\zb;w,\wb) &=-\frac{1}{64\pi}\frac{\wb-w}{w\wb}\frac{\left(\frac{1}{w}+\frac{1}{\wb}+\frac{1}{z}+\frac{1}{\zb}-2\right)(w \wb z \zb)^{\frac{3}{2}}}{[(1-w) (1-\wb) (1-z) (1-\zb)]^{\frac{3}{4}}}\times\\
&\quad\,\times x(z,\zb;w,\wb)^{\frac{3}{2}}{}_2F_1\left(\frac{1}{2},\frac{3}{2};2;1-x(z,\zb;w,\wb)\right)\,,
\label{eq:dispersionKernel2F1}
\ea
where one introduces the combination
\be
x(z,\zb;w,\wb)= \frac{\rho_{w}\rho_{\wb} \rho_{z} \rho_{\zb} \left(1-\rho_{w}^2\right)\left(1-\rho_{\wb}^2\right) \left(1-\rho_{z}^2\right)\left(1-\rho_{\zb}^2\right)}{(\rho_{\wb} \rho_{\zb}-\rho_{w} \rho_{z}) (\rho_{\wb} \rho_{z}-\rho_{w} \rho_{\zb}) (\rho_{z} \rho_{\zb}-\rho_{w} \rho_{\wb}) (1-\rho_{w} \rho_{\wb} \rho_{z} \rho_{\zb})}\,.
\ee

\subsection{Relation to our work}
Having described the s-channel dispersion relation of Carmi and Caron-Huot, let us start making the connection to our results. We start by converting \eqref{eq:DR1}+\eqref{eq:DR2} to a u-channel dispersion relation. Suppose that $\cG(z,\zb)$ is Euclidean single-valued and $\cG(z,\zb)\in\mathcal{U}$, thus in particular super-bounded in the u-channel Regge limit. The cyclic permutation between channels $s\rightarrow u\rightarrow t\rightarrow s$ is implemented by $z\mapsto(z-1)/z$ with inverse $z\mapsto 1/(1-z)$. It follows the u-channel dispersion relation takes the form
\be
\cG(z,\zb) = \cG^{t|u}(z,\zb)+\cG^{s|u}(z,\zb)\,,
\label{eq:DR3}
\ee
where
\ba
\mathcal{G}^{t|u}(z,\zb) &=\!\int\limits_{1}^{\infty}\! dw\!\!\int\limits_{1}^{\infty}\!\! d\wb\,K^{t|u}(z,\zb;w,\wb)\,\dDisc_t[\mathcal{G}(w,\wb)]\\\
\mathcal{G}^{s|u}(z,\zb) &= \!\!\int\limits_{-\infty}^{0}\!\! dw\!\!\!\int\limits_{-\infty}^{0}\!\!\! d\wb\,K^{t|u}(1-z,1-\zb;1-w,1-\wb)\,\dDisc_s[\mathcal{G}(w,\wb)]\,.
\label{eq:DR4}
\ea
The kernel is
\be
K^{t|u}(z,\zb;w,\wb) = K^{t|s}\left(\mbox{$\frac{1}{z}$},\mbox{$\frac{1}{\zb}$};\mbox{$\frac{1}{w}$},\mbox{$\frac{1}{\wb}$}\right)\,.
\label{eq:kSU}
\ee
Equations \eqref{eq:DR3} and \eqref{eq:DR4} are reminiscent of our logic. Indeed, we expect that for Euclidean single-valued functions, the decomposition \eqref{eq:DR3} is the same as \eqref{eq:Gsplit}. However, we expect that \eqref{eq:Gsplit} should hold also for functions in $\mathcal{U}$ which fail to be Euclidean single-valued.

To make the connection more obvious, recall the definitions of the generating functional $\Omega_{z,\zb}$ \eqref{eq:GenFun} and its generating kernel $\mathcal{K}(z,\zb;w,\wb)$ \eqref{eq:GeneratingKernel}. Suppose that the sum over $n,\ell$ in \eqref{eq:GeneratingKernel} converges for some $z,\zb\in\mathbb{C}$, and all $w\in C_-$, $\wb\in C_+$. Then we can write the decomposition \eqref{eq:Gsplit} as
\be
\cG(z,\zb) = \cG^{t}(z,\zb) + \cG^{s}(z,\zb)
\ee
with
\ba
\cG^{t}(z,\zb) &= \int\limits_{C_-}\!\!\frac{dw}{2\pi i} \int\limits_{C_+}\!\!\frac{d\wb}{2\pi i} \,
\mathcal{K}(z,\zb;w,\wb)\, \cG(w,\wb)\\
\cG^{s}(z,\zb) &=\int\limits_{C_-}\!\!\frac{dw}{2\pi i} \int\limits_{C_+}\!\!\frac{d\wb}{2\pi i} \,
\mathcal{K}(1-z,1-\zb;1-w,1-\wb)\, \cG(w,\wb)\,.
\label{eq:DR5}
\ea
In order to relate the first line of \eqref{eq:DR5} to the first line of the u-channel dispersion relation \eqref{eq:DR4}, we should deform both $C_-$ and $C_+$ contours to the right and wrap the right branch cut with both. Assuming we encounter no singularities of $\mathcal{K}(z,\zb;w,\wb)$ in the process, we find the squared discontinuity $\Disc_t\overline{\Disc}_t[\cG(w,\wb)]$, which equals $-2\,\dDisc_t[\cG(w,\wb)]$ if $\cG(w,\wb)$ is Euclidean single-valued (see Section \ref{ssec:dDisc}). Thus, heuristically
\be
\cG^{t}(z,\zb) \stackrel{?}{=} \frac{1}{2\pi^2} \!\!\int\limits_{-\infty}^{0}\!\! dw\!\!\!\int\limits_{-\infty}^{0}\!\!\! d\wb\,\mathcal{K}(z,\zb;w,\wb)\,\dDisc_t[\mathcal{G}(w,\wb)]\,.
\label{eq:KKrelation1}
\ee
So $\mathcal{K}(z,\zb;w,\wb)$ should be related to $K^{t|u}(z,\zb;w,\wb)$ from \eqref{eq:DR4}. To test the relation, recall that we can expand $\mathcal{K}(z,\zb;w,\wb)$ at small $z,\zb$ and the coefficients will be functional kernels, many of which were computed in Sections \ref{ssec:generalBeta}, \ref{ssec:generalAlpha}. Let us then expand $K^{t|u}(z,\zb;w,\wb)$, which is given by \eqref{eq:kTS} and \eqref{eq:kSU}, at small $z,\zb$. We will ignore the contact term in \eqref{eq:kTS} proportional to the delta function and focus on the ``bulk'' term proportional to $\theta\left(\rho_{\frac{1}{z}}\rho_{\frac{1}{\zb}}\rho_{\frac{1}{\wb}}-\rho_{\frac{1}{w}}\right)$. In the limit of small $z,\zb$, the step function becomes
\be
\theta\left(\rho_{\frac{1}{\wb}}-\rho_{\frac{1}{w}}\right) = \theta(w-\wb)\,.
\ee
In other words, the integration on the first line of \eqref{eq:DR4} runs over half of the domain of integration in \eqref{eq:KKrelation1}. This means we should identify
\be
\mathcal{K}(z,\zb;w,\wb) = \pi^2\mathcal{Q}^{t|s}\left(\mbox{$\frac{1}{z}$},\mbox{$\frac{1}{\zb}$};\mbox{$\frac{1}{w}$},\mbox{$\frac{1}{\wb}$}\right)
\label{eq:identification}
\ee
with $\mathcal{Q}^{t|s}(z,\zb;w,\wb)$ given in \eqref{eq:dispersionKernel2F1}. This relation can be tested very precisely. Let us set $z=r e^{i\theta}$, $\zb=r e^{-i\theta}$ and expand the RHS as $r\rightarrow 0$. We will focus on the leading terms. Note that
\be
x\left(\mbox{$\frac{1}{r e^{i\theta}}$},\mbox{$\frac{1}{r e^{-i\theta}}$};\mbox{$\frac{1}{w}$},\mbox{$\frac{1}{\wb}$}\right) = 16\sqrt{\frac{w(1-w)\wb(1-\wb)}{(\wb-w)^4}}r + O(r^2)
\ee
as $r\rightarrow 0$ and
\be
{}_2F_1\left(\frac{1}{2},\frac{3}{2};2;1-y\right) = -\frac{2}{\pi} \left[\log \left(\frac{y}{16}\right)+2\right] + O(y\log y)
\ee
as $y\rightarrow 0$. Therefore, if we pick the correct branch of the square root, we find
\ba
\pi^2\mathcal{Q}^{t|s}&\left(\mbox{$\frac{1}{r e^{i\theta}}$},\mbox{$\frac{1}{r e^{-i\theta}}$};\mbox{$\frac{1}{w}$},\mbox{$\frac{1}{\wb}$}\right) = \frac{2(w+\wb -2)}{(\wb-w)^2}\log(r) + \\
&\quad\,+\frac{(w+\wb -2)}{(w-\wb )^2}\left\{\log\!\left[\frac{w(1-w)\wb(1-\wb)}{(\wb-w)^4}\right]+4\right\}+O(r\log r)\,.
\ea
This agrees precisely with the formula
\be
\pi^2\mathcal{Q}^{t|s}\left(\mbox{$\frac{1}{r e^{i\theta}}$},\mbox{$\frac{1}{r e^{-i\theta}}$};\mbox{$\frac{1}{w}$},\mbox{$\frac{1}{\wb}$}\right) = \widehat{\mathcal{B}}^{s}_{0,0}(w,\wb)\log(r) + \widehat{\mathcal{A}}^{s}_{0,0}(w,\wb) + O(r\log r)
\ee
coming from the identification \eqref{eq:identification}. We used \eqref{eq:BHat001} and \eqref{eq:AHat001} for the functional kernels. We have expanded \eqref{eq:identification} up to $O(r^5)$ in Mathematica and checked that it agrees perfectly with the functional kernels $\widehat{\mathcal{B}}^{s}_{i,j}(w,\wb)$ that can be constructed using the algorithm of Section \ref{ssec:generalBeta}. Conversely, taking the relation \eqref{eq:identification} for granted yields a very efficient method of finding the functional kernels.

We provided ample evidence that there is a close connection between \cite{DispersionRelation} and our work. However, our arguments were quite heuristic and it will be important to make them precise. Our logic suggests the validity of a dispersion relation with a non-standard contour \eqref{eq:DR5} which should apply to all functions in $\mathcal{U}$. It will be important to establish \eqref{eq:DR5} directly using a contour manipulation and relate it rigorously to the dispersion relation of \cite{DispersionRelation}.


\section{Discussion}\label{sec:Discussion}
In this work we have developed a new analytic approach to the  conformal bootstrap, valid for general spacetime dimension. We have proposed a new basis for CFT four-point functions, and given algorithms to construct the dual basis of analytic functionals. These functionals are intimately related to the Polyakov-Regge blocks, which are exchange Witten diagrams with improved Regge behavior. We have also explained the relation between our analytic functionals and the recent discovered CFT dispersion formula \cite{DispersionRelation}.
Our work opens up a number of natural research avenues:
\begin{itemize}
\item Applying the functionals to four-point correlators leads to sum rules which place strong constraints on  CFT data.
 It will be very  interesting to study these constraints in concrete models.  
 A first class of applications will be to to perturbative expansions around mean field theory, such 
as   the Wilson-Fisher fixed point in $4-\epsilon$ dimensions \cite{Alday:2016jfr,Rychkov:2015naa,Basu:2015gpa,Gopakumar:2016wkt,Gopakumar:2016cpb,Gliozzi:2016ysv,Dey:2016mcs,Roumpedakis:2016qcg,Liendo:2017wsn,Dey:2017fab,Gliozzi:2017gzh,Gopakumar:2018xqi,Alday:2017zzv} or $\mathcal{N}=4$ super Yang-Mills theory in the $1/N$ expansion \cite{Rastelli:2016nze,Alday:2017xua,Aprile:2017bgs,Aprile:2017xsp,Rastelli:2017udc,Aprile:2017qoy,Alday:2017vkk,Aprile:2018efk,Caron-Huot:2018kta,Alday:2018kkw,Alday:2018pdi,Binder:2019jwn,Goncalves:2019znr,Drummond:2019odu}.
These cases can already be efficiently handled with the LIF, but
our functionals may provide a more systematic way to proceed to higher orders. A second and more interesting class of applications will be to derive fully non-perturbative sum rules.
In this context it will be important to achieve a detailed understanding of the positivity properties of our functionals.  As an example, we can envision addressing the long-standing question of whether  unitary CFTs exist above six dimensions. Superconformal field theories offer another prime target for non-perturbative studies, as they enjoy improved Regge behavior.
 
\item As we pointed out, the sum rules can alternatively be obtained  by expanding the four-point function in terms of the s- and t-channel Polyakov-Regge blocks, and demanding that double-trace conformal blocks should vanish. This is reminiscent of the Polyakov-Mellin bootstrap. Note however that the Regge behavior played a central role in our construction, and each Polyakov-Regge block is bounded in the u-channel Regge limit. This ensures the Regge boundedness of the correlator, but at the cost of losing manifest crossing symmetry. This should be contrasted with the Polyakov-Mellin bootstrap \cite{Gopakumar:2016wkt,Gopakumar:2016cpb,Gopakumar:2018xqi} where crossing is automatic but Regge boundedness is not. It would be interesting to further investigate the relation between the two approaches.  

\item We presented two complementary methods to obtain the analytic functionals. However, writing the entire basis in a closed form still presents a great technical challenge for both methods. It would be extremely rewarding to streamline the computations and obtain the functionals more efficiently.  Among other things, this would allow to use our basis of analytic functionals  in the numerical bootstrap, as has been done in the $d=1$ case \cite{Paulos:2019fkw} with promising results.  

\item In this work, we mainly focused on four-point functions with equal conformal dimensions. The same logic applies to the case of unequal dimensions, as we have briefly discussed in 
Section \ref{Pgenerald}. The details of the general case should be fleshed out in the future, and it will allow us to consider the bootstrap problem of mixed correlators. A natural next step of our endeavor is to include operators with spins. This will further extend the applicability of our analytic method. 

\item A different functional approach to higher dimensional CFT was explored in a very recent paper \cite{Paulos:2019gtx}. An interesting class of functionals was introduced for $d=2$ by tensoring holomorphic and anti-holomorphic copies of $d=1$ functionals. These $d=2$ functionals were used to prove optimality  of the energy correlator in the two-dimensional Ising model.
Unfortunately this construction does not generalize to $d>2$.  A different class of functionals in any $d$ was introduced: it acts on the crossing antisymmetric vectors and has simple action on generalized free fields.  However, such functionals do not satisfy the orthonormality condition (\ref{eq:dualityC1}, \ref{eq:dualityC2}). It appears that our results are largely orthogonal to the results of \cite{Paulos:2019gtx}, but it would be interesting to understand the connection in more detail.

\end{itemize}

\acknowledgments
We thank Dean Carmi, Simon Caron-Huot, Miguel Paulos, Eric Perlmutter, Charlotte Sleight, David Simmons-Duffin and Massimo Taronna for helpful discussions at various stages of this work. We are especially indebted to Dean Carmi and Simon Caron-Huot for sharing with us the expression of their dispersion kernel before publication, and for graciously agreeing to coordinate submission to the arXiv. This work is supported in part by the Simons Foundation, as part of the Simons Collaboration on the Non-perturbative Bootstrap. The work of L.R.  is supported in part by NSF grant \# PHY1620628. The work of X.Z. is supported in part by the Simons Foundation Grant No. 488653.

\appendix

\bibliography{HigherDFunctionals} 
\bibliographystyle{utphys}

\end{document}